\documentclass{aa}  

\usepackage{graphicx}
\usepackage{txfonts}
\usepackage{xcolor}
\usepackage{subfig}
\usepackage{verbatim}
\usepackage{float}

\begin{document} 

%%%%%%%%%%%%%%%%%%%%%%%%%%%%%%%%%%%%%%%%
%\usepackage[options]{hyperref}
% To add links in your PDF file, use the package "hyperref"
% with options according to your LaTeX or PDFLaTeX drivers.
%

%\addbibresource{bibliography.bib}

\title{Persistent time lags in light curves of Sagittarius A*: evidence of outflow}

%\subtitle{Subtitle here}

\author{Christiaan Brinkerink\inst{1}
\and
Heino Falcke\inst{1}
\and
Andreas Brunthaler\inst{2}
\and 
Casey Law\inst{3,4}}
\institute{Department of Astrophysics, Radboud University, Heyendaalseweg 135, 6525 AJ Nijmegen, the Netherlands
\and
Max Planck Institut für Radioastronomie, Auf dem H\"ugel 69, D-53121 Bonn, Germany
\and
Cahill Center for Astronomy and Astrophysics, MC 249-17 California Institute of Technology, Pasadena, CA 91125, USA
\and
Owens Valley Radio Observatory, California Institute of Technology, Pasadena, CA 91125, USA}

\date{\today}

% \abstract{}{}{}{}{} 
% 5 {} token are mandatory

 \abstract
   {The compact radio source at the center of our Galaxy, Sagittarius\,A* (Sgr\,A*), is the subject of intensive study as it provides a close-up view of an accreting supermassive black hole. Sgr\,A* provides us with a prototype of a low-luminosity active galactic nucleus (LLAGN), but interstellar scattering and the resolution limits of our instruments have limited our understanding of the emission sites in its inner accretion flow.}
   {The temporal variability of Sgr\,A* can help us understand whether we see a plasma outflow or inflow in the region close to the black hole. In this work, we look at a comprehensive set of multi-epoch data recorded with the Karl G. Jansky Very Large Array (VLA) to understand the persistence of the time lag relations that have been found in previous radio observations of Sgr\,A*.}
   {We analyse 8 epochs of data, observed in Spring 2015, each of which has a frequency coverage from 18 to 48 GHz. We cross-correlate the calibrated light curves across twelve frequency subbands. We also generate synthetic data with the appropriate variability characteristics and use it to study the detectability of time lag relations in data with this sampling structure.}
   {We find that the variability amplitude increases with frequency. We see positive time lag slopes across all subbands in five out of eight epochs, with the largest slopes in the cases where a clear extremum in flux density is present. Three epochs show lag slopes close to zero. With the synthetic data analysis we show that these results are explained by a persistent lag relation of $\sim$40 min/cm that covers the bulk of the variability, with at most 2\% of the total flux density in an uncorrelated variability component. Together with the size-frequency relation and inclination constraints this indicates an outflow velocity with $\gamma \beta = 1.5$, consistent with predictions of jet models for Sgr\,A*.}
   {}

\keywords{black hole -- accretion -- interferometry -- AGN -- variability -- radio}

\titlerunning{Persistent time lags in Sgr\,A*}

\maketitle

\section{Introduction}

There is significant evidence and support for the notion that the compact radio source at the center of our Galaxy (Sagittarius\,A*, abbreviated as Sgr\,A*, see \citet{BalickBrown1974}) is associated with plasma flow around a supermassive black hole of approximately 4 million solar masses. Analysis of the orbits of short-period bright stars shows that the central mass is highly concentrated and coincident with the region from which we see the radio emission \citep{ReidBrunthaler2004, Ghez2008, Gillessen2009, GRAVITY2019}. In this introduction we will focus on the different observational aspects of this radio source and discuss the system properties that have been derived from them.

\subsection{Observed spectrum}

Sgr\,A* has a spectral energy distribution (SED) that shows a rising power-law across the radio spectrum, steepening into a submm-bump that starts above $\sim$50 GHz and peaks in the 1 - 2 THz range \citep{Zylka1992, Serabyn1997, Falcke1998, Zhao2003, Bower2019}, before rapidly dropping into the infrared regime. The general shape of this spectrum suggests that we see partially self-obscured synchrotron emission, optically thick at lower frequencies and turning over to optically thin emission at higher frequencies.

The observed spectral shape and flux density of the submm bump allow us to solve for several properties of the inner plasma flow: the electron number density, the magnetic field strength and the temperature of the plasma \citep{Falcke1993, Narayan1995, Falcke1996b, Yuan2003, Goldston2005, Moscibrodzka2009}. These results show that the emission from the submm bump must come from the innermost part of the plasma flow: the lower-frequency part of the spectrum is self-absorbed, indicating that we are seeing into a partially transparent photosphere layer of an otherwise opaque region. This frequency-dependent photosphere shrinks inwards as we consider higher observing frequencies, until we reach the turnover point in the spectrum: there, the photosphere disappears altogether and we see the region closest to the black hole. The fact that the submm spectrum of Sgr\,A* shows this structure means that Sgr\,A* is also a prime candidate for very long baseline interferometry (VLBI) observations of its shadow by the Event Horizon Telescope \citep{EHT2019II}: the attainable angular resolution for a worldwide network of antennas observing at 230\,GHz allows it to resolve the expected scale of the black hole shadow in the case of Sgr\,A* \citep{Falcke2000}. A fundamental open question in this context is: what is the emitting plasma in this inner region doing? Is it part of an inflow or an outflow?

\subsection{Observed morphology}

Studies of the morphology of Sgr\,A* with Very Long Baseline Interferometry (VLBI) at wavelengths from 20\,cm down to 1.3\,mm indicate an elliptical (Gaussian) shape for the source and show that the apparent source size is dominated by interstellar scattering effects at wavelengths longer than $\sim$3\,mm \citep{Langevelde1991, Lo1998, Bower2006}, making it scale according to $\lambda^2$ for wavelengths in that range. At shorter wavelengths, the observed size of Sgr\,A* deviates from this relation: it is larger than the $\lambda^2$ relation predicts \citep{Bower2004, Shen2005, Doeleman2008, OrtizLeon2016, CDB2019}, with the fractional difference becoming more pronounced at progressively shorter wavelengths. This is understood to be a manifestation of the intrinsic source geometry which becomes more clearly visible as the influence from scattering loses its dominance at higher frequencies. Although the measured size of Sgr\,A* at 86\,GHz is larger than predicted from a pure scattering size relation constructed using measurements at lower frequencies, the deviation of its morphology from Gaussianity is modest with only $\sim$1\% of the observed VLBI flux density not matching an elliptical Gaussian brightness distribution \citep{CDB2019}. While the scattered source geometry approximates an elliptical Gaussian very closely at longer wavelengths, substructure at the sub-percent level has also been seen at 1.3\,cm (23 GHz) as reported by \citet{Gwinn2014}.

The observed non-Gaussian substructure manifests as an asymmetry that may either be intrinsic in origin, caused by the effects of interstellar scattering or a combination thereof \citep{OrtizLeon2016, CDB2016}. Observations on the degree to which this observed source asymmetry persists over longer time scales ($\sim$years) should resolve this origin, as the time scale over which the influence from the scattering screen evolves is relatively short ($\sim$weeks).

Algorithmically, disentangling the scattering effects from the contribution of the intrinsic source structure is a challenging task, although in recent years significant progress has been booked in this endeavour \citep{Johnson2015, Johnson2016,Johnson2018,Issaoun2019} which allows for partial reconstruction of intrinsic source geometry from measurements of the scattered source image.

Given modern VLBI capabilities, the argument may be raised that jet morphology should therefore be readily apparent when looking at Sgr\,A* at radio-to-mm wavelengths, and in fact observations at 7\,mm have indeed suggested this \citep{Lo1998}. From the theoretical side, it has been shown that the morphology of a jet outflow of Sgr\,A* may be such that it would appear as a highly compact source, within the size constraints dictated by these VLBI measurements \citep{Markoff2007}. More recent VLBI measurements at 86\,GHz, which suffer less from the effects of interstellar scattering, can still be fitted with models from both jet- and disk-dominated classes \citep{Moscibrodzka2017,Issaoun2019}.

\subsection{Time-domain studies}

Besides the spectral and spatial dimensions, the emission from Sgr\,A* has been studied extensively in the time domain as well. Of particular interest are the different states of activity that Sgr\,A* can exhibit, and the potential temporal correlations between the light curves of Sgr\,A* in different parts of the electromagnetic spectrum.

Analysis by \citet{Falcke1999} showed that a characteristic variability timescale is present at a scale of multiple tens of days, with a suggestion of quasiperiodic behaviour at 57 days. \citet{Herrnstein2004} presented flux density measurements of Sgr\,A* spread out over multiple years at wavelengths of 2.0, 1.3 and 0.7 cm (15, 22 and 43 GHz), and reported tentative evidence for a bi-modal activity pattern. However, the time cadence of those measurements (8 days) was such that the shorter potential variability timescales, at which significant variability power might be concentrated, were not accessible. The short time windows used for each of the observations of Sgr\,A* may even mean that the observed variability over the 8-day cadence is aliased from much shorter timescales. Indeed, \citet{Dexter2014} identified a characteristic timescale of approximately 8 hours in the variability of Sgr\,A* at high observing frequencies (230\,GHz and above). Below this timescale, Sgr\,A* exhibits a variability power spectrum that looks like red noise, with flux density measurements separated closely in time being more strongly correlated than those with a larger time difference between them. Beyond this 8-hour timescale, that analysis shows that the variability has a flat power spectrum and thus indicates no systematic correlation between flux density variations regardless of their separation in time. More recently, a study of the variability of Sgr\,A* in infra-red has shown a coherence time of only 4 hours \citep{Witzel2018} - possibly because of the shorter electron cooling timescales at the associated electron energies and in the relevant emission regions.

The first study reporting the detection of time lags between Sgr\,A* light curves was \citet{YusefZadeh2006}, in which flares at 43 GHz (0.7 cm) were followed by flares at 22 GHz (1.3 cm) approximately 20 to 40 minutes later. This time lag was interpreted in the context of expanding plasma blobs as described by \citet{vdLaan1966}, where the peak of the flare emission shifts to lower frequencies as the plasma blob expands and its optical depth changes. A different interpretation, ascribing the variations in flux density to emission from a compact jet, was put forward by \citet{Falcke2009}, where the frequency-dependent intrinsic size of Sgr\,A* that was derived from VLBI measurements was coupled to the measured time lag to provide support for a compact jet. In that model, the plasma semi-adiabatically expands as it is accelerated along the jet axis. This scenario is also compatible with theoretical models for the outflow from a low-Eddington accreting system.

In a previous paper, we reported seeing a time lag in single-epoch VLA data of Sgr\,A* in which a minor flare was visible, spanning 7 different frequency bands from 100 down to 19 GHz \citep{CDB2015}. The time lag relation was fitted with a linear trend in the wavelength domain, which yielded a slope of $42 \pm 14$ min/cm. Combining this time lag relation with an expression for the intrinsic size of the source and the well-constrained distance to Sgr\,A*, we found that if the variability is indeed associated with an outflow it suggests mildly relativistic outflows with a Lorentz factor of $\sim$2.

Measurements reported by \citet{Miyazaki2013} at higher frequencies (102 and 90 GHz) did not show evidence for time lags between the two light curves. The apparent absence of a non-zero lag in this data set was theorised to either follow from the dynamics of the expanding plasma (the plasma blob might start out optically thin already, therefore not exhibiting a clear peak in its variability at any of the observing frequencies), or alternatively the variability seen in the light curves might be due to some other process besides plasma expansion -- for instance in the form of orbiting hot spots in the accretion disk.

The detection of circular motion in IR flares by GRAVITY \citep{GRAVITY2018} provided a tantalising glimpse into the dynamics of gas close to the black hole (within $\sim$10\,$R_g$), with the emission centroid describing an approximately circular motion on the sky on time scales of 30 minutes to an hour for several observed flares and polarisation EVPAs that also rotate on the sky in tandem with the observed centroid motion. Several models have been put forward to explain these observations, from orbiting plasmoids \citep{Ball2020, GRAVITY2020} to pattern motion of shifting MHD disturbances \citep{Matsumoto2020}. The origins of variability in IR and in radio are expected to be different: for IR, the electron energy is thought to be the most relevant parameter (the IR being optically thin emission for Sgr\,A*), while for radio (optically thick emission) the bulk electron population that is present in different regions of the accretion flow is thought to be the most important factor \citep{Eckart2018}. With the photosphere located at larger radii for radio wavelengths, different regions are expected to feature in IR versus radio flares as well. A tentative connection between IR and radio flaring behaviour has been made \citep{Rauch2017}, although no physically consistent model has yet been framed to connect the two regimes.

\subsection{Questions addressed in this work}

The open questions we wish to address here focus on the statistics of time lags in radio observations of Sagittarius\,A*: how consistent is the measured time lag relation, how much does it vary between epochs? Does it ever reverse sign? Does it correlate with some other property of the source state? Do the source variability statistics work out in such a way that a time lag should be observed for all epochs?\\
In this paper, we present our findings from multiple epochs of data recorded with the VLA in the spring of 2015. We describe the process with which the data was recorded and calibrated in Section 2, and we describe the components of our analysis in Section 3: it contains a discussion on the methods we have used to establish the light curves and their cross-correlations, as well as our synthetic data analysis with which we verify the robustness of our measurements. In Section 4, we present the time lag relations we measure. We connect our results to the broader theoretical context in Section 5.

\section{Observations}

\subsection{Observation epochs, array configuration and spectral setup}

The observations were carried out using the VLA (project code 15A-372, PI: C. Brinkerink). Nine observing blocks were executed between March 15th, 2015 and May 1st, 2015 (see Table \ref{tab:obsblocktimes}). For all of these observations the VLA was in A configuration, with baseline lengths of up to 36.4 km. Although a total of 27 hours of observing time was allocated for this project, schedule planning constraints dictated a fixed scan length and structure so that they could be scheduled as short ($\sim$3-hour), independent observing blocks that could be executed whenever the opportunity arose. All of the observing blocks thus follow the same sequence of configuration steps and scans. For each observing block, we include flux and bandpass calibration observations on standard VLA calibrator 3C286, followed by rapid cycled pointings toward Sgr\,A* (science source), J1744-3116 (gain calibrator) and J1745-283 (check source). Three VLA bands are observed (K, Ka and Q) in LL and RR polarisations, each with 8.192 GHz sky bandwidth using the 3-bit sampling mode for the WIDAR correlator. Each band is covered by 64 contiguous spectral windows (SPWs), where each SPW is 128 MHz wide and contains 128 channels with a bandwidth of 1 MHz each. For each of our three band tunings, we cycle through our three sources where we dwell on each source for 30 seconds (see Table \ref{tab:scanblock} for the structure of one such scan block). Using this scheme, which is repeated 15 times, we get light curves for Sgr\,A* that have a scan cadence of 6.5 minutes for each band. Pointing scans on source NRAO530 are included in each observing block with a 1-hour cadence between these scan blocks. This scan setup gives us the necessary temporal resolution to test for the presence of time lags that follow the relation of $42 \pm 14 \textrm{min}/\textrm{cm}$ as found in \citet{CDB2015}, but also for other time lag relations (see the synthetic data section for details).

\begin{table}
\caption{Structure of one scan block}
\label{tab:scanblock}
\begin{tabular}{l l l}
\hline
Target & Duration & Description\\
\hline
J1744-3116 & 30s & Setup K-band receiver\\
J1744-3116 & 30s & K-band gain calibration scan\\
Sgr\,A*    & 35s & K-band science scan\\
J1745-283  & 35s & K-band check source scan\\
J1745-283  & 30s & Setup Ka-band receiver\\
J1745-283  & 30s & Ka-band check source scan\\
Sgr\,A*    & 35s & Ka-band science scan\\
J1744-3116 & 35s & Ka-band gain calibration scan\\
J1744-3116 & 30s & Setup Q-band receiver\\
J1744-3116 & 30s & Q-band gain calibration scan\\
Sgr\,A*    & 35s & Q-band science scan\\
J1745-283  & 35s & Q-band check source scan\\
\end{tabular}
\end{table}

\begin{table}
\caption{Dates and times of the observations used for this work}
\label{tab:obsblocktimes}
\setlength{\tabcolsep}{3pt}
\begin{tabular}{l|l|l|l}
\hline
No.&Date & Total time range & Time range on Sgr\,A* \\
\hline
1&15 Mar 2015 & 11:06:04 - 14:05:30 & 12:06:45 - 14:04:06 \\
2&30 Mar 2015 & 08:55:45 - 11:55:12 & 09:56:30 - 11:53:51 \\
3&30 Mar 2015 & 11:55:22 - 14:54:45 & 12:56:00 - 14:53:21 \\
4&10 Apr 2015 & 08:12:46 - 11:12:09 & 09:13:27 - 11:10:48 \\
5&11 Apr 2015 & 08:08:36 - 11:08:03 & 09:09:21 - 11:06:42 \\
6&11 Apr 2015 & 11:08:11 - 14:07:36 & 12:08:51 - 14:06:12 \\
\color{gray}7&\color{gray}30 Apr 2015 & \color{gray}06:53:52 - 09:53:21 & \color{gray}07:54:36 - 09:51:57 \\
8&30 Apr 2015 & 09:53:25 - 12:52:51 & 10:54:06 - 12:51:30 \\
9& 1 May 2015 & 09:28:56 - 12:28:21 & 10:29:36 - 12:27:00 \\
\end{tabular}
\end{table}

\subsection{Data calibration and reduction}

\begin{figure*}[h]
    \centering
    \includegraphics[width=0.33\textwidth]{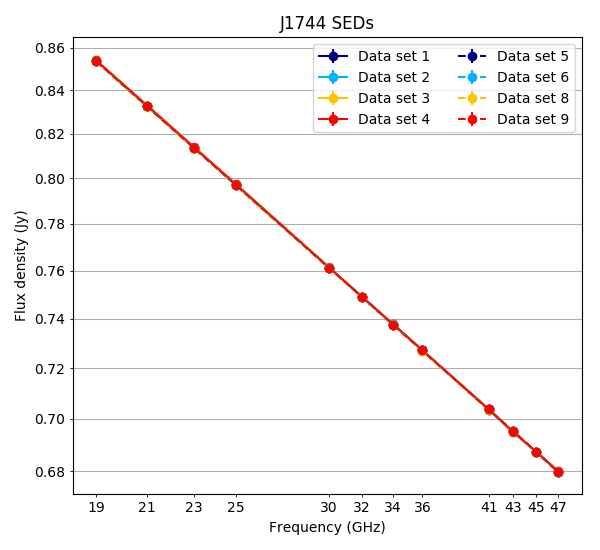}
    \includegraphics[width=0.33\textwidth]{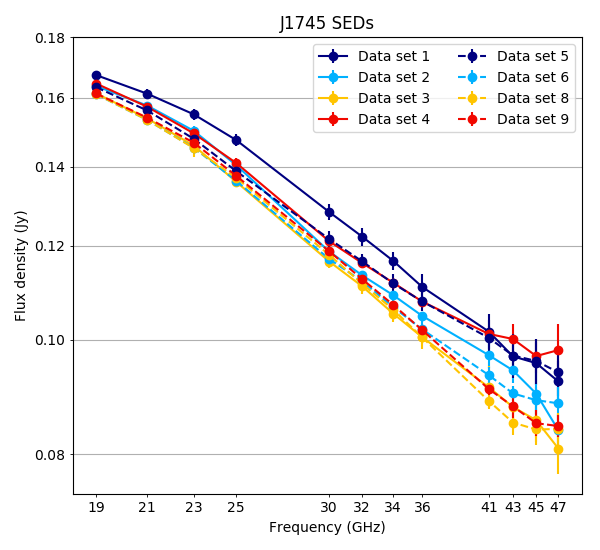}
    \includegraphics[width=0.33\textwidth]{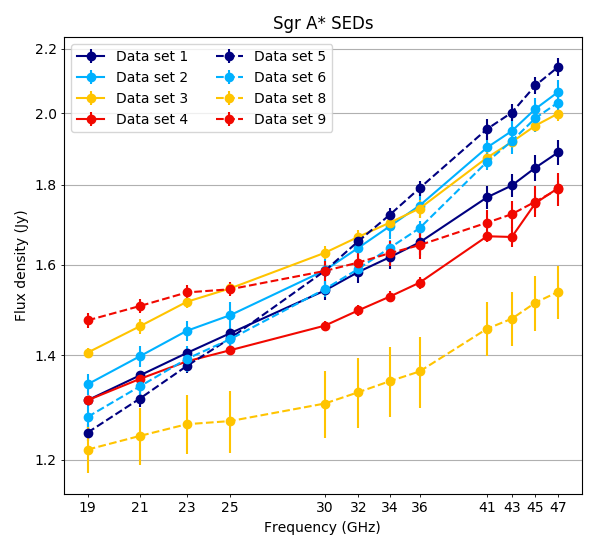}
    \caption{Spectral Energy Distributions plotted in log-log scale for the calibrator J1744-3116 (left), the check source J1745-283 (middle) and Sgr\,A* (right), for all data sets.}
    \label{fig:seds}
\end{figure*}

\begin{figure}
    \centering
    \includegraphics[width=0.49\textwidth]{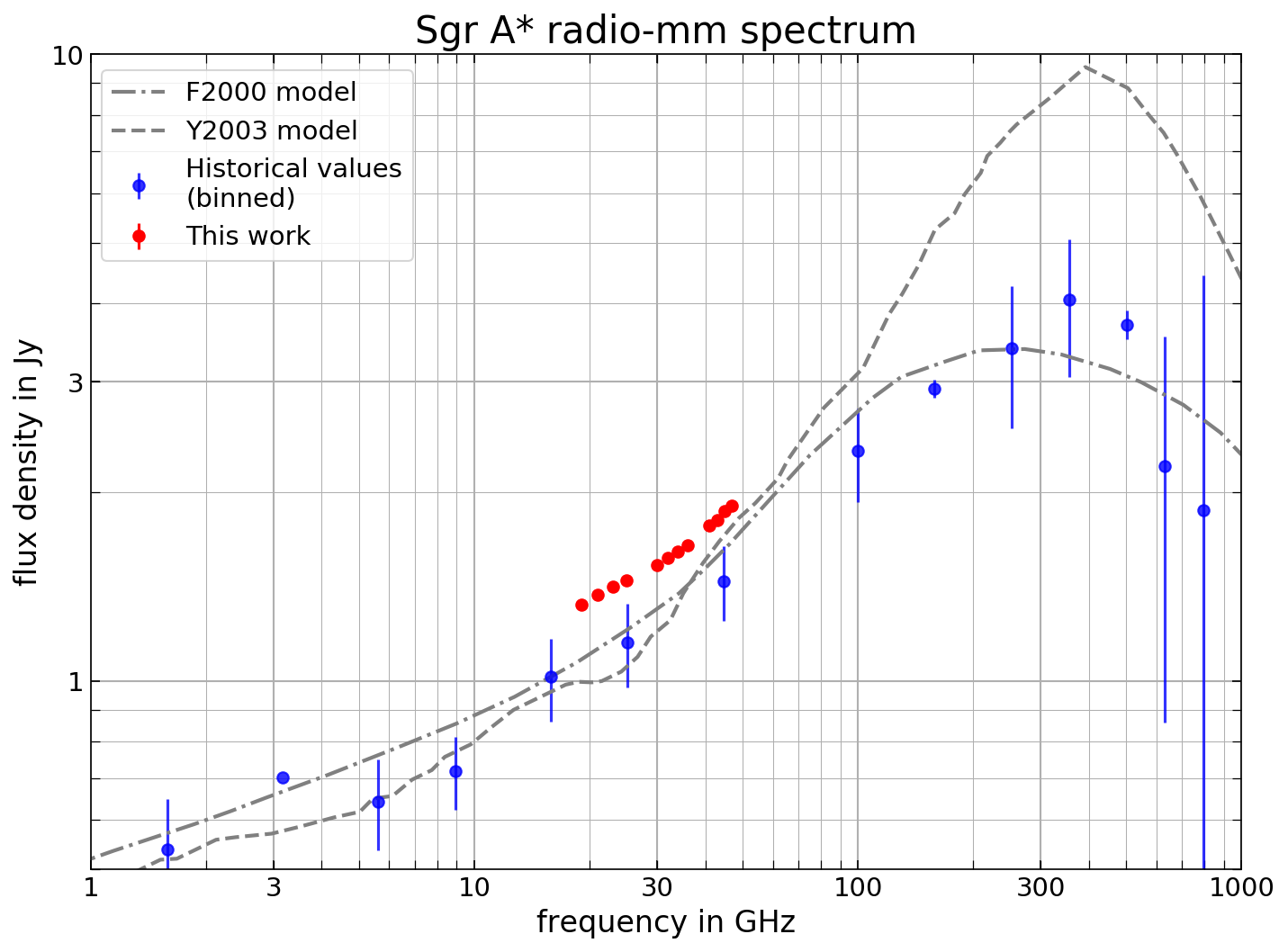}
    \caption{The spectral data from this work, averaged over all epochs (red), and plotted together with binned measurements from the past decades (blue, data from \citet{Zylka1995, Serabyn1997, Falcke1998, Zhao2003, Herrnstein2004, An2005, Marrone2006, Bower2015, CDB2015, Liu2016, Bower2019}). The theoretical jet model from \citet{Falcke2000} is shown, indicating the steepening of the spectrum into the submm bump. The model from \citet{Yuan2003} is shown as well, it reproduces this spectral steepening but overpredicts the magnitude of the submm bump somewhat. The full set of non-binned measurements used in this plot is shown in Figure \ref{fig:sgrafullsed} in Appendix D.}
    \label{fig:sgra-history}
\end{figure}

The general strategy for data reduction was to perform the standard steps of flux and bandpass calibration followed by gain calibration. Data set 7 could not be processed by the VLA pipeline due to missing tables in the downloaded data set. Although the download was done multiple times, the issue kept occurring and so the decision was made not to use the data from that epoch.
When running the VLA pipeline on the data sets (CASA version 4.7.2), a bug was found in the \verb|setjy| task where the model flux density for the gain calibrator source did not get set properly in the model column of the measurement sets. This issue is likely due to the large data volume that needed to be processed for each epoch. As a consequence, the model flux density for the gain calibrator remained fixed at 1 Jy for all frequencies, which is the default value in absence of a supplied model flux density. To remedy this issue, we phase-selfcalibrated the data on each of the three sources after running the VLA pipeline, using all baselines for J1744 and J1745, and using only baselines of >150 k$\lambda$ for Sgr\,A*. After verifying that phases were flat and zero across each spectral window, we then frequency-averaged to yield 1 frequency channel per spectral window. Gain calibration was performed on the resulting frequency-averaged data sets, which were considerably smaller in size and as such posed no issues regarding the setting of the model flux density for J1744, our gain calibrator source. As the last step, a further reduction in data volume was made by frequency-averaging contiguous chunks of 16 SPWs to yield one light curve per 2-GHz frequency band. The resulting gain-calibrated data was verified to yield sensible SEDs for Sgr\,A* and J1745, which are shown in Figure \ref{fig:seds}. If we assume a constant flux density over time for our calibrator source, we see that flux densities for J1745 show a variability between epochs of 6\% at 19 GHz up to 20\% at 47 GHz, while Sgr\,A* shows a larger variability ranging from 25\% at 19 GHz to 40\% at 47 GHz. This indicates that we do see true variability in Sgr\,A* between epochs. The averaged Sgr\,A* spectrum from this work is plotted in the context of previous observations in Figure \ref{fig:sgra-history}. The fact that we see somewhat higher flux densities for Sgr\,A* than from measurements in earlier years is compatible with the finding that Sgr\,A* exhibits a rising trend in flux density over multiple decades, as was remarked in \citet{Bower2015}.

\section{Data analysis}

\subsection{Extracting light curves}

\begin{figure}
    \centering
    \includegraphics[width=0.49\textwidth]{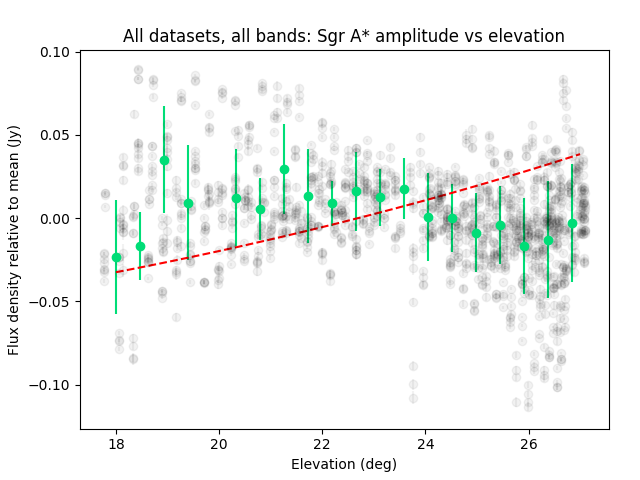}
    \caption{Measured flux density variations plotted against source elevation for Sgr\,A*. For each epoch and each subband, the average flux density for each light curve was subtracted from the measurements. The binned measurements (green points with error bars) show no apparent trend. For reference, the variation in air mass with elevation (plane parallel model) is superimposed with a dashed line.}
    \label{fig:amp-vs-elevation}
\end{figure}

\begin{figure*}[h]
\centering
\includegraphics[width=1.0\textwidth]{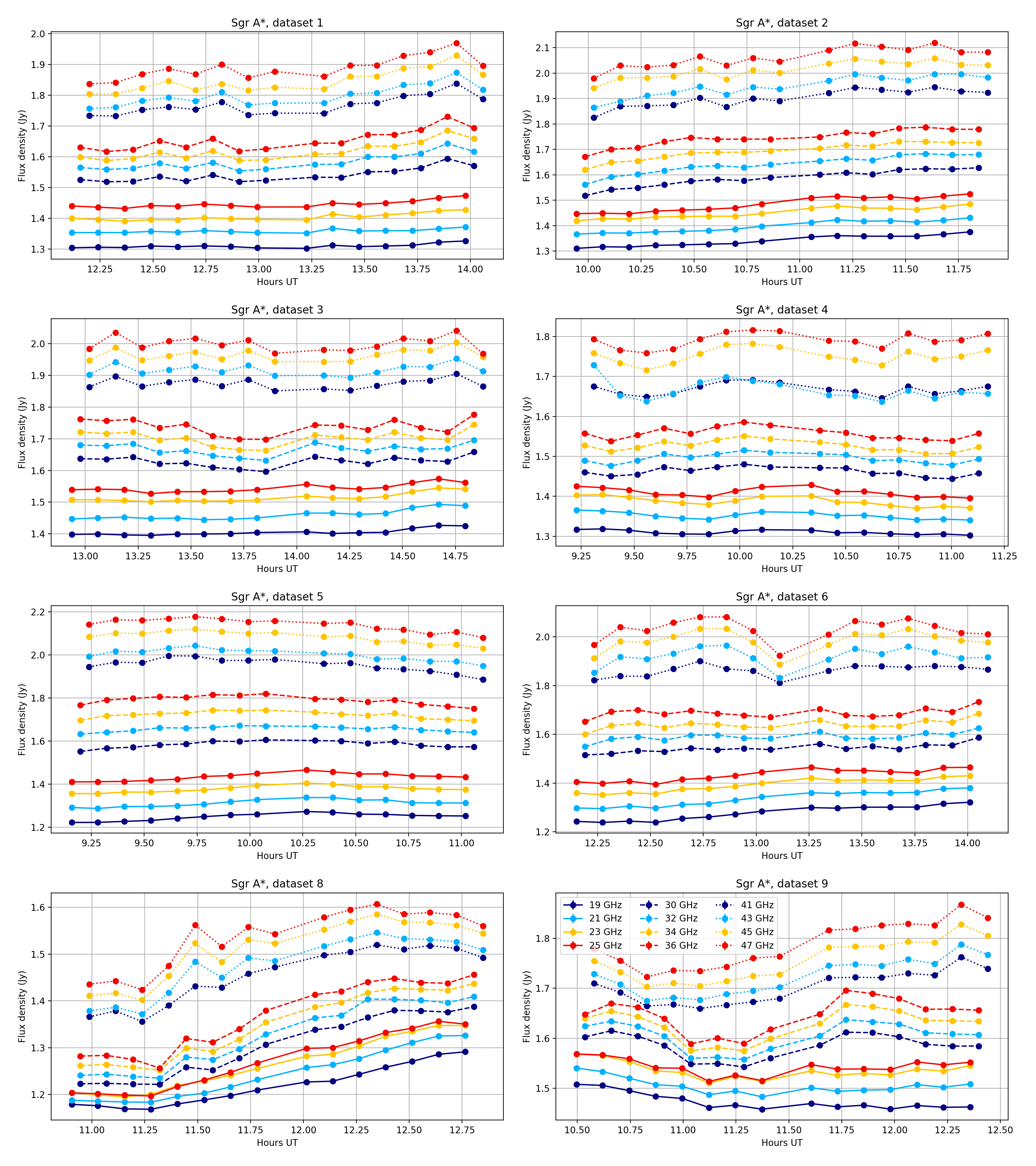}
\caption{Calibrated light curves for Sgr\,A*, data sets 1-6, 8 and 9. All data is plotted with error bars, which are so small as to be obscured by the data markers in most cases.}
\label{fig:lightcurves-science}
\end{figure*}

As the visibility data show constant amplitudes per scan over the range of $(u,v)$-distances we consider for Sgr\,A*, the Sgr\,A* light curves are calculated directly from calibrated visibility data with no intermediate imaging steps involved. To this end, the calibrated visibilities from the longer baselines ($\geq 150$ k$\lambda$) are averaged together per scan. These long baselines resolve out all the non-compact structure in the field of view, so that Sgr\,A* itself is the only source with a meaningful contribution to the total flux density for these baselines. Because VLA was in A-configuration for these observations, the majority of baselines are longer than this chosen cutoff length at all times and all observing frequencies. The resulting Sgr\,A* light curves for each epoch are plotted in Figure \ref{fig:lightcurves-science}, while a sample of light curves for the calibrator source and the check source are shown in Figure \ref{fig:cal-check-lc-sample} (the full set of light curves for these sources is included in Appendix A). In order to check any correlation of measured variations in flux density with elevation for Sgr\,A*, we plot the flux density excursion with respect to the mean for all subbands and epochs against elevation in Figure \ref{fig:amp-vs-elevation}. Binning the measured values shows no apparent dependence on elevation.

The theoretical thermal noise for VLA ranges from $\sim$90 microJy per 30s scan for low K-band to $\sim$300 microJy per 30s scan for high Q-band\footnote{From the VLA exposure calculator at https://obs.vla.nrao.edu/ect/, with the settings appropriate for our observations: low elevation, 27 antennas, 3-bit sampling, natural weighting, dual polarisation, 2 GHz bandwidth, spring weather}. In these calibrated light curves, the measured variance per 2 GHz averaged spectral window and per 30s scan ranges from $\sim$1 mJy for K band to $\sim$2.5mJy for Q band. Considering possible causes for this variability, the integrated flux density from the Galactic Center within the VLA primary beams is not expected to have a significant effect: the total emission from Sgr A West, which is comparable in angular extent to the VLA primary beam, is expected to be several tens of Jy in this part of the radio spectrum \citep{Law2008}. The System Equivalent Flux Density for the VLA between 20 and 40\,GHz is $\sim$500\,Jy\footnote{from https://science.nrao.edu/facilities/vla/docs/manuals/oss/performance/sensitivity, consulted July 2021}, which means that the thermal noise of the system is only expected to vary by at most 10\% as a consequence of this emission. The larger difference that we find between the theoretical noise predictions and the spread in the measured values can therefore be attributed to minor variations in compact source flux density within a single scan. However, the scan-to-scan variations in flux density are considerably larger than this (by about an order of magnitude), and hence dominate the total variability at each frequency. For the remainder of this paper, the twelve subbands for each data set will be indicated as K1 thru K4 (19, 21, 23 and 25 GHz), Ka1 thru Ka4 (30, 32, 34 and 36 GHz) and Q1 thru Q4 (41, 43, 45 and 47 GHz), so in order of increasing frequency.

\begin{figure*}[h]
    \centering
    \includegraphics[width=0.48\textwidth]{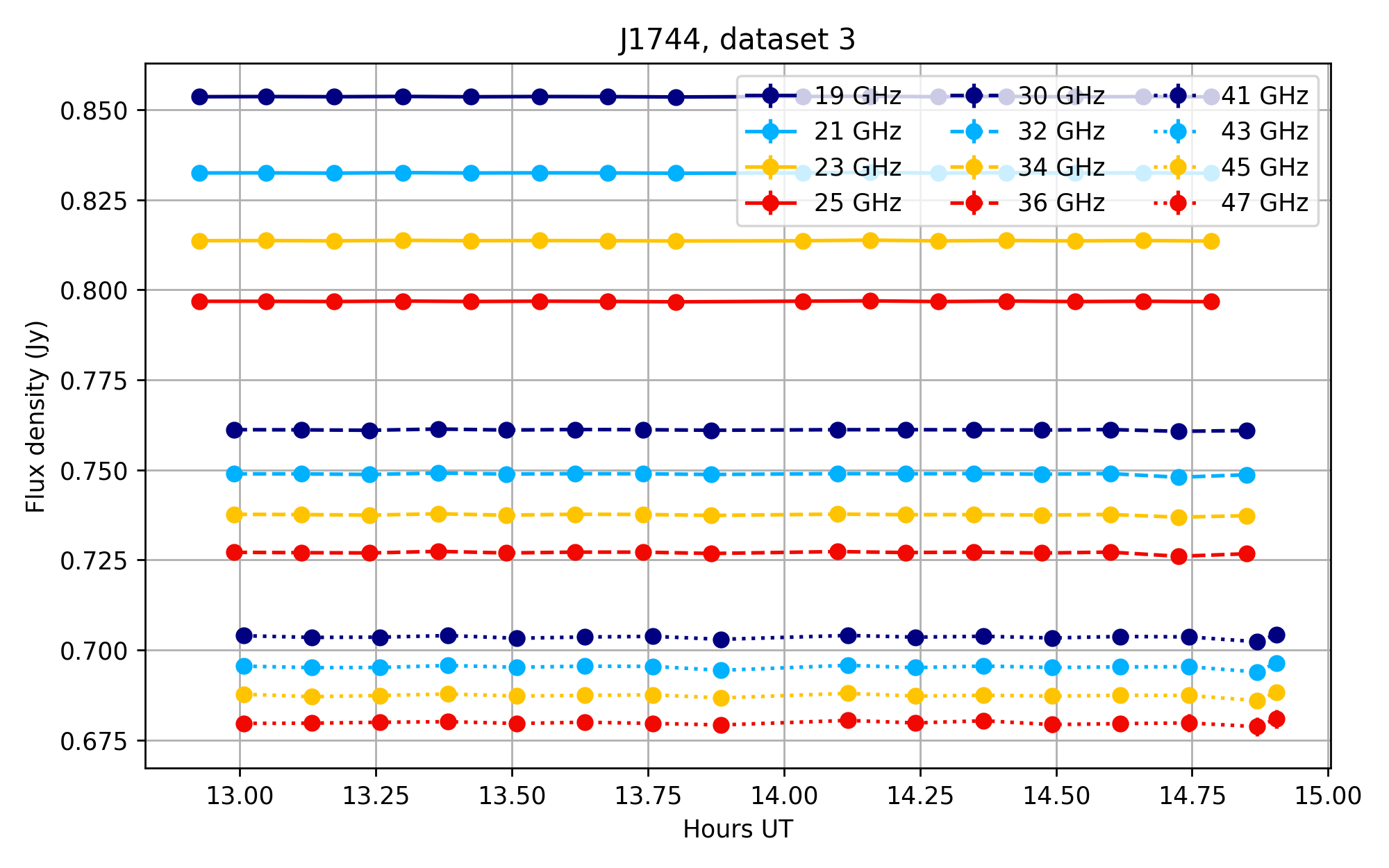}
    \includegraphics[width=0.48\textwidth]{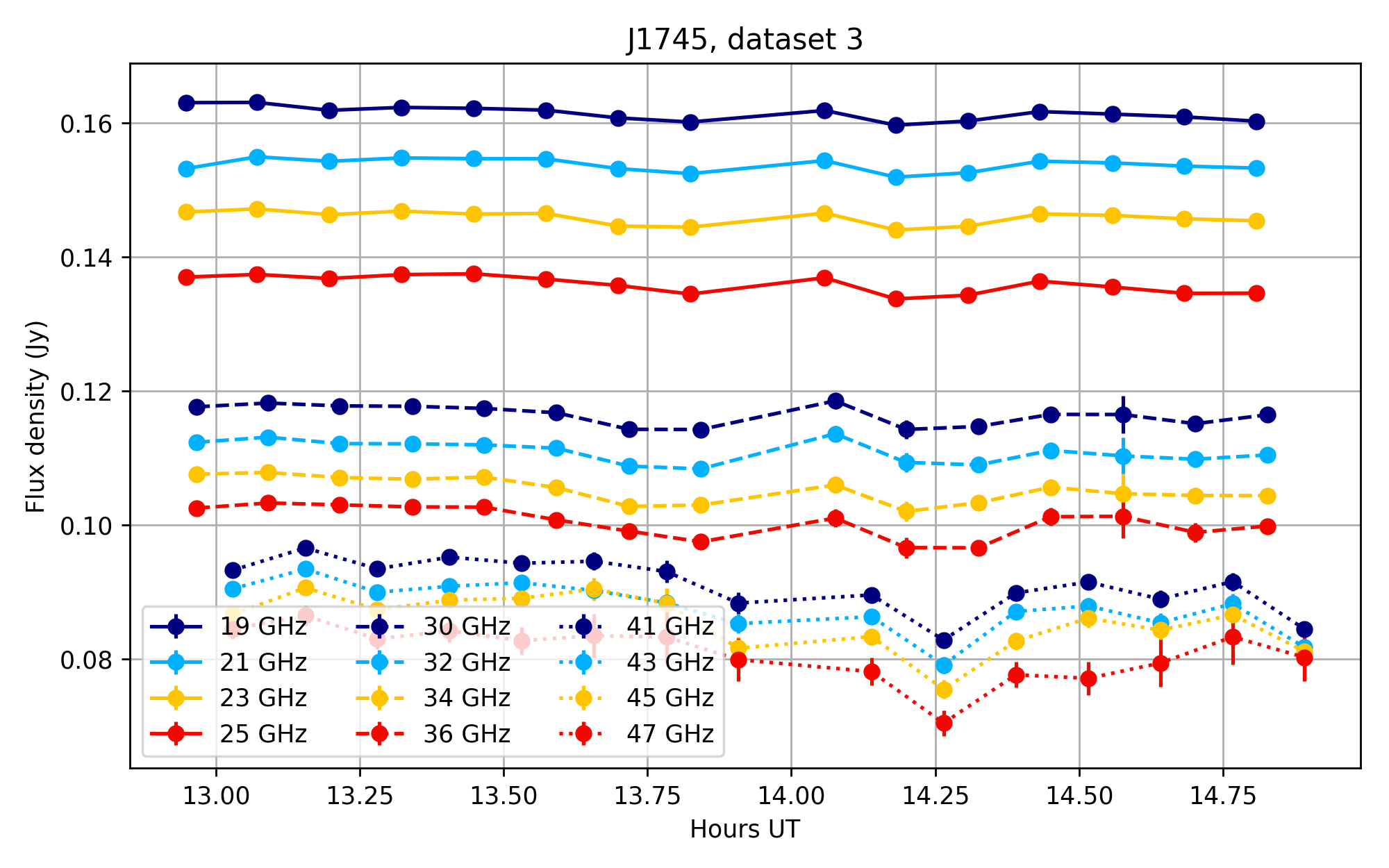}
    \caption{Examples of calibrated light curves for J1744-3116 (calibrator) and J1745-283 (check source) for epoch 3. All data is plotted with error bars, which are obscured by the data markers in most cases. The full set of light curves for the calibrator and the check source for all epochs can be found in Appendix A.}
\label{fig:cal-check-lc-sample}
\end{figure*}

\begin{figure*}[h]
\centering
\includegraphics[width=0.48\textwidth]{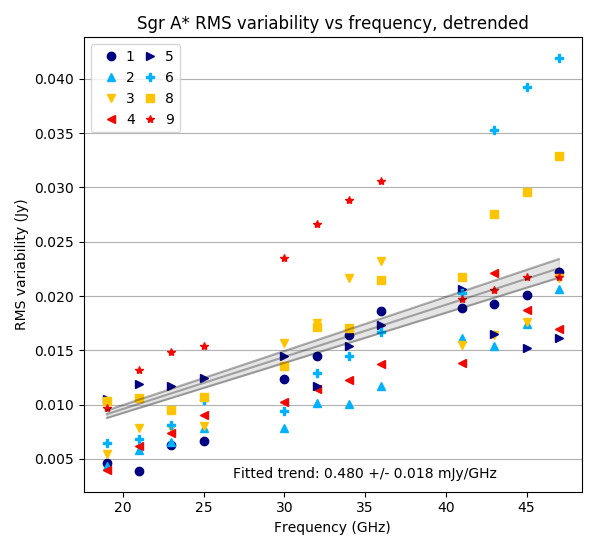}
\includegraphics[width=0.48\textwidth]{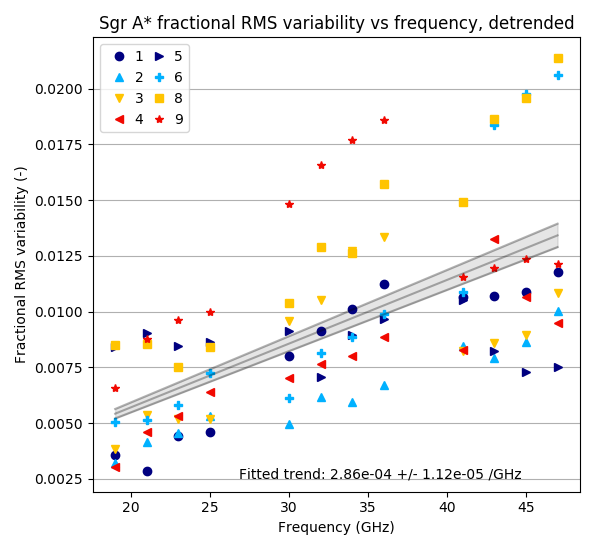}
\centering
\includegraphics[width=0.48\textwidth]{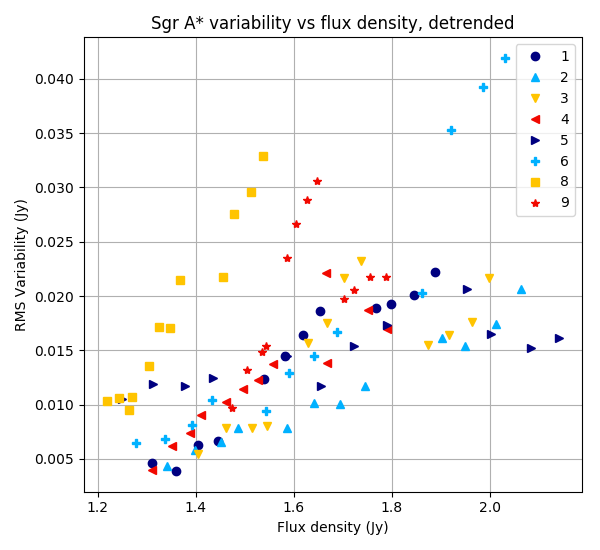}
\includegraphics[width=0.48\textwidth]{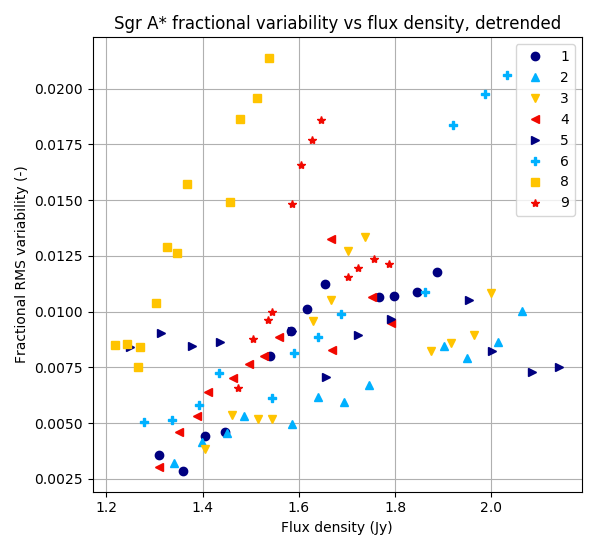}
\caption{Top left: the de-trended RMS variability of Sgr\,A* flux density versus frequency, for all bands (horizontal axis) and all epochs (symbol series). Top right: same, but variability is expressed in terms of the fraction of the average flux density per epoch at that frequency. Both plots show linear fits to the variability versus frequency with a 1-sigma uncertainty band. Bottom left: de-trended RMS variability of flux density per subband versus average flux density per epoch, for all epochs (symbol series). Bottom right: same, but variability is expressed in terms of the fraction of the average flux density.}
\label{fig:SgrA-rms}
\end{figure*}

An initial inspection of the Sgr\,A* light curves shows clear variability over time within epochs, where several data sets exhibit a curvature with well-localised flux density maxima or minima across all bands in their general trends (most notable in this regard are epoch numbers 5, 8 and 9). We also see examples of single-scan offsets in flux density that show strong correlation across the 4 subbands within one band but which do not have a counterpart in the other two bands, which suggests the presence of residual calibration errors of up to $\sim$0.03 Jy in the light curves. 
Because of the limited time interval covered by each of the light curves, they typically do not show variation around a clear stationary flux density but rather exhibit trends over the full epoch where the flux density changes as part of a pattern of longer-timescale variation. As a result, this longer-timescale variation (which is not completely characterised) tends to dominate the total variability of a light curve and also induces a large spread in flux density values. To be able to characterise the variability of these light curves in a meaningful way, we chose to focus solely on the short time scales by de-trending the light curves using linear least-squares fits and looking at the residual variability.
In Figure \ref{fig:SgrA-rms}, the de-trended RMS flux density variability per epoch is shown for all twelve frequency bands. In the plots, the RMS variability of the residual is shown. Generally, variability is larger for higher frequencies and appears to scale roughly linearly with flux density but with considerable variance. We find a modulation index (RMS variability as fraction of flux density) on the order of 1\%, but with a growing trend towards higher flux densities. \citet{Bower2015} reported a larger modulation index of 8\% for this frequency range, but that figure included longer-term variability. The fact that we do not see a clear linear relation between flux density and trend-subtracted RMS variability for any subband frequency, as would be expected from the RMS-flux relation \citep{Uttley2001, Abuter2020}, may be due to the fact that we are not sensitive to this longer-term contribution to variability.

\subsection{Detection of time lags}

In our previous time lag paper \citep{CDB2015}, the time lag calculations were done by fitting Fast-Rise Exponential Decay (FRED) curves to the measured light curves in the different frequency bands. This was done because of the clear and consistent presence of a flux density maximum across multiple light curves, of which the shape lent itself well to FRED fitting. For the data sets presented in this paper, the flux density evolution shows different characteristics for each epoch and so a more general way to look for lag relations between bands is needed.

To this end, multiple cross-correlation methods were considered that can deal with a non-constant time sampling cadence for the data series. The Z-transformed Discrete Correlation Function (ZDCF) by Tal Alexander \citep{Alexander1997} is such an implementation, as is the Local Cross-Correlation Function (LCCF) as described by \citet{Welsh1999}. We found that both of these methods yield highly similar auto- and cross-correlation functions for our data sets. Out of these two methods the LCCF was chosen for the main analysis in this paper, because of its simple and portable implementation.

For the cross-correlation parameters we chose a minimum number of 11 measurements per time lag bin (which is the same as the recommended minimum number for the ZDCF, and advised from a statistics point of view), and a minimum time lag bin width of 30 seconds (the length of a single scan). Using these parameter values, the cross-correlation functions for all twelve light curves in each data set were calculated against themselves as well as each of the other eleven light curves in the same data set. Each cross-correlation between two different light curves from the same data set is thus represented twice in the correlation results.

\begin{figure*}[h]
    \centering
    \includegraphics[width=0.33\textwidth]{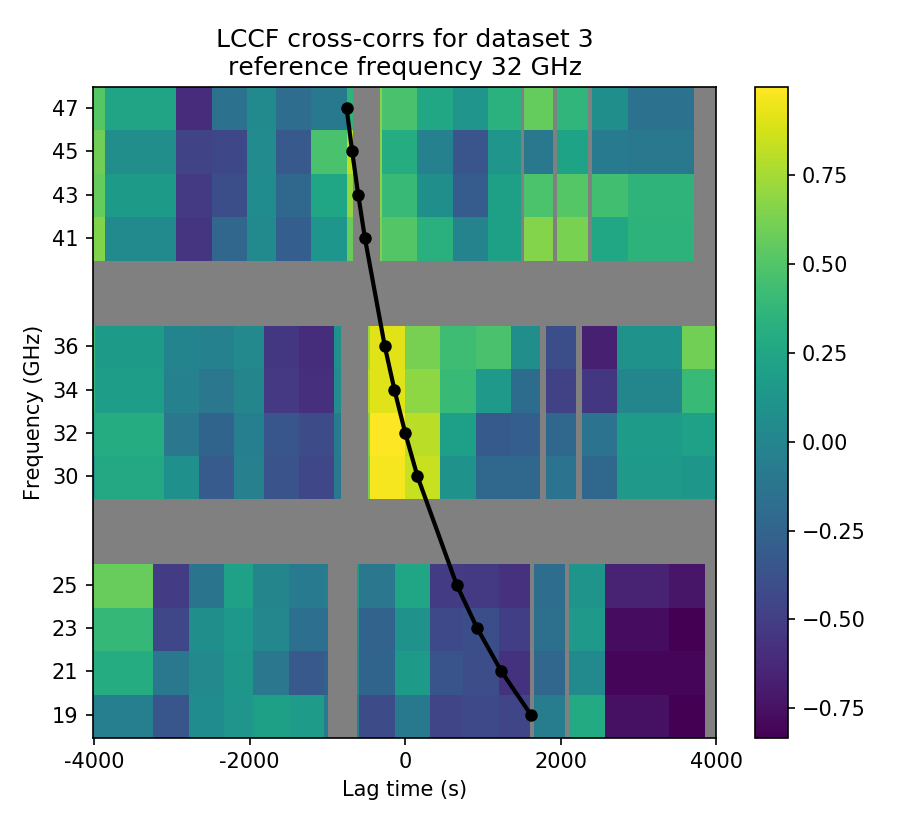}
    \includegraphics[width=0.33\textwidth]{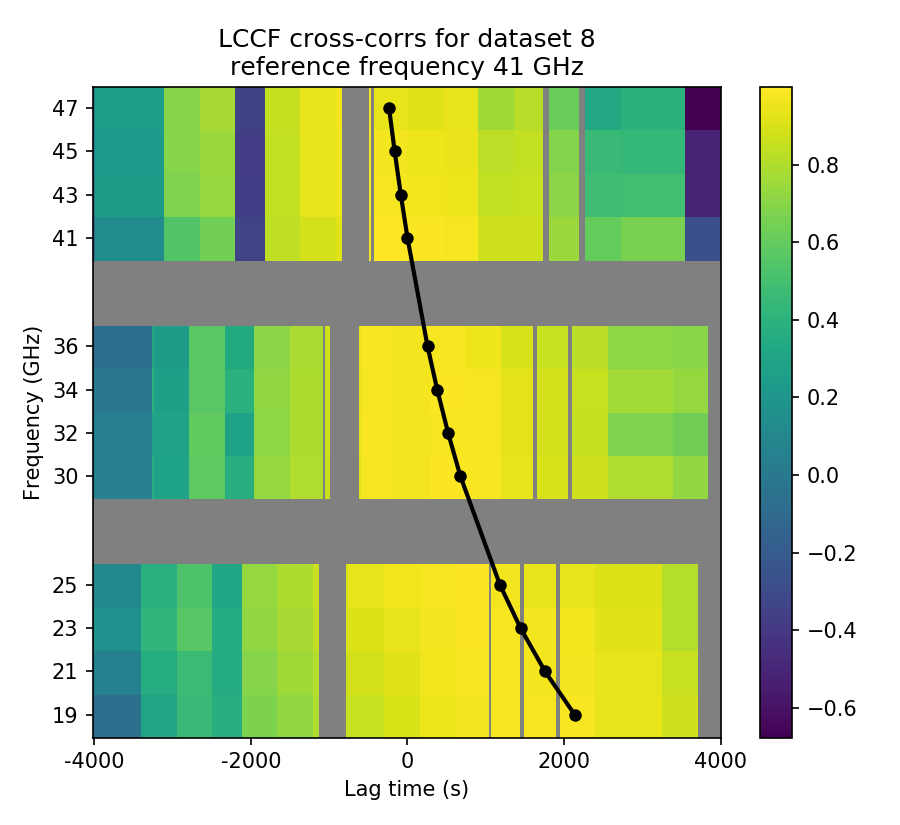}
    \includegraphics[width=0.33\textwidth]{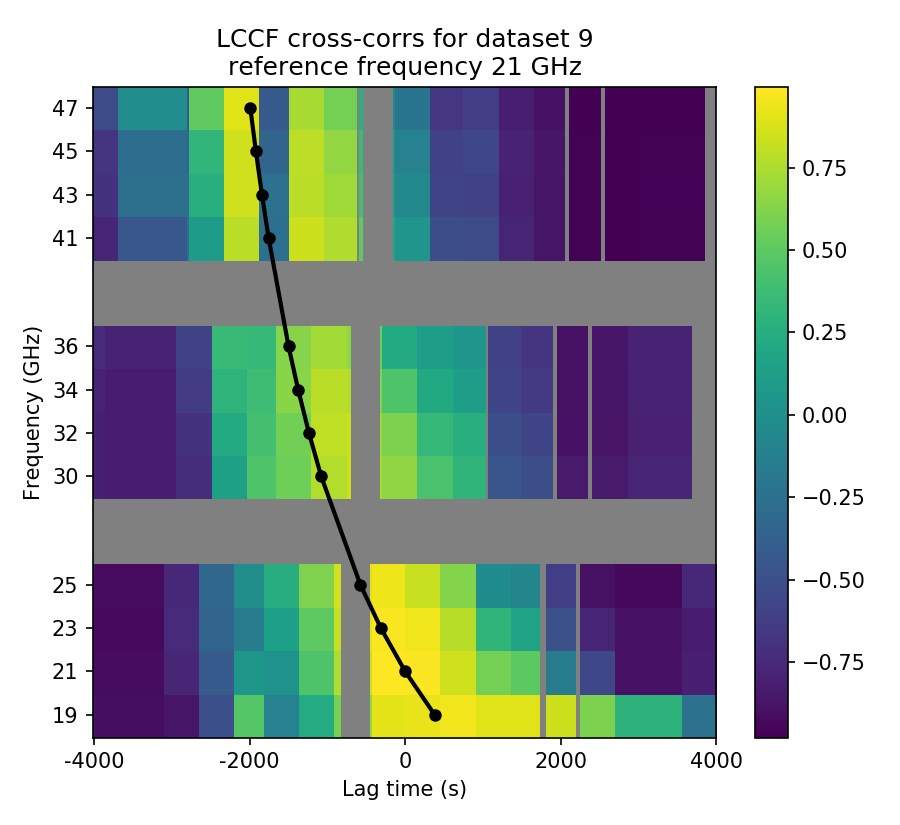}
    \caption{Three examples of correlation products. Left: poor correlation between bands for data set 3, using reference band Ka2. Middle: wide correlation for data set 8, reference band Q1. Right: strong correlation with time lag for data set 9, reference band K2. The black curved line indicates the lag/frequency relation found in \citet{CDB2015}, and is not a fit to the correlation data shown here.}
    \label{fig:corrs}
\end{figure*}

Examples of the intermediate results from this cross-correlation step are shown in Figure \ref{fig:corrs}. In the cross-correlation maps shown there, the time lag relation found in \citet{CDB2015} is overplotted as a black line with marker dots. Note that it is not a fit to the maxima of the plotted cross-correlation data, but it is included to provide a comparison. Each of the figures uses a reference light curve at one of the 12 subband frequencies, and is thus one of the twelve cross-correlation plots that was generated for that particular data set. The plots in Figure \ref{fig:corrs} were selected to show a representative range of behaviours encountered in the cross-correlation products: for data set 3 there is no obvious feature that correlates across all frequencies and total variability is limited. For data set 8 there is a strong climbing trend in flux density present in the light curves, with a maximum being reached for the higher frequencies but not as clearly for the lower ones. For data set 9, there is a clear and consistent minimum visible in the light curves at all frequencies. We see that localised features in the light curves that correlate across all bands tend to occur over timescales corresponding to one hour or more, matching the flaring timescale reported by \citet{YusefZadeh2006}. In the leftmost plot, we see a case where the light curves are only weakly correlated across the three main frequency bands (data set 3). The light curves in the same band as the reference light curve show a relatively strong correlation peak, but this pattern is abruptly broken for the correlation functions in the other two bands. The light curves for this epoch do not show clear trends or strong variability, and calibration uncertainties can therefore cause correlated flux density excursions that are limited to a single band as they are part of the same set of scans. In the middle plot, we see an example of a relatively broad cross-correlation function, where the location of its maximum along the lag axis is not tightly constrained (data set 8). The observed cross-correlation pattern may be compatible with the overplotted lag relation (which is taken from our previously found result in \citet{CDB2015}), but it is also consistent with zero time lag -- depending on the reference light curve used for cross-correlation, both of these trends can appear. The rightmost plot shows a case in which there is a robust and monotonic lag trend visible across all bands (data set 9), which appears to lie close to the lag relation we found in our previous paper. The associated light curves show a clearly defined minimum in their flux density evolutions, which improves the clarity of the cross-correlation peaks.

In order to distill potential time lag trends from the maxima in the cross-correlation graphs generated from the data, the choice was made to employ a piecewise 4th-order spline fitting to the calculated cross-correlation functions (taking their error bars into account) and to perform a local maximum search on this spline fit. This approach allowed us to avoid bias when identifying the dominant peaks in the cross-correlation functions. We chose to reject any maximum at time lags greater than 3000 seconds or below -3000 seconds, as the cross-correlation functions have poor statistics at those largest lag values because of the small overlap between light curves at large lags, and as such are too noisy to use there. Our method was found to be robust in the context of the various shapes and degrees of smoothness that the cross-correlation functions can take across our data sets. Most importantly, this method is agnostic to any expected value of the time lag for any cross-correlation function. In the cases where no clear maximum can be identified, the peak finding algorithm returns an empty result. It should be noted that although the peak-finding method generally agrees quite well with a human assessment of the cross-correlation function, there are cases where the peak of the function remains poorly constrained. The typical causes of this behaviour are either a plateau-shaped correlation curve, for which the peak position is extremely sensitive to details of the fitting parameters, or a complex shape of the cross-correlation function with many local maxima - none of which is clearly dominant. In those cases, other heuristics for peak finding are likely to encounter the same issues.

For each choice of reference frequency, we thus get one 'trend' containing at most twelve different times at which the maximum value of the cross-correlation function is detected, one for each of the subbands in the data set. Using twelve different reference frequencies, we get twelve of these trends. As we cannot presuppose any temporal relation between the zero-lag autocorrelation peaks for these different trends (that, is after all, the very relation we are trying to determine), we need to apply global shifts to these trends so that we minimise their mutual average offsets while respecting their internal structure. Using a least-squares fit and picking one time lag trend as the reference trend, we obtain co-registered time lag trends that are ready for trend fitting.

\begin{figure*}[h]
    \centering
    \includegraphics[width=0.33\textwidth]{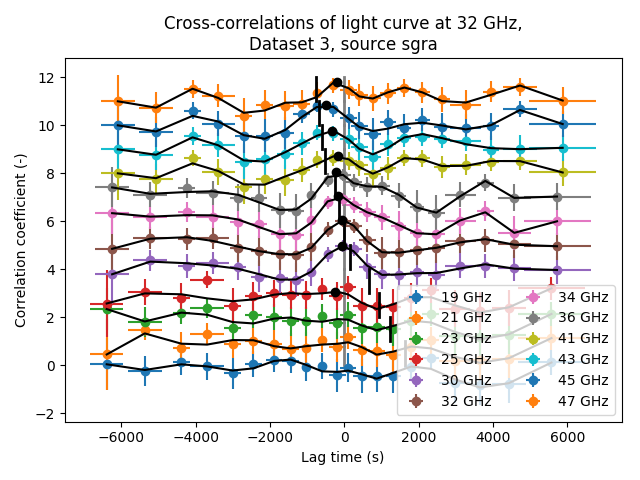}
    \includegraphics[width=0.33\textwidth]{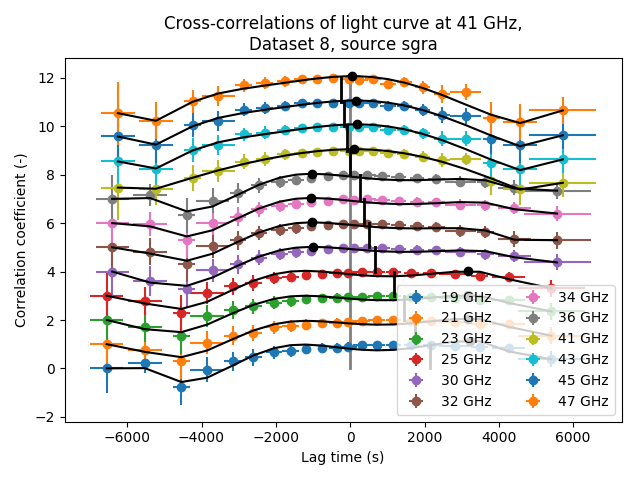}
    \includegraphics[width=0.33\textwidth]{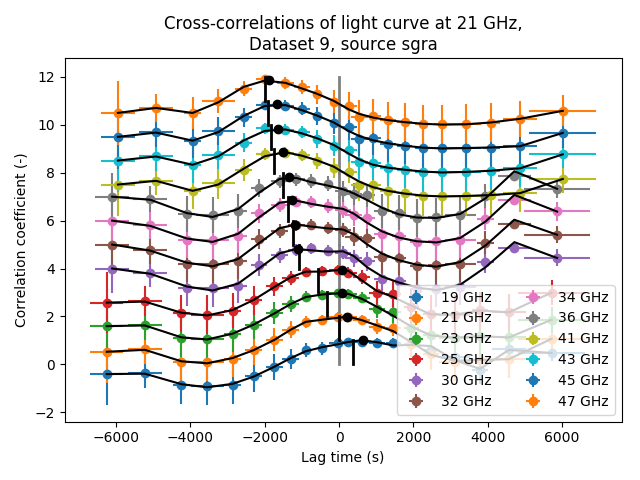}
    \caption{Fitted splines and detected cross-correlation peaks, for the same data sets and reference bands as shown in Figure \ref{fig:corrs} where cross-correlation functions for progressively higher frequency bands are vertically offset by steps of 1. Vertical black lines indicate the expected time lags based on the previously found lag relation from \citet{CDB2015}, while black dots indicate the detected maxima of the cross-correlation spline fits. Note that when the cross-correlation function has a broad plateau, the detected peak is poorly constrained (middle plot).}
    \label{fig:peakfits}
\end{figure*}

Upon initial inspection, the patterns seen in the time lag trends for the different data sets show considerable variance. We see weakly correlated lag trends showing large outlier values for three data sets (data sets 1, 2, and 3), one data set that has no clear trend pattern, because of poor cross-correlations (data set 6), a data set that shows consistent time lag trends in some subbands but not in others (data set 8) and data sets with trends having a positive slope that indicate the presence of time lags (data sets 4, 5 and 9).

We note that the data sets with cross-correlation functions that have more clearly defined peaks also show a better consistency among their individual time lag trends for the different reference subbands than the other data sets do. Furthermore no examples of a consistently present time lag trend with a negative slope, where low-frequency variability would lead high-frequency variability, is seen in any of the data sets.

\begin{figure*}[ht]
\centering
\includegraphics[width=0.48\textwidth]{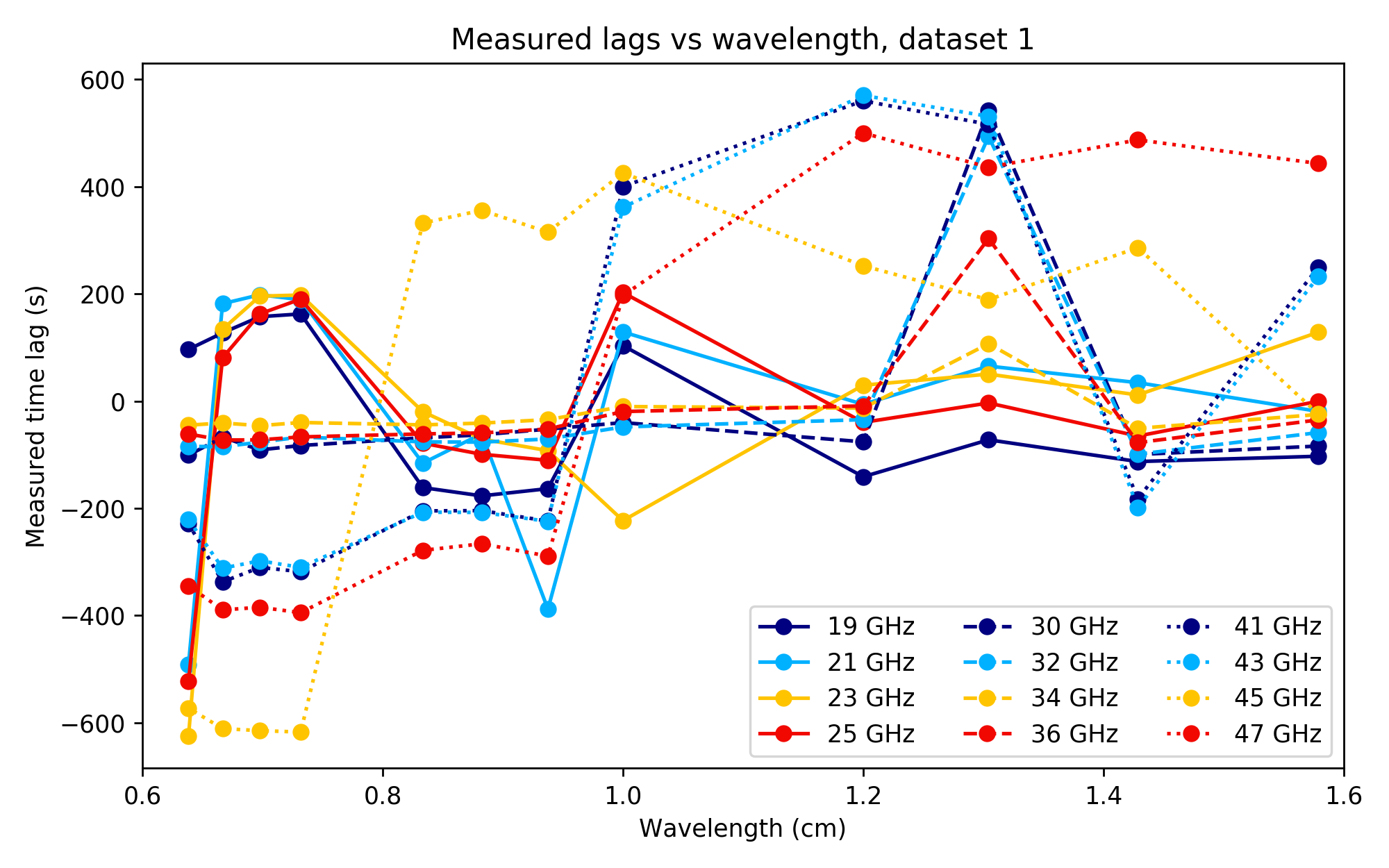}
\includegraphics[width=0.48\textwidth]{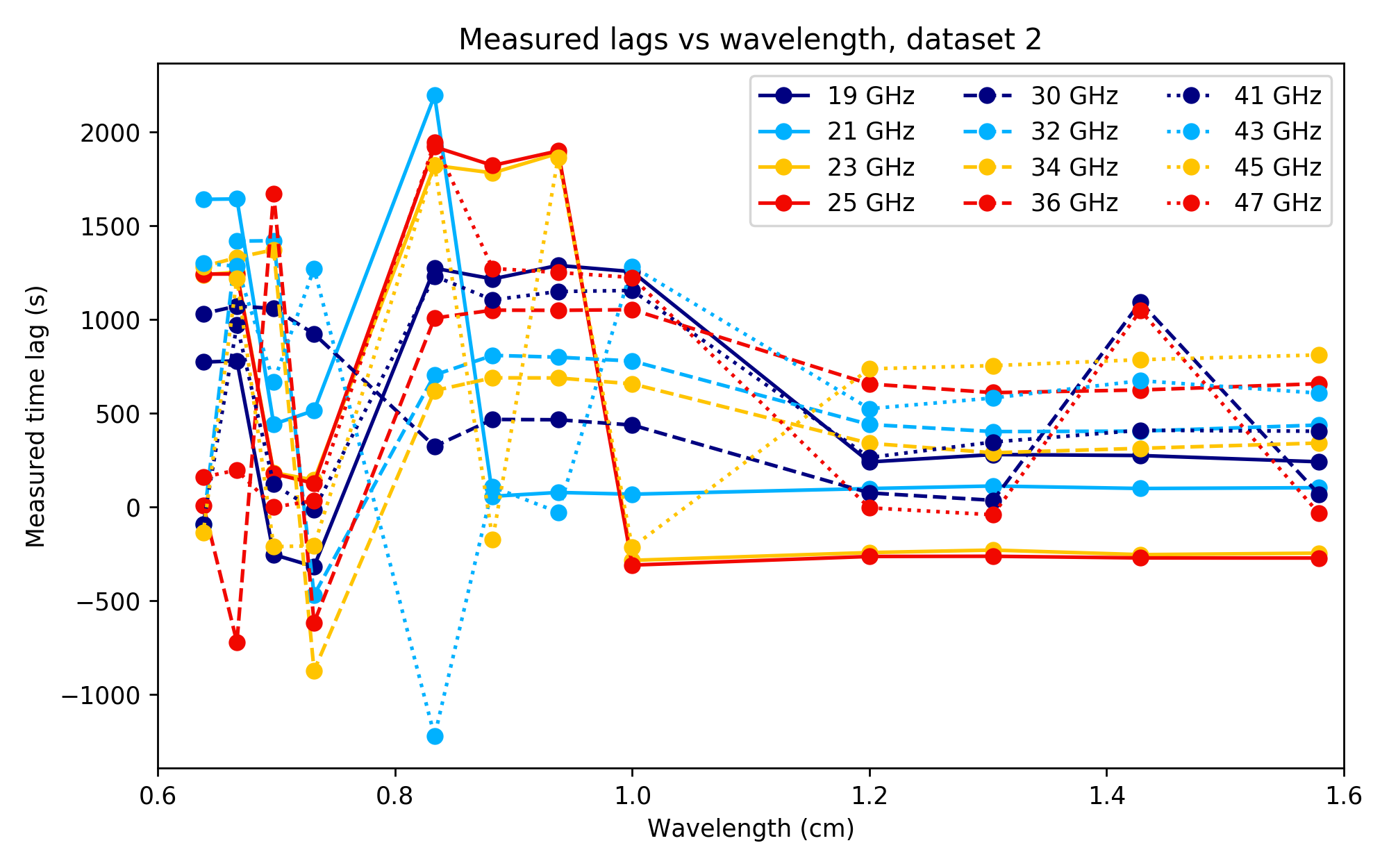}
\includegraphics[width=0.48\textwidth]{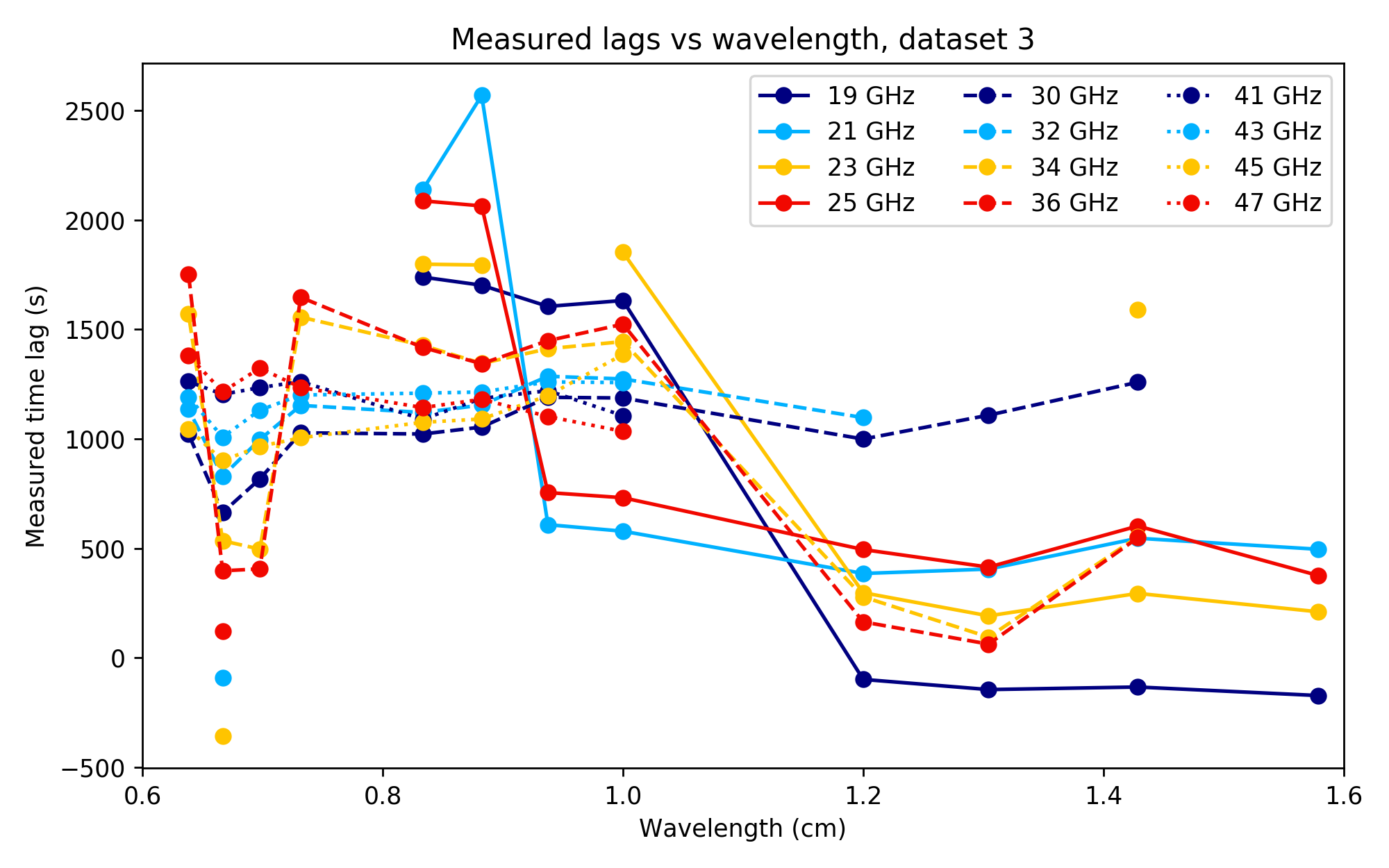}
\includegraphics[width=0.48\textwidth]{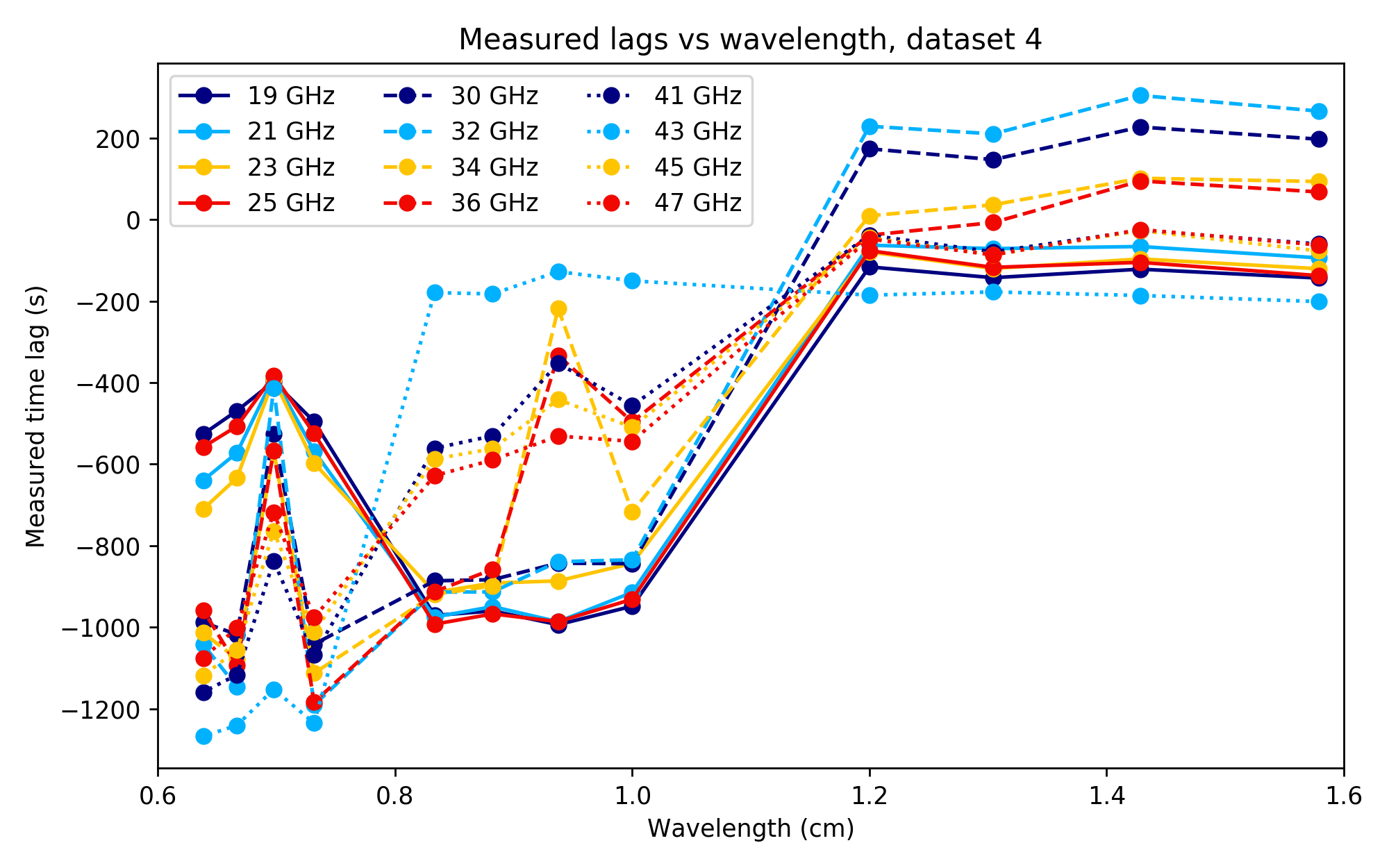}
\includegraphics[width=0.48\textwidth]{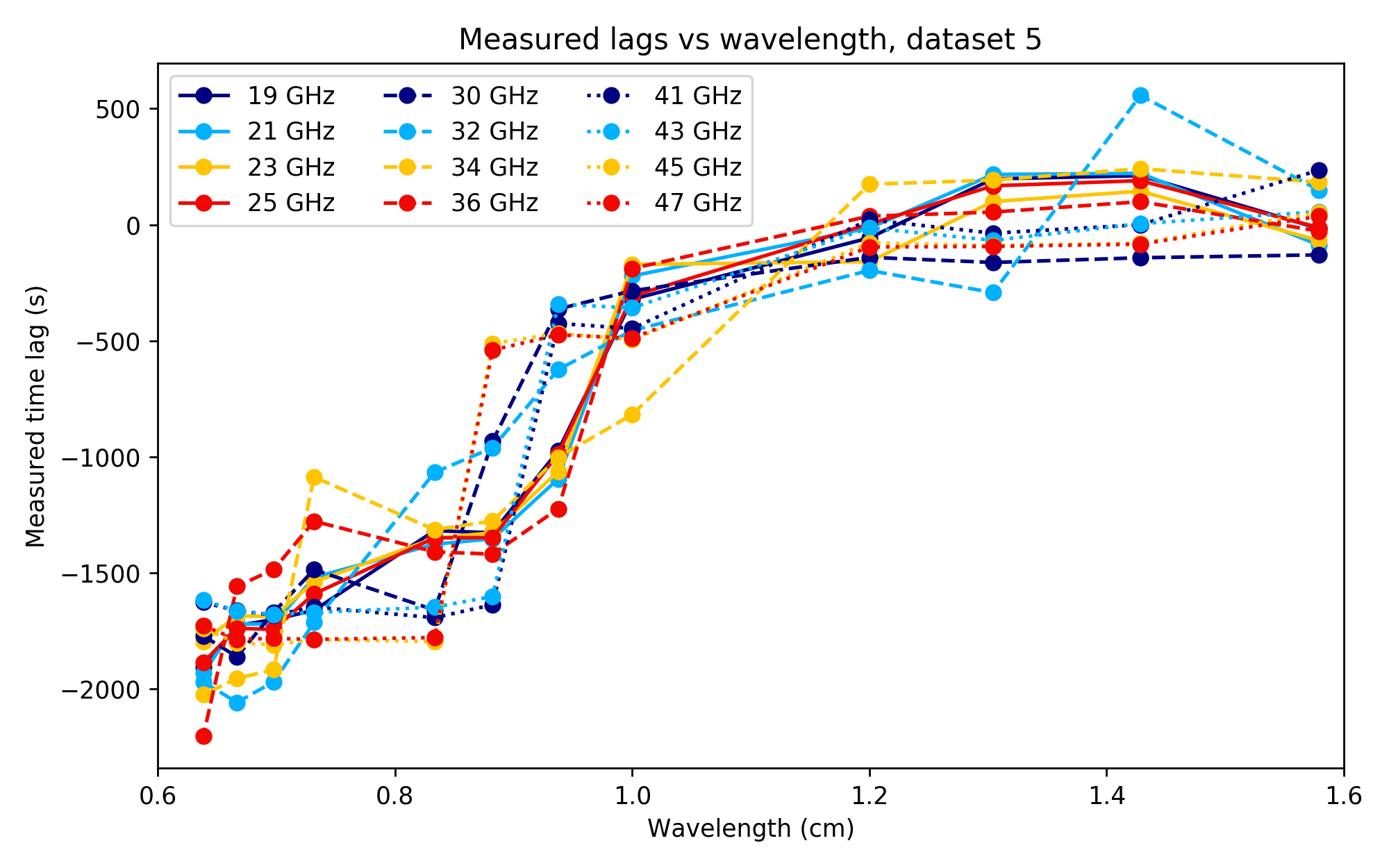}
\includegraphics[width=0.48\textwidth]{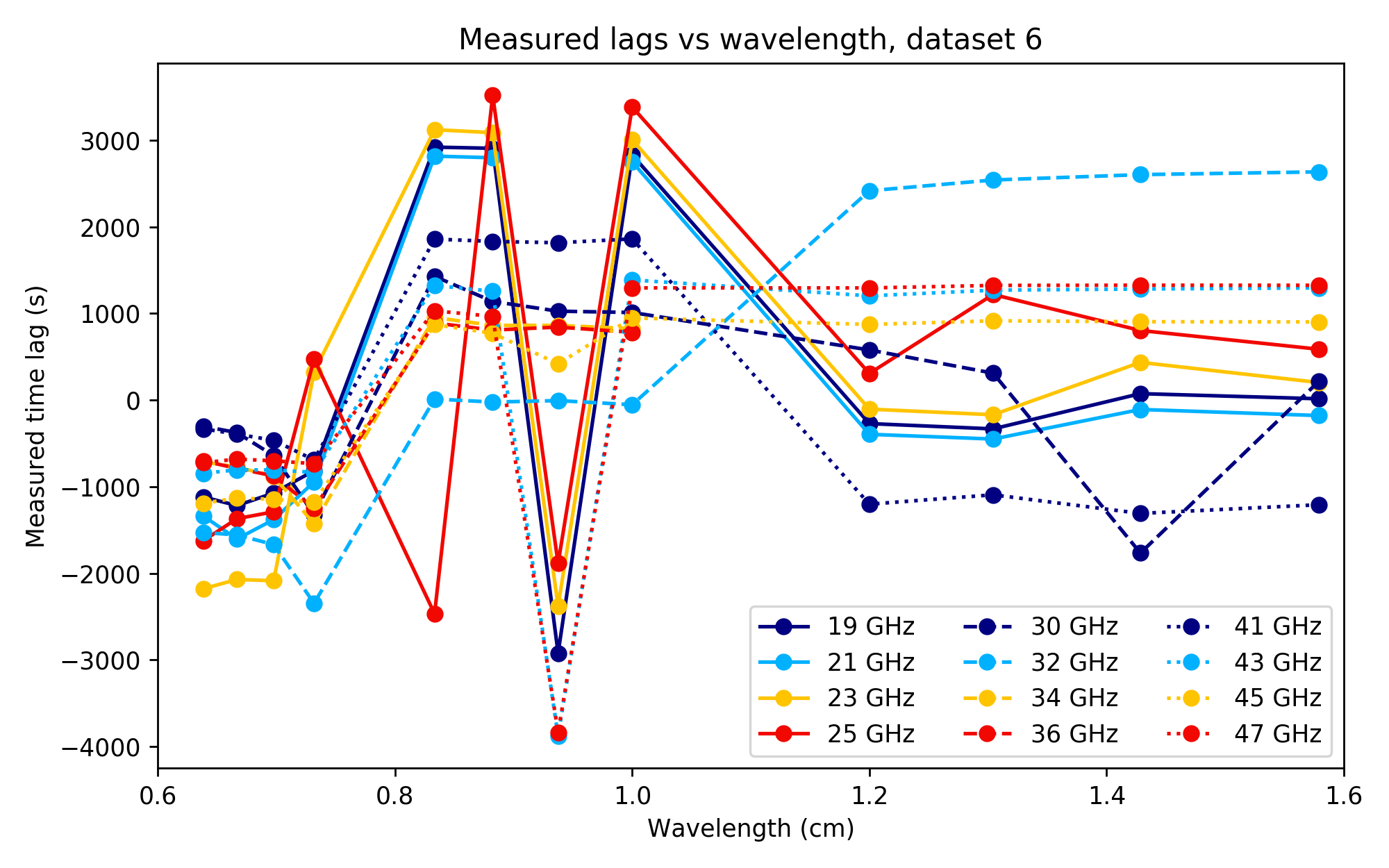}
\includegraphics[width=0.48\textwidth]{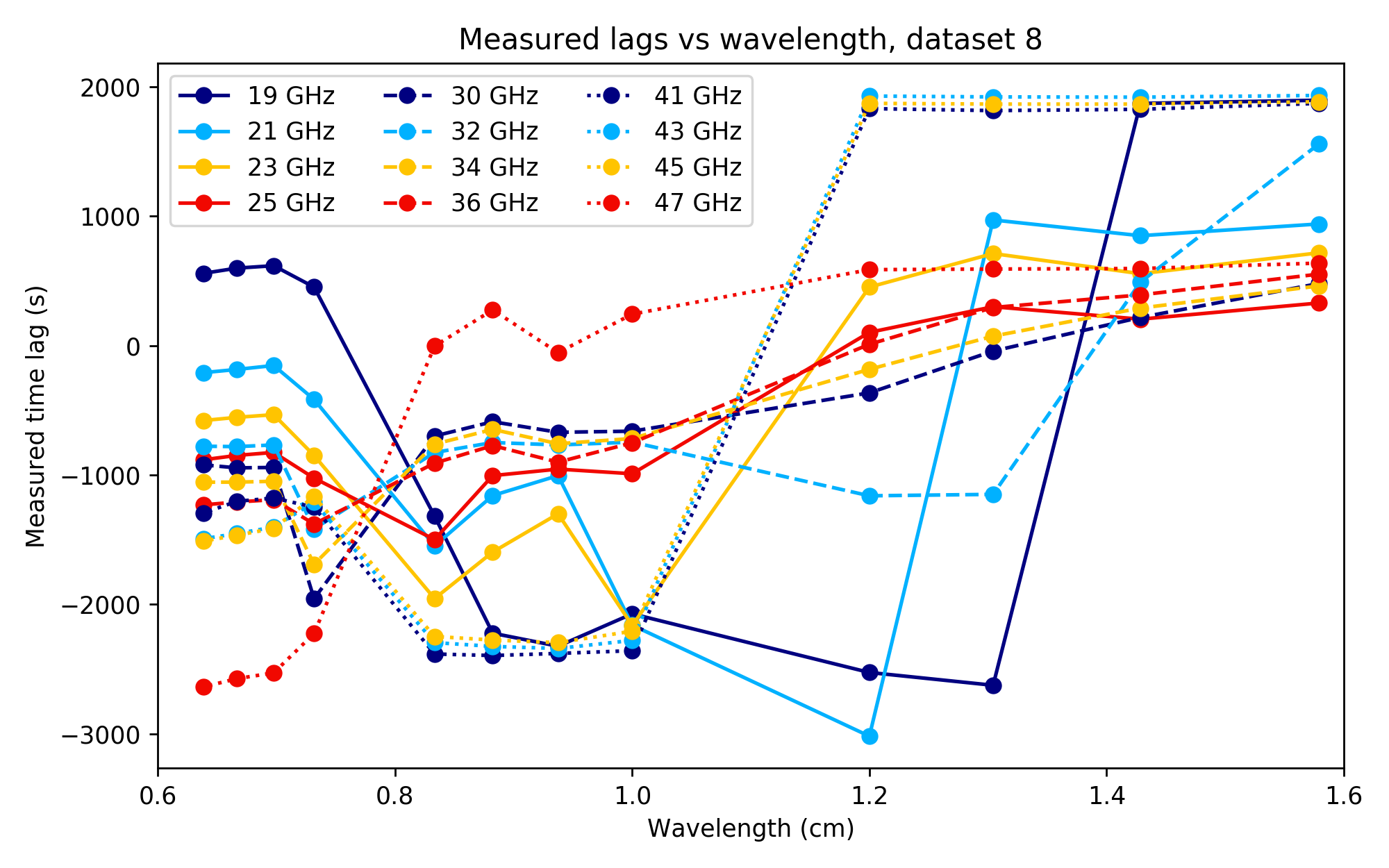}
\includegraphics[width=0.48\textwidth]{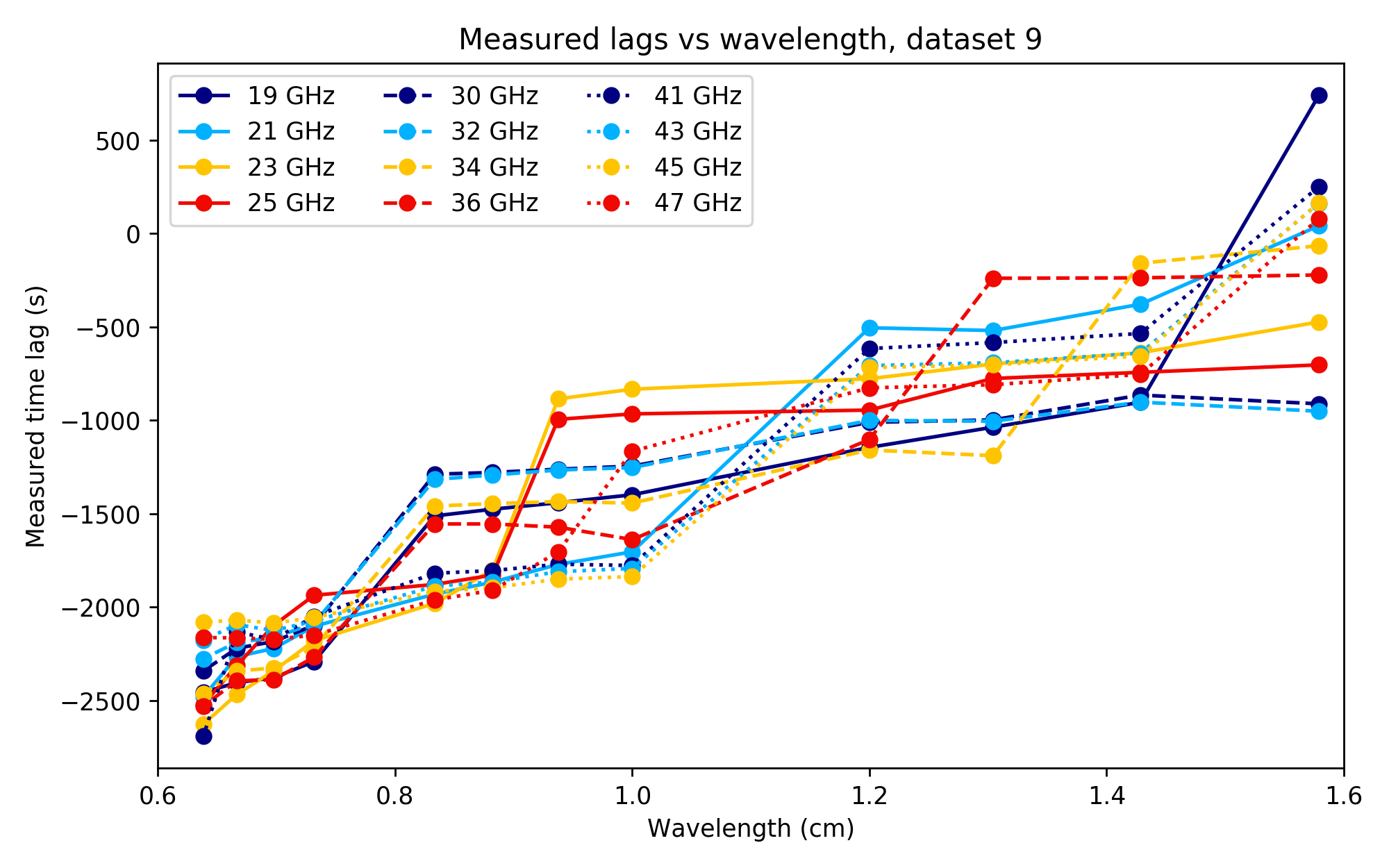}
\caption{The co-registered maxima of the cross-correlation functions, collectively forming the observed time lag trends in the data sets. The trends have been shifted to minimise mutual spread.}
\label{fig:measuredlags}
\end{figure*}

\begin{figure*}[ht]
\centering
\includegraphics[width=0.48\textwidth]{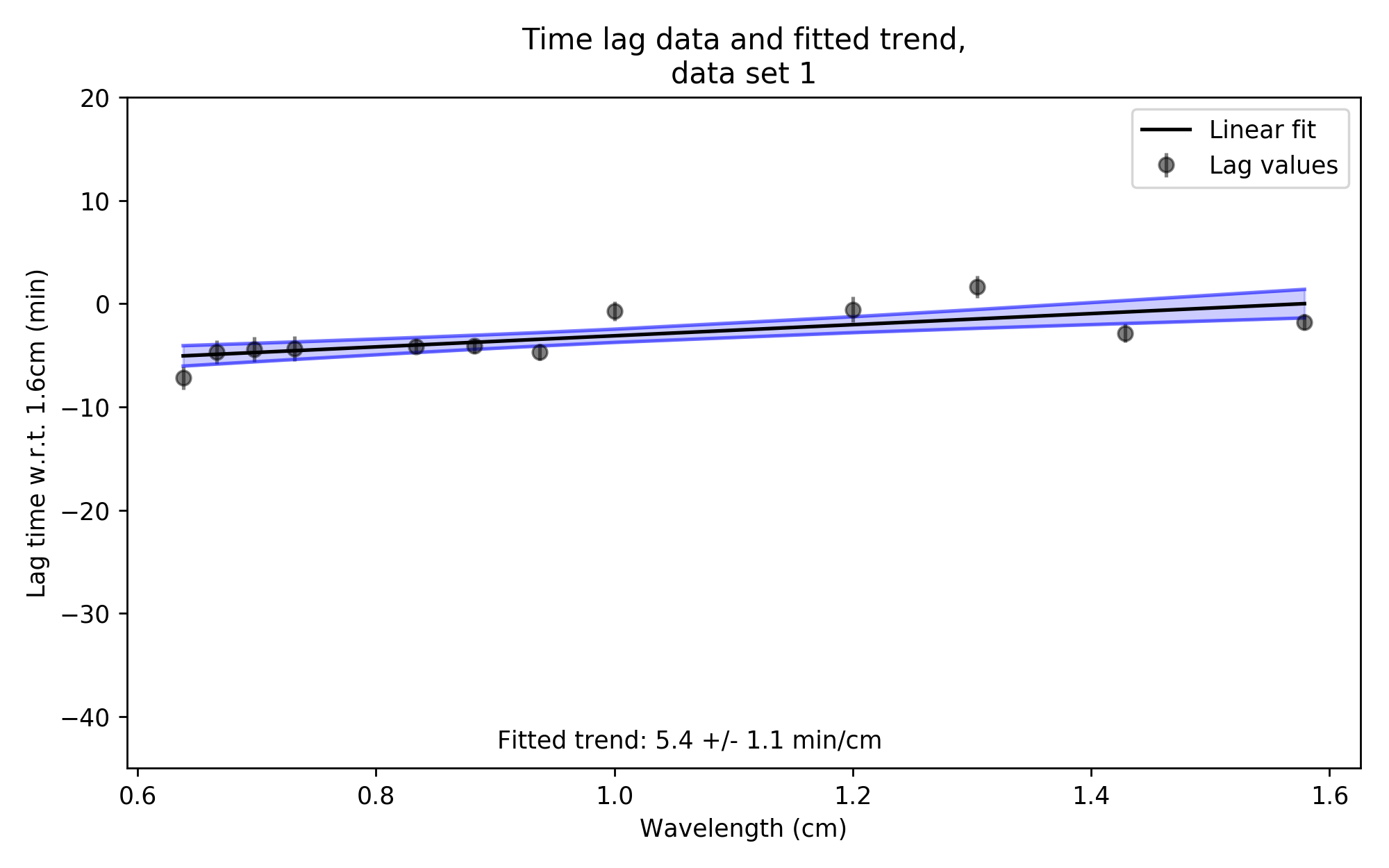}
\includegraphics[width=0.48\textwidth]{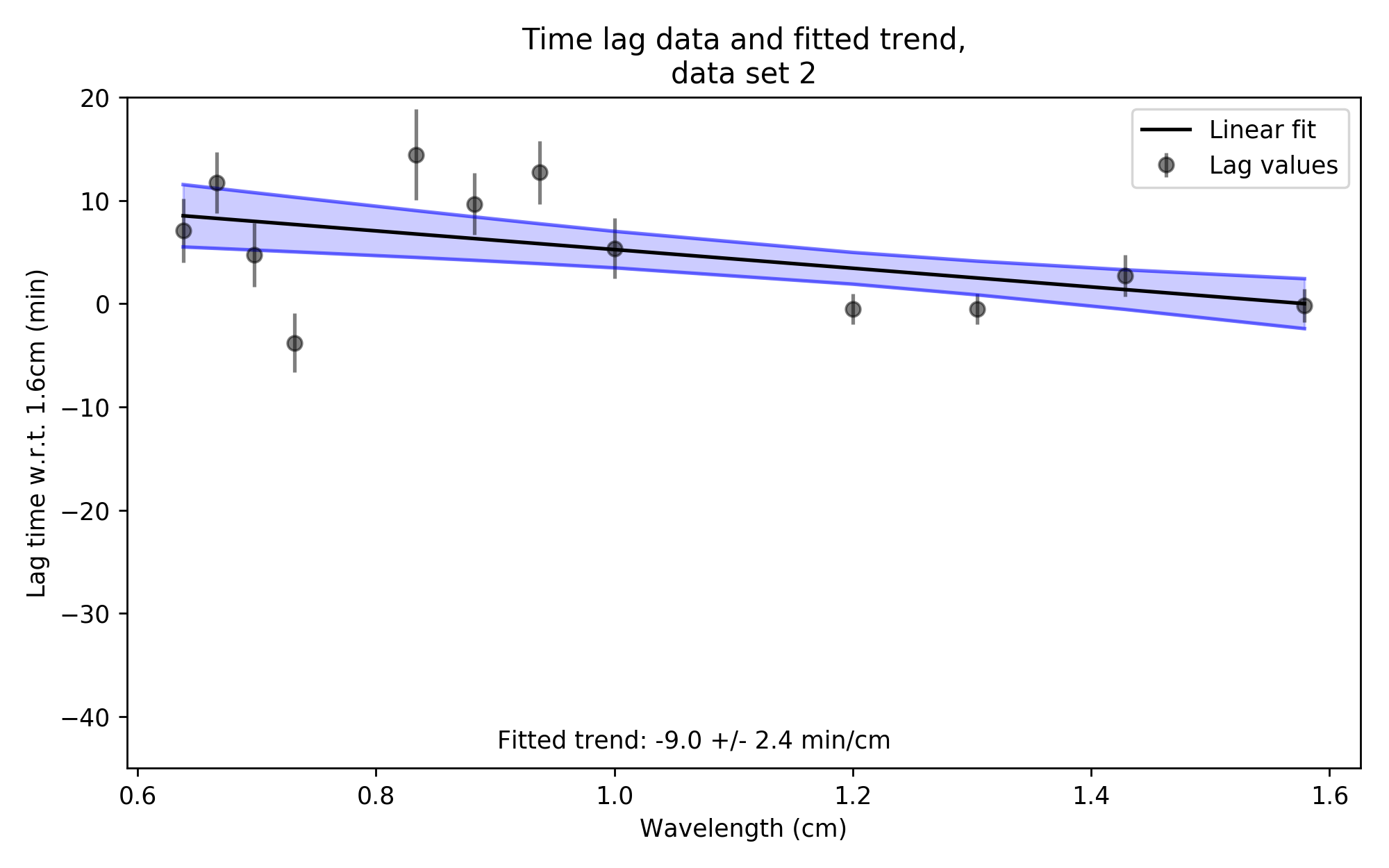}
\includegraphics[width=0.48\textwidth]{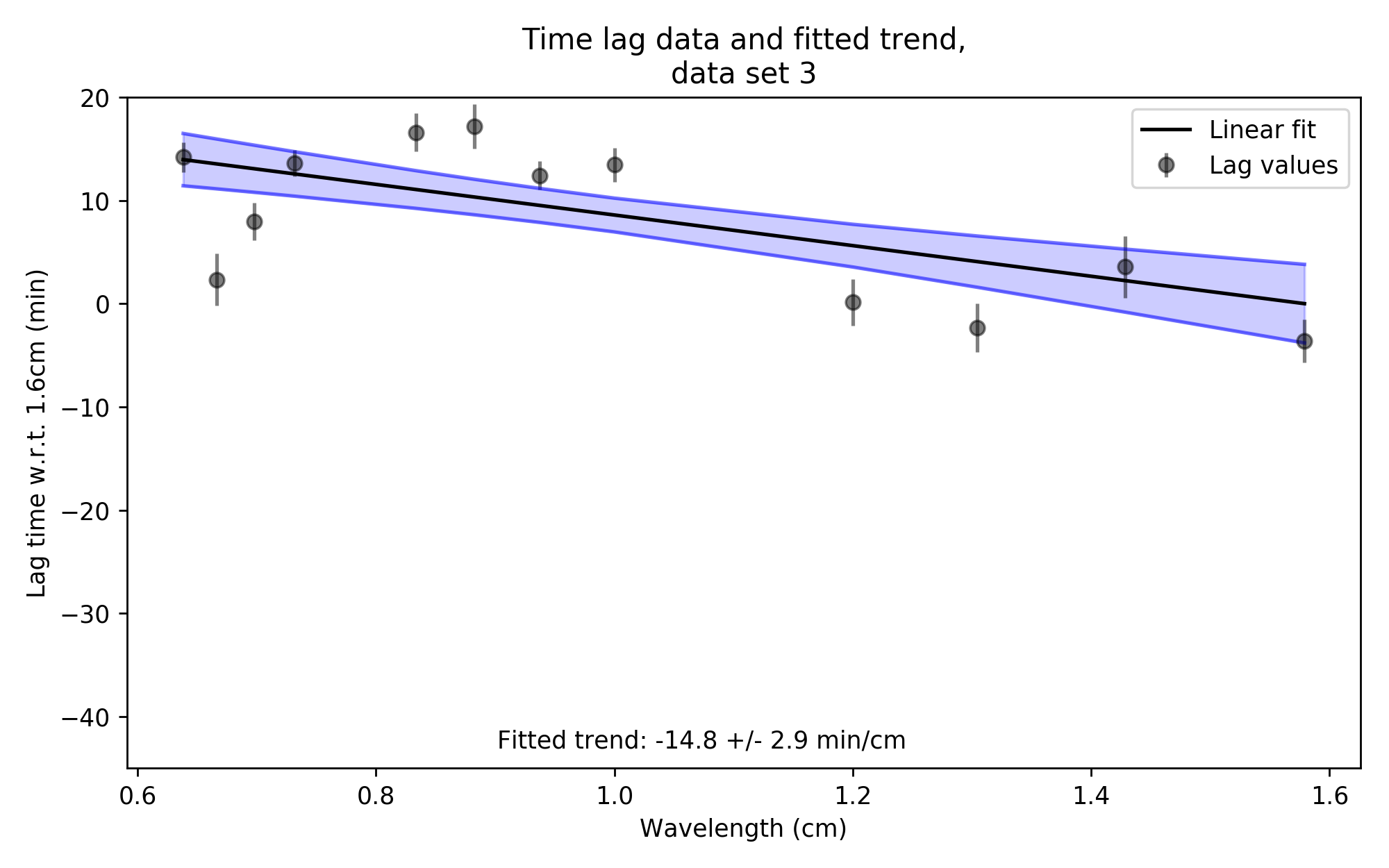}
\includegraphics[width=0.48\textwidth]{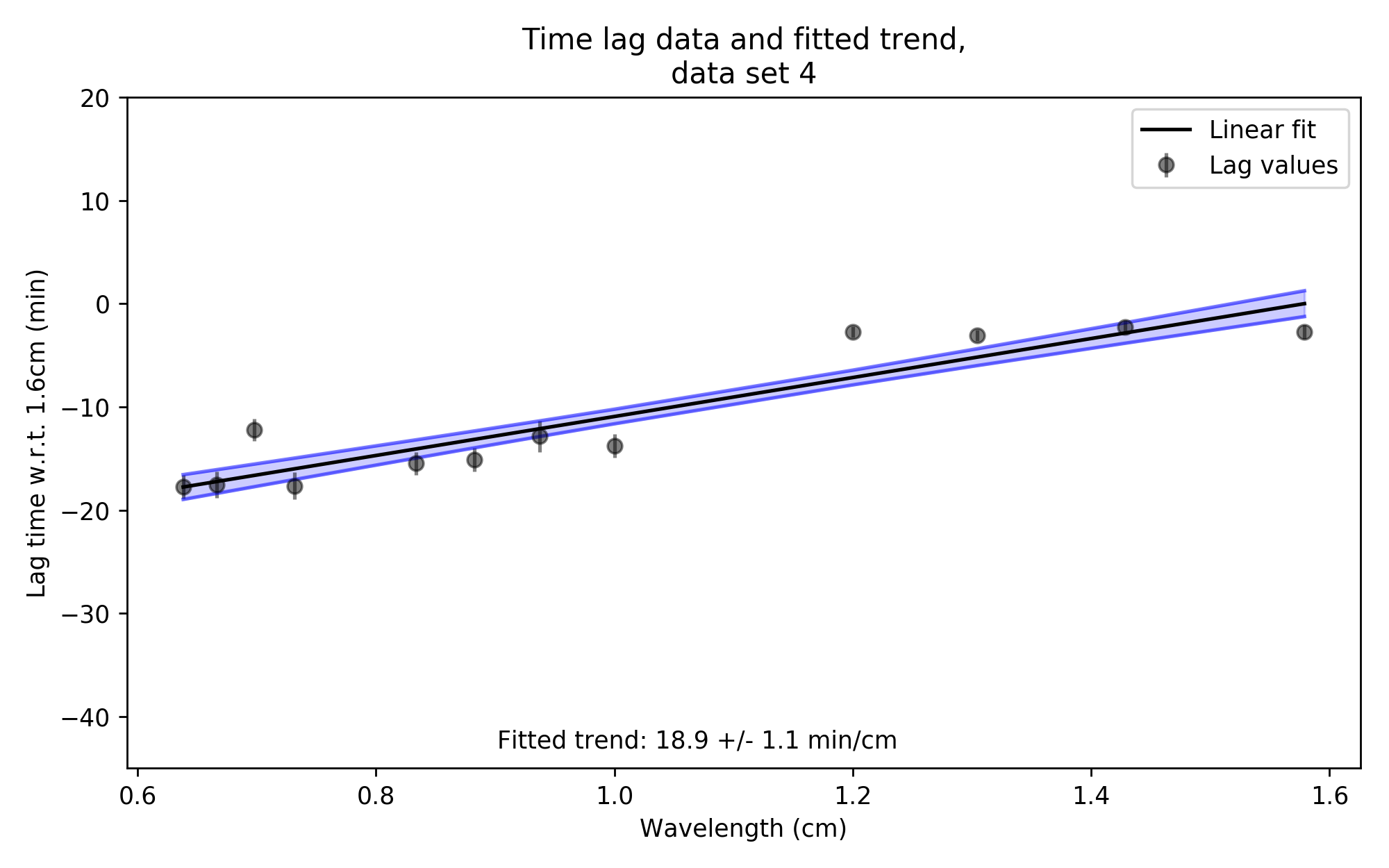}
\includegraphics[width=0.48\textwidth]{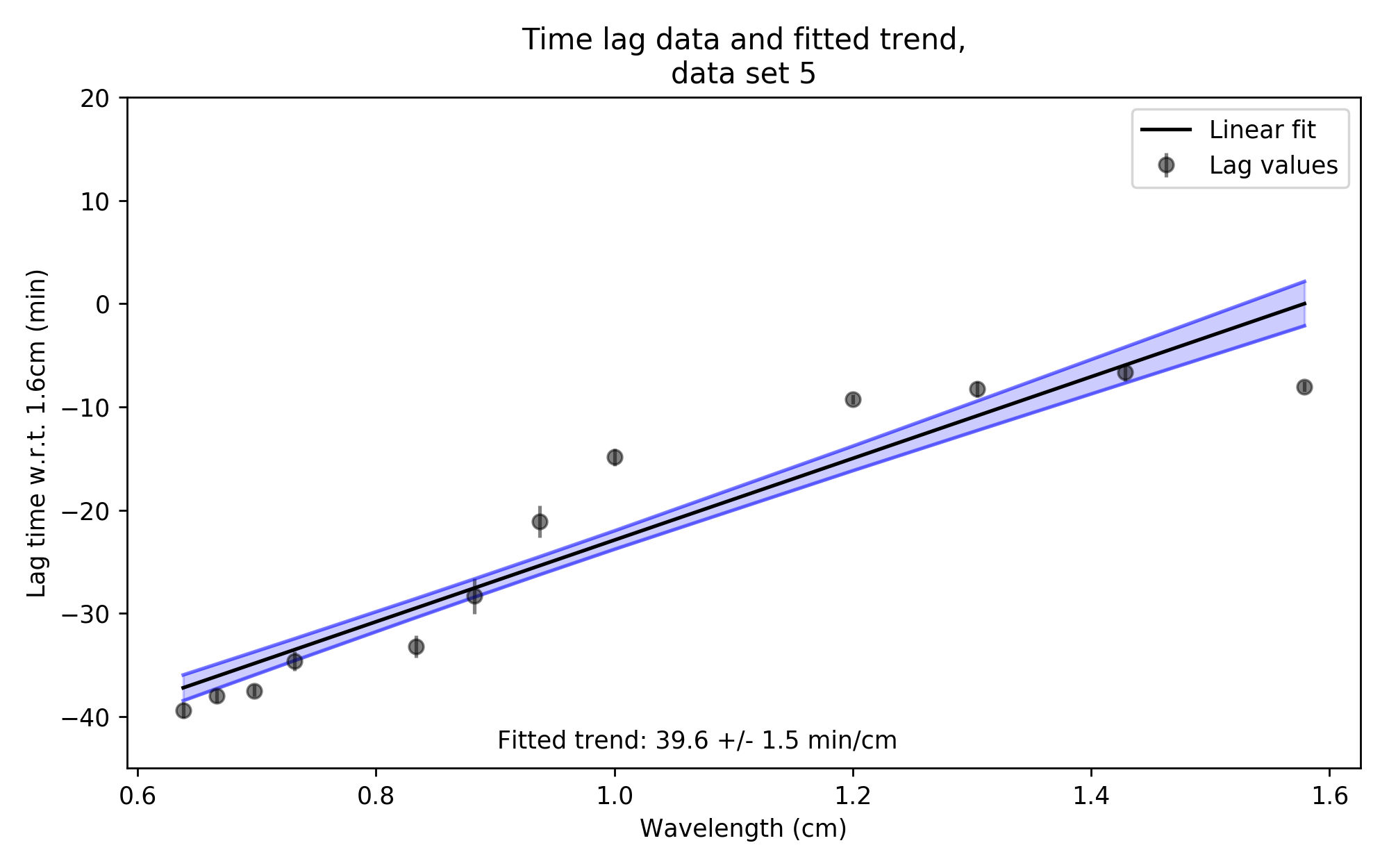}
\includegraphics[width=0.48\textwidth]{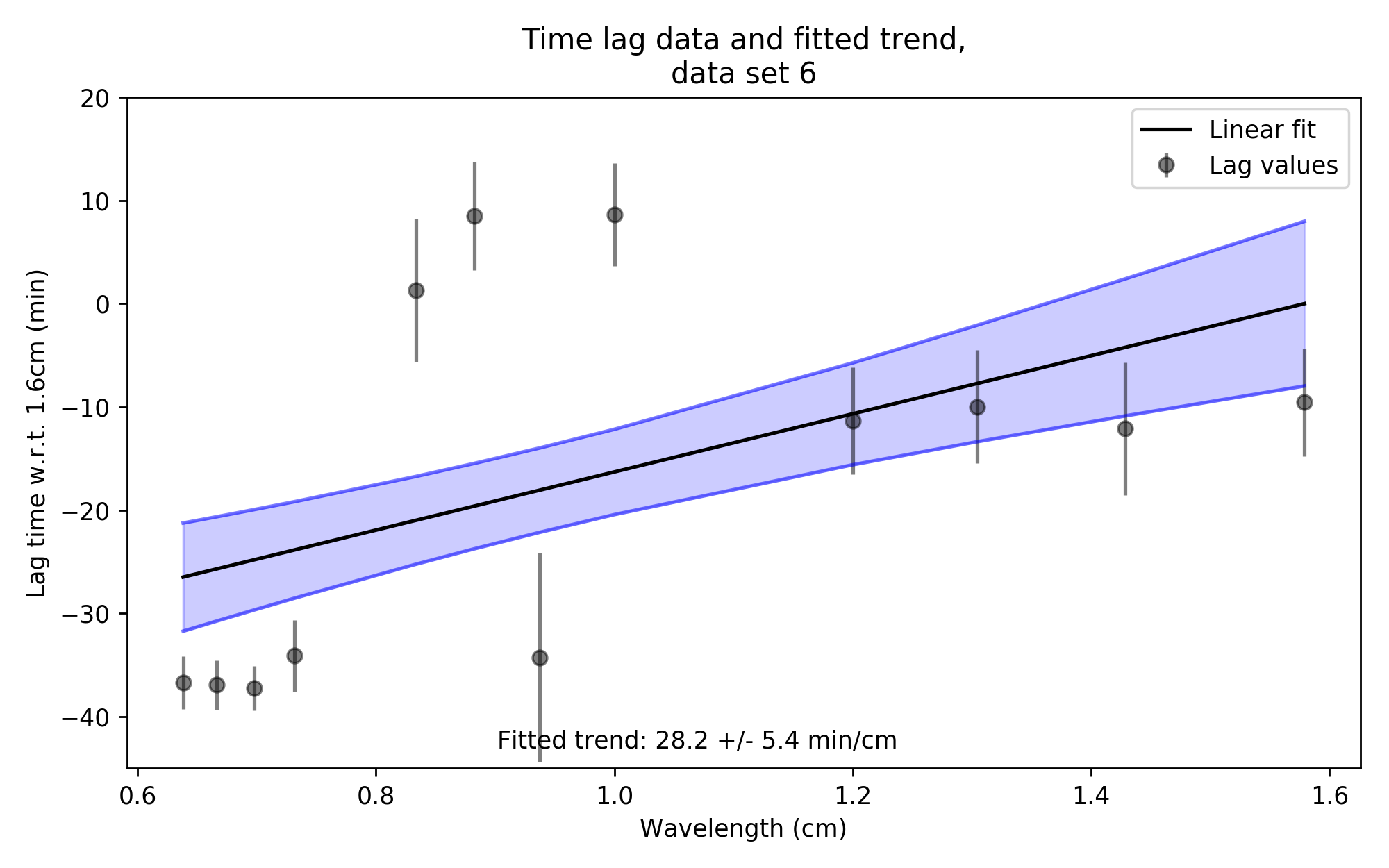}
\includegraphics[width=0.48\textwidth]{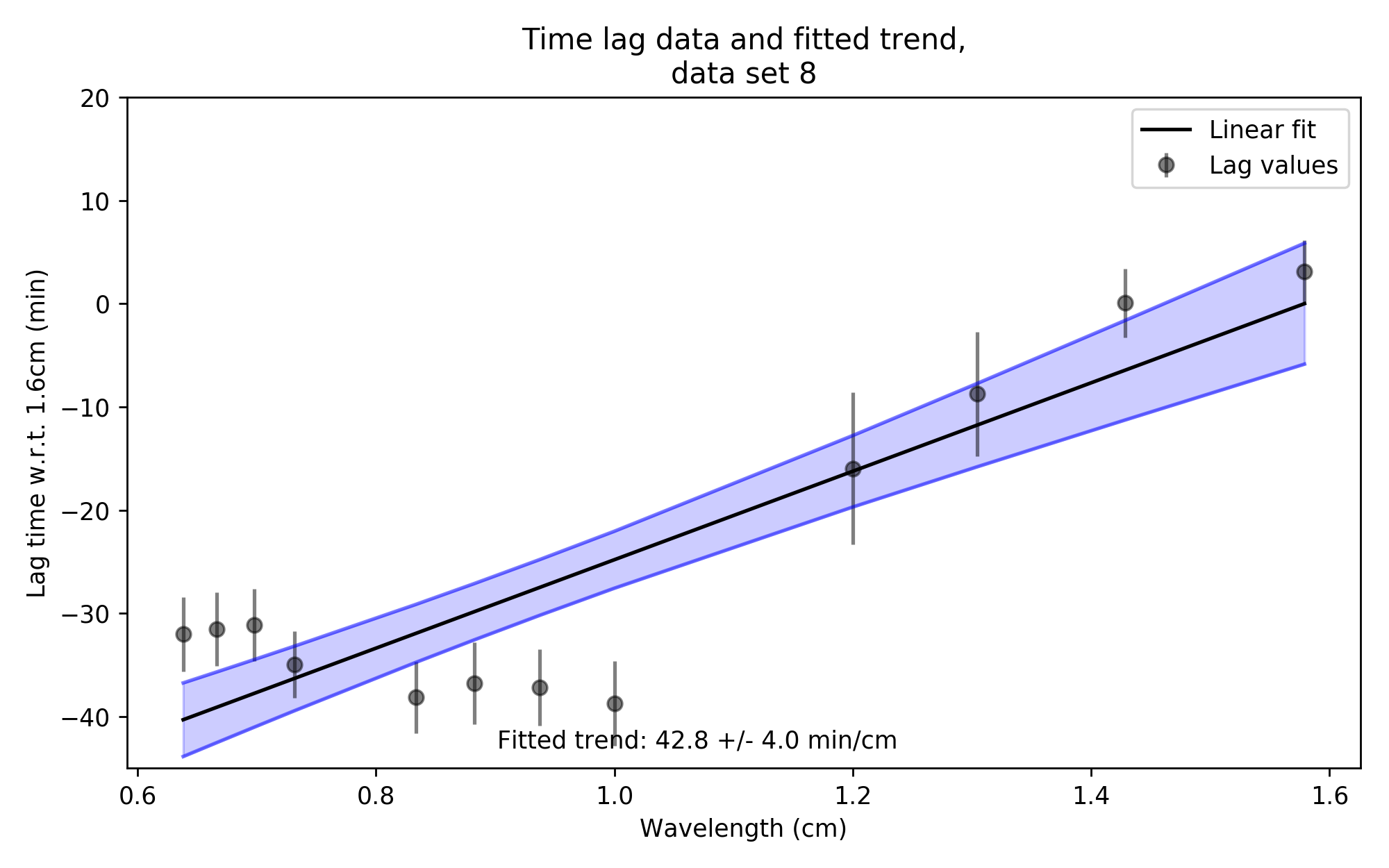}
\includegraphics[width=0.48\textwidth]{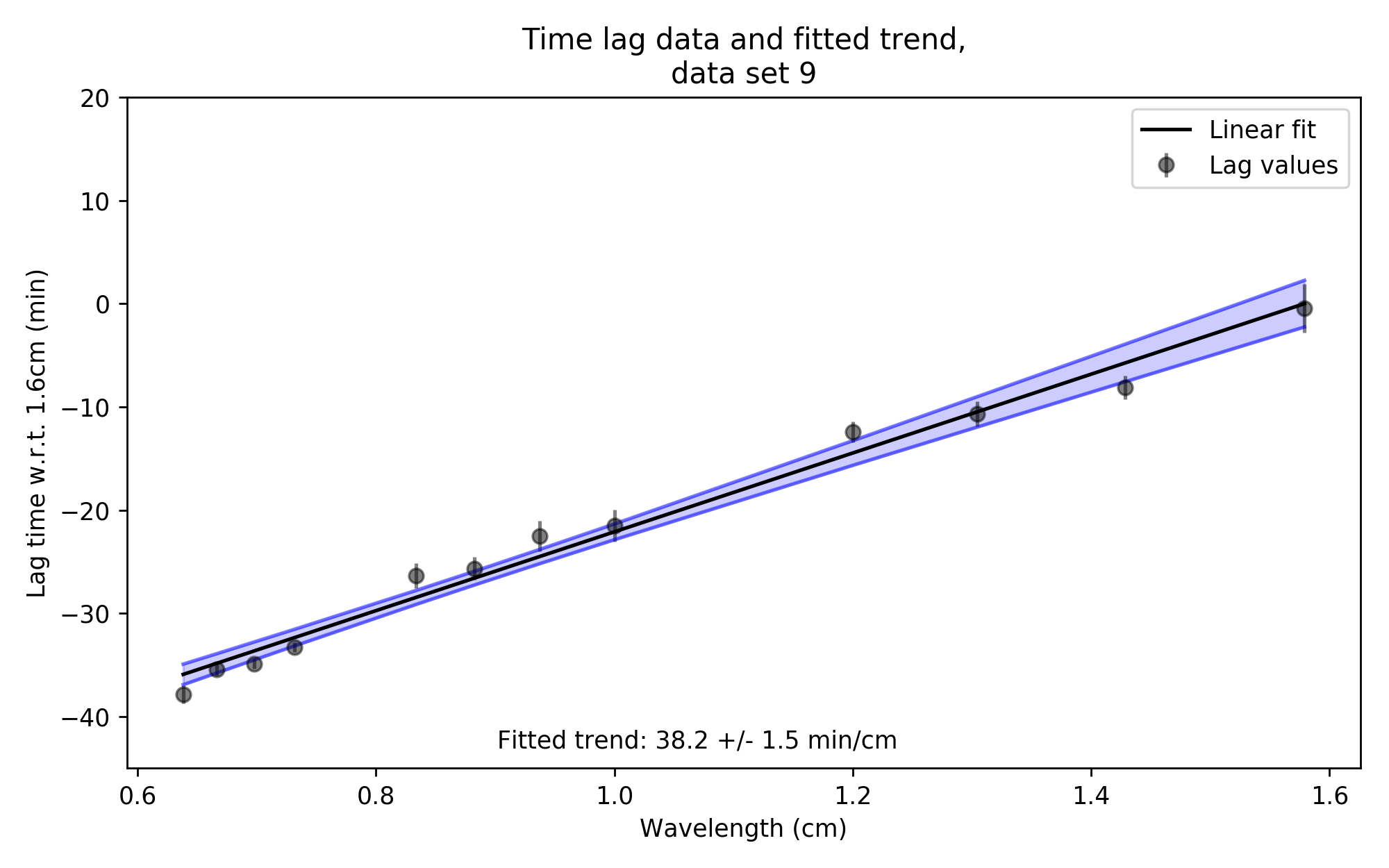}
\caption{Linear time lag versus wavelength trends, fitted to the data shown in Figure \ref{fig:measuredlags}.}
\label{fig:lagfitsdata}
\end{figure*}

\subsection{Checks with synthetic data}

While we do see clear time lags in several of our data sets, the question that immediately arises is why we do not always see them. The answer to this question comes down to three possibilities: 1) the time lags are always present due to the fundamental behaviour of the source, but the statistics of variability does not always let us detect them, 2) the time lags are intrinsic but do not account for all source variability (so that they only sometimes dominate the observed variability), or 3) they are pure chance occurrences that have no physical mechanism or pattern behind them. Despite us now having multiple data sets available, there is still room for different interpretations of these results.

To address this question and investigate the likelihood of these scenarios, we must try to assess the probability with which time lags are detected if they are indeed a fundamental property of the source, as well as the biases we are subject to. Conversely, we also need to assess how likely it is that we appear to see time lags in a data set where no such intrinsic relation between the light curves is present.

To this end, we made use of the formalism of Gaussian processes as described in  \citet{Roberts2013} to model the statistical properties of the variability of Sgr\,A* on the time scales we have access to with these light curves. Specifically, we construct a multidimensional Gaussian distribution with the appropriate covariances from which realisations of synthetic light curves are drawn. To illustrate the fundamental thought behind this formalism, consider for example the task of constructing 10 samples from a light curve, not necessarily equally spaced in time. We know beforehand that these sampled flux density values should be correlated with one another to a certain degree. For one thing they should all have positive values, but more specifically we expect that flux density samples spaced closely together in time should be correlated more strongly than samples with a larger temporal separation between them. This degree of correlation is captured in the auto-correlation function of our measured light curves: using this relation we see how the degree of correlation between samples depends on their separation in time. We fit a functional form to this auto-correlation function and use it as a covariance kernel function. Using this kernel function, we then construct a covariance matrix that is valid for a given set of sampling times.

To make matters simpler, we consider our simulated light curve as a 'standard' function by shifting and scaling it to have zero mean, a variance of unity, and be fully characterised by its auto-correlation function. In this case, the autocorrelation and the covariance are identical. Our sampled values thus become a draw from a 10-dimensional joint Gaussian distribution, the shape of which is defined by the values in our covariance matrix. Every realisation of our example light curve is a single point in this 10-dimensional space. Many separate realisations of the light curve are calculated, and we then study the impact that our sampling cadence and limited epoch length have on the light curve time lag properties we derive.

In our case, we start by modeling the flux density of Sgr\,A* in a single subband as a function of time. As mentioned above, we assume the variability to be stochastic in nature, with a well-defined mean value and an auto-correlation function that does not change with time. With knowledge of this auto-correlation function we generate synthetic light curves that have, on average, the same auto-correlation function as our original data does. Taking a specific realisation we then clone this light curve for our 12 subbands, scale the copies appropriately in flux density and shift them in time according to the time lag relation we wish to investigate. These light curves are then sampled with the same cadence as our observed data sets, with thermal noise and estimated calibration errors added for each virtual measurement. The resulting synthetic data is treated in exactly the same way as our observed data: we correlate these light curves against one another and check if we see the time lag relations that we initially put in. The process by which we search for the appropriate auto-correlation function to use in generating the synthetic light curves is described in full in Appendix E.

With this replication of the behaviour of the measured light curves, we now investigate the detectability of time lag trends using different intrinsic lag trends in our synthetic data. To set up the synthetic sampled light curves for all 12 subbands, we first generate a 'master' light curve using the appropriate auto-correlation function. For each of the 12 subbands we add an additional time shift to this master light curve, according to a time lag relation we define. We then scale the synthetic light curve flux density values according to the average flux density for the appropriate frequency band (see Figure \ref{fig:sgra-history}). The resulting shifted, scaled light curve is then sampled following the scan setup for one epoch of the measured data sets, including the scan timing offsets that pertain to the subband in question (as is visible in Figure \ref{fig:lightcurves-science}).

\begin{figure*}[h]
\centering
\includegraphics[width=0.49\textwidth]{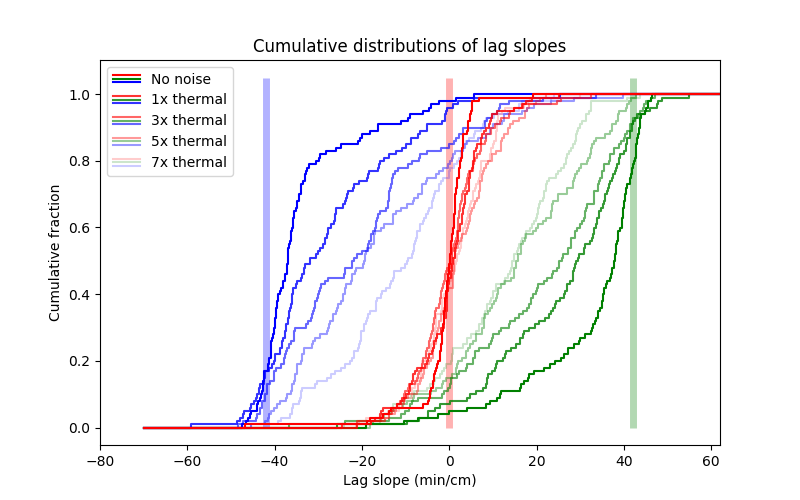}
\includegraphics[width=0.49\textwidth]{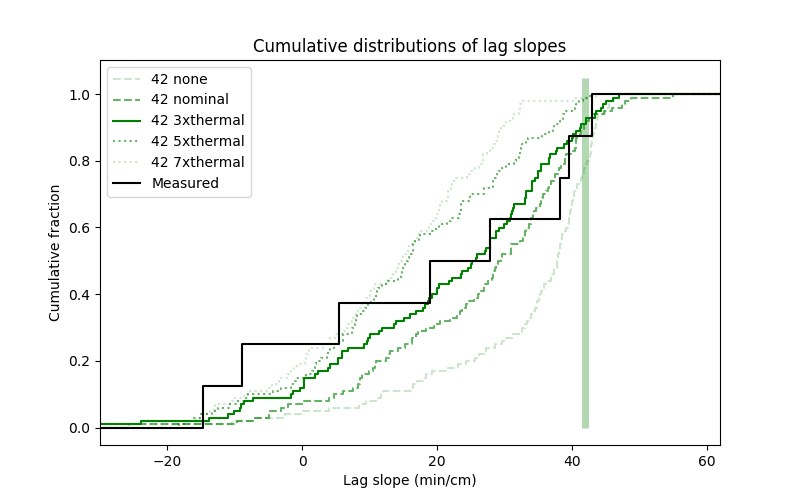}    
\caption{Left: the cumulative distributions of time lag slopes for synthetic light curves, for different input lag relations (blue: -42 min/cm, red: 0 min/cm, green: 42 min/cm), and different contributions of noise (indicated by shading). We see that lag slope distributions creep closer to zero for higher noise contributions, but that the largest recovered values stay close to the input slopes. Right: the cumulative distributions for only the 42 min/cm synthetic lag relation, plotted for different noise contributions and compared to the measured cumulative distribution.}
\label{fig:lagdistributions-synthetic}
\end{figure*}

In our analysis of the synthetic light curves we test three different time lag relations. The first one is the 'nominal' time lag relation which we found in \citet{CDB2015}, $dt/d\lambda = 42$ min/cm, where the variability at lower frequencies lags behind the higher frequencies. The second relation is the inverse of this, $dt/d\lambda = -42$ min/cm, as we wish to check if there is a bias introduced by the scan timing offsets between K, Ka and Q bands. The third relation uses zero time lag, where no shifts in time are applied to any of the light curves and the intrinsic variability is contemporaneous across all frequencies. For each time lag relation, we assess the influence of measurement errors. We compare the distribution of lag slopes we find when no noise is present to cases where the random uncorrelated variations in flux density (i.e. flux density variations that are realised independently per light curve) correspond to our estimated value of $\sim10$mJy on a scan-to-scan timescale, as well as to cases where this random variability component is larger by a factor of 3, 5 or 7. For the cases where we include noise, we also include a realistic calibration error due to the combined influence of unmodeled minor variations in the flux density of our gain calibrator and atmospheric effects. These errors are strongly correlated across each frequency band, meaning that all four subbands within one frequency band show coherent flux density offsets due to this calibration uncertainty. We estimate the magnitude of this calibration error to be $\sim$30 mJy, from the cross-band trend breaks in our measured data (see Figures \ref{fig:lightcurves-science} and \ref{fig:cal-check-lc-sample}, as well as the check source light curves in Appendix A). For each combination of parameters (time lag relation, thermal noise strength) we generate 100 realisations from which we reconstruct the observed time lag slope in the same way as was done for our measured data, and so we get a distribution of lag slopes for each choice of parameters. A sample of generated light curves, their cross-correlations and the recovered lag trends are shown in Appendix C. An overview showing the distributions of the recovered time lag slopes for all considered cases is shown in Figure \ref{fig:lagdistributions-synthetic}.

\section{Results}

In our observed data sets, we reconstruct strong time lag trends in two epochs with high internal consistency ($39.6 \pm 1.5$ min/cm and $38.2 \pm 1.4$ min/cm respectively for data sets 5 and 9), another significant but less consistent time lag in one epoch ($43.1 \pm 4.0$ min/cm, for data set 8), one weaker time lag in another epoch ($18.9 \pm 1.1$ min/cm, for data set 4) and no clear or internally consistent results for the other epochs. Although the linear time lag fits for these other data sets show relatively high confidence in the trend uncertainties (small errors in the fitted trend slopes), we consider these reconstructions less reliable as the individual lag relations show very low internal consistency across all 12 subbands with a large residual scatter after their co-registration. Slope errors obtained from bootstrapping carry the implicit assumption that the data follow the presupposed linear relation, which is not clearly present in certain cases and manifests itself as a discrepancy between the bootstrapping errors and the spread between recovered lag trends.

An identical analysis was performed on the light curves from our check source J1745, the results of which exhibit no time lag patterns similar to those of Sgr\,A*. This shows that the variability we measure in Sgr\,A* is not correlated to that of the check source, which would have suggested data calibration artefacts. The cross-correlation plots generated for J1745 are shown in Appendix B, with a selection of plots for every epoch.

We wish to understand how likely it is that a measurable time lag is detected in any given epoch for a source where a persistent time lag relation is present between the light curves at different frequencies. To this end, we generated synthetic data with the same general variability properties as those that we observed in the measured light curves. As we see from Figure \ref{fig:lagdistributions-synthetic}, the magnitude of the time lag trends from synthetic data tend to be underpredicted by the reconstructions, even in the case where zero-error measurements are modeled. The true lag slopes used as input for the synthetic light curves are plotted using vertical coloured bars in the left panel, and it is apparent that for the non-zero time lag relations the distribution of the corresponding recovered lag slope is such that the input lag slope value is close to an extremum of the distribution.

For both the tested positive and negative time lag trends, we see an asymmetric distribution in the recovered trends with longer tails towards zero time lag (and in a few cases even showing recovered lags with a sign opposite from the input lag relation). For the zero-lag case, we see a symmetrical spread in the recovered trends without any discernible bias to positive or negative slopes. Overprediction of the lag trend slope in the reconstructions, on the other hand, is much more rare and never extends far away from the input lag slope values. The asymmetry in reconstructed lag slope distributions is more severe when calibration errors are included, causing a larger bias towards smaller measured lag trend slopes. Thus, we see that a time lag relation that is consistently present in the data will tend to be underestimated from measurements using the cadence of our observations, and that among many measurements of the time lag relation the larger reconstructed slopes are a better estimate of the true time lag relation that is present. The set of synthetic light curves that best reproduces the distribution of time lag slopes we find in our observations incorporate a stronger uncorrelated noise component than our initial estimate (see the right panel of Figure \ref{fig:lagdistributions-synthetic}) - an uncorrelated random flux density component with a standard deviation of $\sim$30 mJy per scan (plotted as '3xthermal') gives the best match to the observed lag distribution. Whether this is due to an underestimation of measurement error or to an unmodeled random component of source variability is an open issue.

\section{Discussion and Conclusions}

In this paper, we have presented the most comprehensive and systematic study of time lags in Sagittarius\,A* radio observations performed to date. We find positive time lags (between 20 and 40 min/cm) over a frequency range from 47 down to 19 GHz, where low-frequency variability lags behind high-frequency variability, for five out of eight epochs in our data. Three out of eight epochs show time lags close to zero. The larger time lags are found for epochs where the light curves show more pronounced local maxima or minima.

In our synthetic data analysis, we find that the reconstructed time lag slopes are typically underestimated with the measurement cadence and length that we have used for our observations, particularly when variability is low and no recognisable flaring occurs. The distribution of recovered synthetic time lag slopes closely matches the observed distribution when a lag relation of 42 min/cm is assumed, together with a 30\,mJy uncorrelated variability component.

The lag relation found in our previous paper \citep{CDB2015} was calculated from a set of light curves that showed a clear temporal maximum for all of them, and we see comparable lag slopes for the light curves that have a clear flux density extremum here -- both for the  measured light curves and for the synthetically generated light curves. We are therefore confident that we have observed a persistent and consistent time lag in these results. The consistency of the time lag slope over multiple years of measurement (covering the time range from 2005 to 2015) indicates that we are sampling some kind of characteristic velocity, scale or expansion speed.

\begin{figure}
\centering
\includegraphics[width=0.49\textwidth]{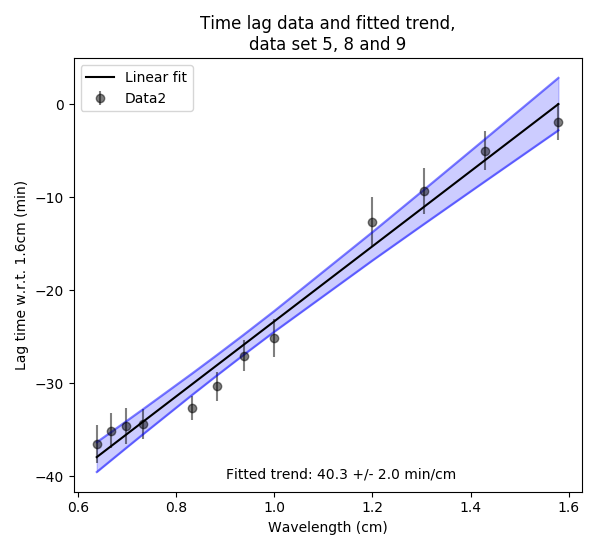}
\caption{The aggregate time lag trend fitted to the measured time lags from data sets 5, 8 and 9 after co-registration.}
\label{fig:lag589}
\end{figure}

Combining the time lag values found in the three epochs with the largest time lag slopes, which lie close together in value, gives us a fitted slope of $40.3 \pm 2.0$ min/cm, which is plotted in Figure \ref{fig:lag589}.

\begin{figure}
\centering
\includegraphics[width=0.48\textwidth]{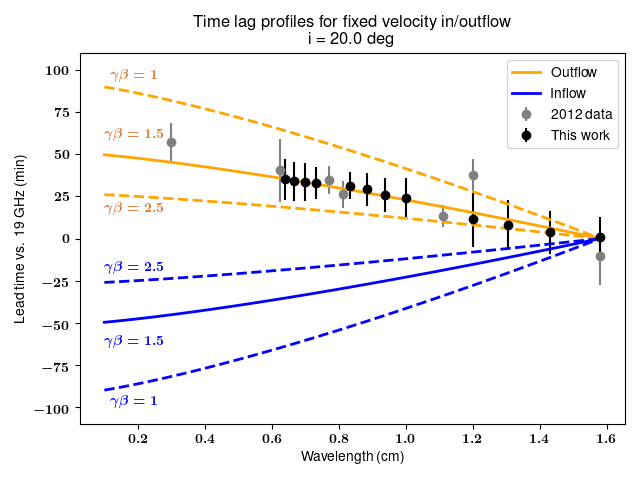}
\caption{Constant-velocity lead times w.r.t. the 19 GHz peak (calculated). Coloured curves show the expected time lag relations for an inclination of 20$^\circ$ when a size-frequency relation of $\theta[\textrm{mas}] = 0.52 \cdot \lambda[\textrm{cm}]^{1.3}$ is assumed. Data from this work and from \citet{CDB2015} is overplotted.}
\label{fig:constveloutflow}
\end{figure}

When we consider the constraints on the inclination angle of a putative compact jet from Sgr\,A* that follow from recently published results, we see that small values are favoured: \citet{Issaoun2019} report good fits for jet models with $i \leq 20^\circ$, while \citet{GRAVITY2020} indicate that BH spin axis inclinations larger than 140$^\circ$ fit well for circular hotspot motion, with the value increasing to 175$^\circ$ for helical hotspot motion. Both of these scenarios correspond to a small angle between our line of sight and the jet direction pointing closest to us. Combining this information with the size-frequency relation for Sgr\,A* \citep{Bower2006, Falcke2009} and our data, like we did in \citet{CDB2015}, results in an estimate for the associated outflow speed (see Figure \ref{fig:constveloutflow}). We find that the plasma flows out at moderately relativistic speeds with $\gamma\beta \approx 1.5$, which is well within the range predicted by both analytic jet models \citep{Falcke1996b} and jets in GRMHD simulations \citep{Moscibrodzka2013}. Here, the assumption is made that the variability we see at the various wavelengths is associated with the angular scales given by the size-frequency relation, and not limited to any smaller sub-region of the source. There is some uncertainty associated with this, as the size of the flaring component has not been directly measured. However, given that stronger flares can reach a significant fraction of the total flux density, it seems unlikely that they are emitted at scales widely separated from the average structure probed by VLBI. The size-frequency relation of Sgr\,A* is well-constrained if a particular functional form is assumed, but shows a larger uncertainty once different forms of the relation are considered \citep{CDB2019}. In general, the fact that the photosphere lies further out at lower frequencies suggests that flares at these frequencies involve some temporal evolution of the shape of this photosphere. With time lags appearing to always be present in the emission from Sgr\,A* across radio frequencies, the scenario where an organised outflow is present is favoured.

From our investigation using synthetic data with a known intrinsic time lag trend, we see that there is a clear bias towards smaller reconstructed time lag slopes when the employed measurement strategy and data calibration quality are considered. For future observations, an observing strategy where light curves are sampled over longer time scales (i.e., over 5 or 6 hours) may aid in a more precise measurement and characterisation of these time lags, for two reasons: the larger number of data points will reduce the impact of spurious flux density variations from calibration uncertainties on the measured cross-correlation functions, and the likelihood of catching and fully tracing flux density excursions also increases with longer epochs.

The persistent time lag relation we observe between different radio observing frequencies provides another benchmark by which to test theoretical emission models for Sgr\,A*, next to the other properties of the source that have been taken into account in model fitting so far (spectral shape, variability strength, angular size and polarisation properties).

\begin{acknowledgements}
This work was supported by the ERC Synergy Grant 610058: "BlackHoleCam: Imaging the Event Horizon of a Black Hole". The author wishes to thank Federica Palla for her initial analysis of the data, which motivated this work.
\end{acknowledgements}


\bibliographystyle{aa}
\bibliography{bibliography}

\begin{thebibliography}{61}
\expandafter\ifx\csname natexlab\endcsname\relax\def\natexlab#1{#1}\fi

\bibitem[{Abuter {et~al.}(2020)Abuter, Amorim, Bauböck, Berger, Bonnet,
  Brandner, Cardoso, Clénet, de~Zeeuw, \& et~al.}]{Abuter2020}
Abuter, R., Amorim, A., Bauböck, M., {et~al.} 2020, \aap, 638, A2

\bibitem[{{Alexander}(1997)}]{Alexander1997}
{Alexander}, T. 1997, Astrophysics and Space Science Library, Vol. 218, {Is AGN
  Variability Correlated with Other AGN Properties? ZDCF Analysis of Small
  Samples of Sparse Light Curves} (Springer), 163

\bibitem[{{An} {et~al.}(2005){An}, {Goss}, {Zhao}, {Hong}, {Roy}, {Rao}, \&
  {Shen}}]{An2005}
{An}, T., {Goss}, W.~M., {Zhao}, J.-H., {et~al.} 2005, \apjl, 634, L49

\bibitem[{{Balick} \& {Brown}(1974)}]{BalickBrown1974}
{Balick}, B. \& {Brown}, R.~L. 1974, \apj, 194, 265

\bibitem[{{Ball} {et~al.}(2020){Ball}, {{\"O}zel}, {Christian}, {Chan}, \&
  {Psaltis}}]{Ball2020}
{Ball}, D., {{\"O}zel}, F., {Christian}, P., {Chan}, C.-K., \& {Psaltis}, D.
  2020, arXiv e-prints, arXiv:2005.14251

\bibitem[{{Bower} {et~al.}(2019){Bower}, {Dexter}, {Asada}, {Brinkerink},
  {Falcke}, {Ho}, {Inoue}, {Markoff}, {Marrone}, {Matsushita}, {Moscibrodzka},
  {Nakamura}, {Peck}, \& {Rao}}]{Bower2019}
{Bower}, G.~C., {Dexter}, J., {Asada}, K., {et~al.} 2019, \apjl, 881, L2

\bibitem[{{Bower} {et~al.}(2004){Bower}, {Falcke}, {Herrnstein}, {Zhao},
  {Goss}, \& {Backer}}]{Bower2004}
{Bower}, G.~C., {Falcke}, H., {Herrnstein}, R.~M., {et~al.} 2004, Science, 304,
  704

\bibitem[{{Bower} {et~al.}(2006){Bower}, {Goss}, {Falcke}, {Backer}, \&
  {Lithwick}}]{Bower2006}
{Bower}, G.~C., {Goss}, W.~M., {Falcke}, H., {Backer}, D.~C., \& {Lithwick}, Y.
  2006, \apjl, 648, L127

\bibitem[{{Bower} {et~al.}(2015){Bower}, {Markoff}, {Dexter}, {Gurwell},
  {Moran}, {Brunthaler}, {Falcke}, {Fragile}, {Maitra}, {Marrone}, {Peck},
  {Rushton}, \& {Wright}}]{Bower2015}
{Bower}, G.~C., {Markoff}, S., {Dexter}, J., {et~al.} 2015, \apj, 802, 69

\bibitem[{{Brinkerink} {et~al.}(2015){Brinkerink}, {Falcke}, {Law}, {Barkats},
  {Bower}, {Brunthaler}, {Gammie}, {Impellizzeri}, {Markoff}, {Menten},
  {Moscibrodzka}, {Peck}, {Rushton}, {Schaaf}, \& {Wright}}]{CDB2015}
{Brinkerink}, C.~D., {Falcke}, H., {Law}, C.~J., {et~al.} 2015, \aap, 576, A41

\bibitem[{{Brinkerink} {et~al.}(2016){Brinkerink}, {M{\"u}ller}, {Falcke},
  {Bower}, {Krichbaum}, {Castillo}, {Deller}, {Doeleman}, {Fraga-Encinas},
  {Goddi}, {Hern{\'a}ndez-G{\'o}mez}, {Hughes}, {Kramer}, {L{\'e}on-Tavares},
  {Loinard}, {Monta{\~n}a}, {Mo{\'s}cibrodzka}, {Ortiz-Le{\'o}n},
  {Sanchez-Arguelles}, {Tilanus}, {Wilson}, \& {Zensus}}]{CDB2016}
{Brinkerink}, C.~D., {M{\"u}ller}, C., {Falcke}, H., {et~al.} 2016, \mnras,
  462, 1382

\bibitem[{{Brinkerink} {et~al.}(2019){Brinkerink}, {M{\"u}ller}, {Falcke},
  {Issaoun}, {Akiyama}, {Bower}, {Krichbaum}, {Deller}, {Castillo}, {Doeleman},
  {Fraga-Encinas}, {Goddi}, {Hern{\'a}ndez-G{\'o}mez}, {Hughes}, {Kramer},
  {L{\'e}on-Tavares}, {Loinard}, {Monta{\~n}a}, {Mo{\'s}cibrodzka},
  {Ortiz-Le{\'o}n}, {Sanchez-Arguelles}, {Tilanus}, {Wilson}, \&
  {Zensus}}]{CDB2019}
{Brinkerink}, C.~D., {M{\"u}ller}, C., {Falcke}, H.~D., {et~al.} 2019, \aap,
  621, A119

\bibitem[{Dexter {et~al.}(2014)Dexter, Kelly, Bower, Marrone, Stone, \&
  Plambeck}]{Dexter2014}
Dexter, J., Kelly, B., Bower, G.~C., {et~al.} 2014, Monthly Notices of the
  Royal Astronomical Society, 442, 2797–2808

\bibitem[{{Doeleman} {et~al.}(2008){Doeleman}, {Weintroub}, {Rogers},
  {Plambeck}, {Freund}, {Tilanus}, {Friberg}, {Ziurys}, {Moran}, {Corey},
  {Young}, {Smythe}, {Titus}, {Marrone}, {Cappallo}, {Bock}, {Bower},
  {Chamberlin}, {Davis}, {Krichbaum}, {Lamb}, {Maness}, {Niell}, {Roy},
  {Strittmatter}, {Werthimer}, {Whitney}, \& {Woody}}]{Doeleman2008}
{Doeleman}, S.~S., {Weintroub}, J., {Rogers}, A. E.~E., {et~al.} 2008, \nat,
  455, 78

\bibitem[{{Eckart} {et~al.}(2018){Eckart}, {Zajacek}, {Parsa}, {Fazeli},
  {Busch}, {Shahzamanian}, {Subroweit}, {Peissker}, {Sabha}, {Valencia-S.},
  {Horrobin}, {Straubmeier}, {Rost}, {Borkar}, {Karas}, {Britzen}, {Zensus}, \&
  {Kamali}}]{Eckart2018}
{Eckart}, A., {Zajacek}, M., {Parsa}, M., {et~al.} 2018, arXiv e-prints,
  arXiv:1806.00284

\bibitem[{{Event Horizon Telescope Collaboration} {et~al.}(2019){Event Horizon
  Telescope Collaboration}, {Akiyama}, {Alberdi}, {Alef}, {Asada}, {Azulay},
  {Baczko}, {Ball}, {Balokovi{\'c}}, {Barrett}, {Bintley}, {Blackburn},
  {Boland}, {Bouman}, {Bower}, {Bremer}, {Brinkerink}, {Brissenden}, {Britzen},
  {Broderick}, {Broguiere}, {Bronzwaer}, {Byun}, {Carlstrom}, {Chael}, {Chan},
  {Chatterjee}, {Chatterjee}, {Chen}, {Chen}, {Cho}, {Christian}, {Conway},
  {Cordes}, {Crew}, {Cui}, {Davelaar}, {De Laurentis}, {Deane}, {Dempsey},
  {Desvignes}, {Dexter}, {Doeleman}, {Eatough}, {Falcke}, {Fish}, {Fomalont},
  {Fraga-Encinas}, {Friberg}, {Fromm}, {G{\'o}mez}, {Galison}, {Gammie},
  {Garc{\'\i}a}, {Gentaz}, {Georgiev}, {Goddi}, {Gold}, {Gu}, {Gurwell},
  {Hada}, {Hecht}, {Hesper}, {Ho}, {Ho}, {Honma}, {Huang}, {Huang}, {Hughes},
  {Ikeda}, {Inoue}, {Issaoun}, {James}, {Jannuzi}, {Janssen}, {Jeter}, {Jiang},
  {Johnson}, {Jorstad}, {Jung}, {Karami}, {Karuppusamy}, {Kawashima},
  {Keating}, {Kettenis}, {Kim}, {Kim}, {Kim}, {Kino}, {Koay}, {Koch}, {Koyama},
  {Kramer}, {Kramer}, {Krichbaum}, {Kuo}, {Lauer}, {Lee}, {Li}, {Li},
  {Lindqvist}, {Liu}, {Liuzzo}, {Lo}, {Lobanov}, {Loinard}, {Lonsdale}, {Lu},
  {MacDonald}, {Mao}, {Markoff}, {Marrone}, {Marscher}, {Mart{\'\i}-Vidal},
  {Matsushita}, {Matthews}, {Medeiros}, {Menten}, {Mizuno}, {Mizuno}, {Moran},
  {Moriyama}, {Moscibrodzka}, {M{\"u}ller}, {Nagai}, {Nagar}, {Nakamura},
  {Narayan}, {Narayanan}, {Natarajan}, {Neri}, {Ni}, {Noutsos}, {Okino},
  {Olivares}, {Ortiz-Le{\'o}n}, {Oyama}, {{\"O}zel}, {Palumbo}, {Patel}, {Pen},
  {Pesce}, {Pi{\'e}tu}, {Plambeck}, {PopStefanija}, {Porth}, {Prather},
  {Preciado-L{\'o}pez}, {Psaltis}, {Pu}, {Ramakrishnan}, {Rao}, {Rawlings},
  {Raymond}, {Rezzolla}, {Ripperda}, {Roelofs}, {Rogers}, {Ros}, {Rose},
  {Roshanineshat}, {Rottmann}, {Roy}, {Ruszczyk}, {Ryan}, {Rygl},
  {S{\'a}nchez}, {S{\'a}nchez-Arguelles}, {Sasada}, {Savolainen}, {Schloerb},
  {Schuster}, {Shao}, {Shen}, {Small}, {Sohn}, {SooHoo}, {Tazaki}, {Tiede},
  {Tilanus}, {Titus}, {Toma}, {Torne}, {Trent}, {Trippe}, {Tsuda}, {van
  Bemmel}, {van Langevelde}, {van Rossum}, {Wagner}, {Wardle}, {Weintroub},
  {Wex}, {Wharton}, {Wielgus}, {Wong}, {Wu}, {Young}, {Young}, {Younsi},
  {Yuan}, {Yuan}, {Zensus}, {Zhao}, {Zhao}, {Zhu}, {Algaba}, {Allardi},
  {Amestica}, {Bach}, {Beaudoin}, {Benson}, {Berthold}, {Blanchard},
  {Blundell}, {Bustamente}, {Cappallo}, {Castillo-Dom{\'\i}nguez}, {Chang},
  {Chang}, {Chang}, {Chen}, {Chilson}, {Chuter}, {C{\'o}rdova Rosado},
  {Coulson}, {Crawford}, {Crowley}, {David}, {Derome}, {Dexter}, {Dornbusch},
  {Dudevoir}, {Dzib}, {Eckert}, {Erickson}, {Everett}, {Faber}, {Farah},
  {Fath}, {Folkers}, {Forbes}, {Freund}, {G{\'o}mez-Ruiz}, {Gale}, {Gao},
  {Geertsema}, {Graham}, {Greer}, {Grosslein}, {Gueth}, {Halverson}, {Han},
  {Han}, {Hao}, {Hasegawa}, {Henning}, {Hern{\'a}ndez-G{\'o}mez},
  {Herrero-Illana}, {Heyminck}, {Hirota}, {Hoge}, {Huang}, {Impellizzeri},
  {Jiang}, {Kamble}, {Keisler}, {Kimura}, {Kono}, {Kubo}, {Kuroda}, {Lacasse},
  {Laing}, {Leitch}, {Li}, {Lin}, {Liu}, {Liu}, {Lu}, {Marson},
  {Martin-Cocher}, {Massingill}, {Matulonis}, {McColl}, {McWhirter}, {Messias},
  {Meyer-Zhao}, {Michalik}, {Monta{\~n}a}, {Montgomerie}, {Mora-Klein},
  {Muders}, {Nadolski}, {Navarro}, {Nguyen}, {Nishioka}, {Norton}, {Nystrom},
  {Ogawa}, {Oshiro}, {Oyama}, {Padin}, {Parsons}, {Paine}, {Pe{\~n}alver},
  {Phillips}, {Poirier}, {Pradel}, {Primiani}, {Raffin}, {Rahlin}, {Reiland},
  {Risacher}, {Ruiz}, {S{\'a}ez-Mada{\'\i}n}, {Sassella}, {Schellart}, {Shaw},
  {Silva}, {Shiokawa}, {Smith}, {Snow}, {Souccar}, {Sousa}, {Sridharan},
  {Srinivasan}, {Stahm}, {Stark}, {Story}, {Timmer}, {Vertatschitsch},
  {Walther}, {Wei}, {Whitehorn}, {Whitney}, {Woody}, {Wouterloot}, {Wright},
  {Yamaguchi}, {Yu}, {Zeballos}, \& {Ziurys}}]{EHT2019II}
{Event Horizon Telescope Collaboration}, {Akiyama}, K., {Alberdi}, A., {et~al.}
  2019, \apjl, 875, L2

\bibitem[{{Falcke}(1999)}]{Falcke1999}
{Falcke}, H. 1999, in Astronomical Society of the Pacific Conference Series,
  Vol. 186, The Central Parsecs of the Galaxy, ed. H.~{Falcke}, A.~{Cotera},
  W.~J. {Duschl}, F.~{Melia}, \& M.~J. {Rieke}, 113

\bibitem[{{Falcke} \& {Biermann}(1996)}]{Falcke1996b}
{Falcke}, H. \& {Biermann}, P.~L. 1996, \aap, 308, 321

\bibitem[{{Falcke} {et~al.}(1998){Falcke}, {Goss}, {Matsuo}, {Teuben}, {Zhao},
  \& {Zylka}}]{Falcke1998}
{Falcke}, H., {Goss}, W.~M., {Matsuo}, H., {et~al.} 1998, \apj, 499, 731

\bibitem[{{Falcke} {et~al.}(1993){Falcke}, {Mannheim}, \&
  {Biermann}}]{Falcke1993}
{Falcke}, H., {Mannheim}, K., \& {Biermann}, P.~L. 1993, \aap, 278, L1

\bibitem[{{Falcke} \& {Markoff}(2000)}]{Falcke2000}
{Falcke}, H. \& {Markoff}, S. 2000, \aap, 362, 113

\bibitem[{{Falcke} {et~al.}(2009){Falcke}, {Markoff}, \& {Bower}}]{Falcke2009}
{Falcke}, H., {Markoff}, S., \& {Bower}, G.~C. 2009, \aap, 496, 77

\bibitem[{{Ghez} {et~al.}(2008){Ghez}, {Salim}, {Weinberg}, {Lu}, {Do}, {Dunn},
  {Matthews}, {Morris}, {Yelda}, {Becklin}, {Kremenek}, {Milosavljevic}, \&
  {Naiman}}]{Ghez2008}
{Ghez}, A.~M., {Salim}, S., {Weinberg}, N.~N., {et~al.} 2008, \apj, 689, 1044

\bibitem[{{Gillessen} {et~al.}(2009){Gillessen}, {Eisenhauer}, {Trippe},
  {Alexand er}, {Genzel}, {Martins}, \& {Ott}}]{Gillessen2009}
{Gillessen}, S., {Eisenhauer}, F., {Trippe}, S., {et~al.} 2009, \apj, 692, 1075

\bibitem[{{Goldston} {et~al.}(2005){Goldston}, {Quataert}, \&
  {Igumenshchev}}]{Goldston2005}
{Goldston}, J.~E., {Quataert}, E., \& {Igumenshchev}, I.~V. 2005, \apj, 621,
  785

\bibitem[{{Gravity Collaboration} {et~al.}(2018){Gravity Collaboration},
  {Abuter}, {Amorim}, {Baub{\"o}ck}, {Berger}, {Bonnet}, {Brand ner},
  {Cl{\'e}net}, {Coud{\'e} Du Foresto}, {de Zeeuw}, {Deen}, {Dexter}, {Duvert},
  {Eckart}, {Eisenhauer}, {F{\"o}rster Schreiber}, {Garcia}, {Gao}, {Gendron},
  {Genzel}, {Gillessen}, {Guajardo}, {Habibi}, {Haubois}, {Henning}, {Hippler},
  {Horrobin}, {Huber}, {Jim{\'e}nez-Rosales}, {Jocou}, {Kervella}, {Lacour},
  {Lapeyr{\`e}re}, {Lazareff}, {Le Bouquin}, {L{\'e}na}, {Lippa}, {Ott},
  {Panduro}, {Paumard}, {Perraut}, {Perrin}, {Pfuhl}, {Plewa}, {Rabien},
  {Rodr{\'\i}guez-Coira}, {Rousset}, {Sternberg}, {Straub}, {Straubmeier},
  {Sturm}, {Tacconi}, {Vincent}, {von Fellenberg}, {Waisberg}, {Widmann},
  {Wieprecht}, {Wiezorrek}, {Woillez}, \& {Yazici}}]{GRAVITY2018}
{Gravity Collaboration}, {Abuter}, R., {Amorim}, A., {et~al.} 2018, \aap, 618,
  L10

\bibitem[{{Gravity Collaboration} {et~al.}(2019){Gravity Collaboration},
  {Abuter}, {Amorim}, {Baub{\"o}ck}, {Berger}, {Bonnet}, {Brand ner},
  {Cl{\'e}net}, {Coud{\'e} Du Foresto}, {de Zeeuw}, {Dexter}, {Duvert},
  {Eckart}, {Eisenhauer}, {F{\"o}rster Schreiber}, {Garcia}, {Gao}, {Gendron},
  {Genzel}, {Gerhard}, {Gillessen}, {Habibi}, {Haubois}, {Henning}, {Hippler},
  {Horrobin}, {Jim{\'e}nez-Rosales}, {Jocou}, {Kervella}, {Lacour},
  {Lapeyr{\`e}re}, {Le Bouquin}, {L{\'e}na}, {Ott}, {Paumard}, {Perraut},
  {Perrin}, {Pfuhl}, {Rabien}, {Rodriguez Coira}, {Rousset}, {Scheithauer},
  {Sternberg}, {Straub}, {Straubmeier}, {Sturm}, {Tacconi}, {Vincent}, {von
  Fellenberg}, {Waisberg}, {Widmann}, {Wieprecht}, {Wiezorrek}, {Woillez}, \&
  {Yazici}}]{GRAVITY2019}
{Gravity Collaboration}, {Abuter}, R., {Amorim}, A., {et~al.} 2019, \aap, 625,
  L10

\bibitem[{{Gravity Collaboration} {et~al.}(2020){Gravity Collaboration},
  {Baub{\"o}ck}, {Dexter}, {Abuter}, {Amorim}, {Berger}, {Bonnet}, {Brandner},
  {Cl{\'e}net}, {Coud{\'e} Du Foresto}, {de Zeeuw}, {Duvert}, {Eckart},
  {Eisenhauer}, {F{\"o}rster Schreiber}, {Gao}, {Garcia}, {Gendron}, {Genzel},
  {Gerhard}, {Gillessen}, {Habibi}, {Haubois}, {Henning}, {Hippler},
  {Horrobin}, {Jim{\'e}nez-Rosales}, {Jocou}, {Kervella}, {Lacour},
  {Lapeyr{\`e}re}, {Le Bouquin}, {L{\'e}na}, {Ott}, {Paumard}, {Perraut},
  {Perrin}, {Pfuhl}, {Rabien}, {Rodriguez Coira}, {Rousset}, {Scheithauer},
  {Stadler}, {Sternberg}, {Straub}, {Straubmeier}, {Sturm}, {Tacconi},
  {Vincent}, {von Fellenberg}, {Waisberg}, {Widmann}, {Wieprecht}, {Wiezorrek},
  {Woillez}, \& {Yazici}}]{GRAVITY2020}
{Gravity Collaboration}, {Baub{\"o}ck}, M., {Dexter}, J., {et~al.} 2020, \aap,
  635, A143

\bibitem[{{Gwinn} {et~al.}(2014){Gwinn}, {Kovalev}, {Johnson}, \&
  {Soglasnov}}]{Gwinn2014}
{Gwinn}, C.~R., {Kovalev}, Y.~Y., {Johnson}, M.~D., \& {Soglasnov}, V.~A. 2014,
  \apjl, 794, L14

\bibitem[{Herrnstein {et~al.}(2004)Herrnstein, Zhao, Bower, \&
  Goss}]{Herrnstein2004}
Herrnstein, R.~M., Zhao, J.-H., Bower, G.~C., \& Goss, W.~M. 2004, The
  Astronomical Journal, 127, 3399

\bibitem[{{Issaoun} {et~al.}(2019){Issaoun}, {Johnson}, {Blackburn},
  {Brinkerink}, {Mo{\'s}cibrodzka}, {Chael}, {Goddi}, {Mart{\'\i}-Vidal},
  {Wagner}, {Doeleman}, {Falcke}, {Krichbaum}, {Akiyama}, {Bach}, {Bouman},
  {Bower}, {Broderick}, {Cho}, {Crew}, {Dexter}, {Fish}, {Gold}, {G{\'o}mez},
  {Hada}, {Hern{\'a}ndez-G{\'o}mez}, {Jan{\ss}en}, {Kino}, {Kramer}, {Loinard},
  {Lu}, {Markoff}, {Marrone}, {Matthews}, {Moran}, {M{\"u}ller}, {Roelofs},
  {Ros}, {Rottmann}, {Sanchez}, {Tilanus}, {de Vicente}, {Wielgus}, {Zensus},
  \& {Zhao}}]{Issaoun2019}
{Issaoun}, S., {Johnson}, M.~D., {Blackburn}, L., {et~al.} 2019, \apj, 871, 30

\bibitem[{{Johnson}(2016)}]{Johnson2016}
{Johnson}, M.~D. 2016, \apj, 833, 74

\bibitem[{{Johnson} \& {Gwinn}(2015)}]{Johnson2015}
{Johnson}, M.~D. \& {Gwinn}, C.~R. 2015, \apj, 805, 180

\bibitem[{{Johnson} {et~al.}(2018){Johnson}, {Narayan}, {Psaltis}, {Blackburn},
  {Kovalev}, {Gwinn}, {Zhao}, {Bower}, {Moran}, {Kino}, {Kramer}, {Akiyama},
  {Dexter}, {Broderick}, \& {Sironi}}]{Johnson2018}
{Johnson}, M.~D., {Narayan}, R., {Psaltis}, D., {et~al.} 2018, \apj, 865, 104

\bibitem[{{Law} {et~al.}(2008){Law}, {Yusef-Zadeh}, {Cotton}, \&
  {Maddalena}}]{Law2008}
{Law}, C.~J., {Yusef-Zadeh}, F., {Cotton}, W.~D., \& {Maddalena}, R.~J. 2008,
  \apjs, 177, 255

\bibitem[{{Liu} {et~al.}(2016){Liu}, {Wright}, {Zhao}, {Brinkerink}, {Ho},
  {Mills}, {Mart{\'\i}n}, {Falcke}, {Matsushita}, \&
  {Mart{\'\i}-Vidal}}]{Liu2016}
{Liu}, H.~B., {Wright}, M. C.~H., {Zhao}, J.-H., {et~al.} 2016, \aap, 593, A107

\bibitem[{{Lo} {et~al.}(1998){Lo}, {Shen}, {Zhao}, \& {Ho}}]{Lo1998}
{Lo}, K.~Y., {Shen}, Z.-Q., {Zhao}, J.-H., \& {Ho}, P. T.~P. 1998, \apjl, 508,
  L61

\bibitem[{{Markoff} {et~al.}(2007){Markoff}, {Bower}, \&
  {Falcke}}]{Markoff2007}
{Markoff}, S., {Bower}, G.~C., \& {Falcke}, H. 2007, \mnras, 379, 1519

\bibitem[{{Marrone}(2006)}]{Marrone2006}
{Marrone}, D.~P. 2006, PhD thesis, Harvard University

\bibitem[{{Matsumoto} {et~al.}(2020){Matsumoto}, {Chan}, \&
  {Piran}}]{Matsumoto2020}
{Matsumoto}, T., {Chan}, C.-H., \& {Piran}, T. 2020, \mnras, 497, 2385

\bibitem[{Miyazaki {et~al.}(2013)Miyazaki, Tsuboi, \& Tsutsumi}]{Miyazaki2013}
Miyazaki, A., Tsuboi, M., \& Tsutsumi, T. 2013, Publications of the
  Astronomical Society of Japan, 65, L6

\bibitem[{{Mo{\'s}cibrodzka}(2017)}]{Moscibrodzka2017}
{Mo{\'s}cibrodzka}, M. 2017, in IAU Symposium, Vol. 322, The Multi-Messenger
  Astrophysics of the Galactic Centre, ed. R.~M. {Crocker}, S.~N. {Longmore},
  \& G.~V. {Bicknell}, 43--49

\bibitem[{{Mo{\'s}cibrodzka} \& {Falcke}(2013)}]{Moscibrodzka2013}
{Mo{\'s}cibrodzka}, M. \& {Falcke}, H. 2013, \aap, 559, L3

\bibitem[{{Mo{\'s}cibrodzka} {et~al.}(2009){Mo{\'s}cibrodzka}, {Gammie},
  {Dolence}, {Shiokawa}, \& {Leung}}]{Moscibrodzka2009}
{Mo{\'s}cibrodzka}, M., {Gammie}, C.~F., {Dolence}, J.~C., {Shiokawa}, H., \&
  {Leung}, P.~K. 2009, \apj, 706, 497

\bibitem[{{Narayan} {et~al.}(1995){Narayan}, {Yi}, \&
  {Mahadevan}}]{Narayan1995}
{Narayan}, R., {Yi}, I., \& {Mahadevan}, R. 1995, \nat, 374, 623

\bibitem[{{Ortiz-Le{\'o}n} {et~al.}(2016){Ortiz-Le{\'o}n}, {Johnson},
  {Doeleman}, {Blackburn}, {Fish}, {Loinard}, {Reid}, {Castillo}, {Chael},
  {Hern{\'a}ndez-G{\'o}mez}, {Hughes}, {Le{\'o}n-Tavares}, {Lu}, {Monta{\~n}a},
  {Narayanan}, {Rosenfeld}, {S{\'a}nchez}, {Schloerb}, {Shen}, {Shiokawa},
  {SooHoo}, \& {Vertatschitsch}}]{OrtizLeon2016}
{Ortiz-Le{\'o}n}, G.~N., {Johnson}, M.~D., {Doeleman}, S.~S., {et~al.} 2016,
  \apj, 824, 40

\bibitem[{{Rauch} {et~al.}(2017){Rauch}, {Ros}, {Krichbaum}, {Eckart},
  {Zensus}, {Lu}, {Shahzamanian}, {Mu{\v{z}}i{\'c}}, \&
  {Pei{\ss}ker}}]{Rauch2017}
{Rauch}, C., {Ros}, E., {Krichbaum}, T.~P., {et~al.} 2017, in The
  Multi-Messenger Astrophysics of the Galactic Centre, ed. R.~M. {Crocker},
  S.~N. {Longmore}, \& G.~V. {Bicknell}, Vol. 322, 52--53

\bibitem[{{Reid} \& {Brunthaler}(2004)}]{ReidBrunthaler2004}
{Reid}, M.~J. \& {Brunthaler}, A. 2004, \apj, 616, 872

\bibitem[{Roberts {et~al.}(2013)Roberts, Osborne, Ebden, Reece, Gibson, \&
  Aigrain}]{Roberts2013}
Roberts, S., Osborne, M., Ebden, M., {et~al.} 2013, Philosophical Transactions
  of the Royal Society (Part A

\bibitem[{{Serabyn} {et~al.}(1997){Serabyn}, {Carlstrom}, {Lay}, {Lis},
  {Hunter}, {Lacy}, \& {Hills}}]{Serabyn1997}
{Serabyn}, E., {Carlstrom}, J., {Lay}, O., {et~al.} 1997, \apjl, 490, L77

\bibitem[{{Shen} {et~al.}(2005){Shen}, {Lo}, {Liang}, {Ho}, \&
  {Zhao}}]{Shen2005}
{Shen}, Z.-Q., {Lo}, K.~Y., {Liang}, M.~C., {Ho}, P. T.~P., \& {Zhao}, J.~H.
  2005, \nat, 438, 62

\bibitem[{{Uttley} \& {McHardy}(2001)}]{Uttley2001}
{Uttley}, P. \& {McHardy}, I.~M. 2001, \mnras, 323, L26

\bibitem[{{van der Laan}(1966)}]{vdLaan1966}
{van der Laan}, H. 1966, \nat, 211, 1131

\bibitem[{{van Langevelde} \& {Diamond}(1991)}]{Langevelde1991}
{van Langevelde}, H.~J. \& {Diamond}, P.~J. 1991, \mnras, 249, 7P

\bibitem[{{Welsh}(1999)}]{Welsh1999}
{Welsh}, W.~F. 1999, \pasp, 111, 1347

\bibitem[{{Witzel} {et~al.}(2018){Witzel}, {Martinez}, {Hora}, {Willner},
  {Morris}, {Gammie}, {Becklin}, {Ashby}, {Baganoff}, {Carey}, {Do}, {Fazio},
  {Ghez}, {Glaccum}, {Haggard}, {Herrero-Illana}, {Ingalls}, {Narayan}, \&
  {Smith}}]{Witzel2018}
{Witzel}, G., {Martinez}, G., {Hora}, J., {et~al.} 2018, \apj, 863, 15

\bibitem[{{Yuan} {et~al.}(2003){Yuan}, {Quataert}, \& {Narayan}}]{Yuan2003}
{Yuan}, F., {Quataert}, E., \& {Narayan}, R. 2003, \apj, 598, 301

\bibitem[{{Yusef-Zadeh} {et~al.}(2006){Yusef-Zadeh}, {Roberts}, {Wardle},
  {Heinke}, \& {Bower}}]{YusefZadeh2006}
{Yusef-Zadeh}, F., {Roberts}, D., {Wardle}, M., {Heinke}, C.~O., \& {Bower},
  G.~C. 2006, \apj, 650, 189

\bibitem[{{Zhao} {et~al.}(2003){Zhao}, {Young}, {Herrnstein}, {Ho}, {Tsutsumi},
  {Lo}, {Goss}, \& {Bower}}]{Zhao2003}
{Zhao}, J.-H., {Young}, K.~H., {Herrnstein}, R.~M., {et~al.} 2003, \apjl, 586,
  L29

\bibitem[{{Zylka} {et~al.}(1992){Zylka}, {Mezger}, \& {Lesch}}]{Zylka1992}
{Zylka}, R., {Mezger}, P.~G., \& {Lesch}, H. 1992, \aap, 261, 119

\bibitem[{{Zylka} {et~al.}(1995){Zylka}, {Mezger}, {Ward-Thompson}, {Duschl},
  \& {Lesch}}]{Zylka1995}
{Zylka}, R., {Mezger}, P.~G., {Ward-Thompson}, D., {Duschl}, W.~J., \& {Lesch},
  H. 1995, \aap, 297, 83

\end{thebibliography}

%\appendix
\clearpage
\onecolumn
\section*{Appendix A: Calibrator and check source light curves}

Light curves for both the calibrator source (J1744-3116) and the check source (J1745-283) are included for all epochs in this appendix.

\vspace{0cm}

\begin{figure*}[h]
\centering
\includegraphics[width=0.95\textwidth]{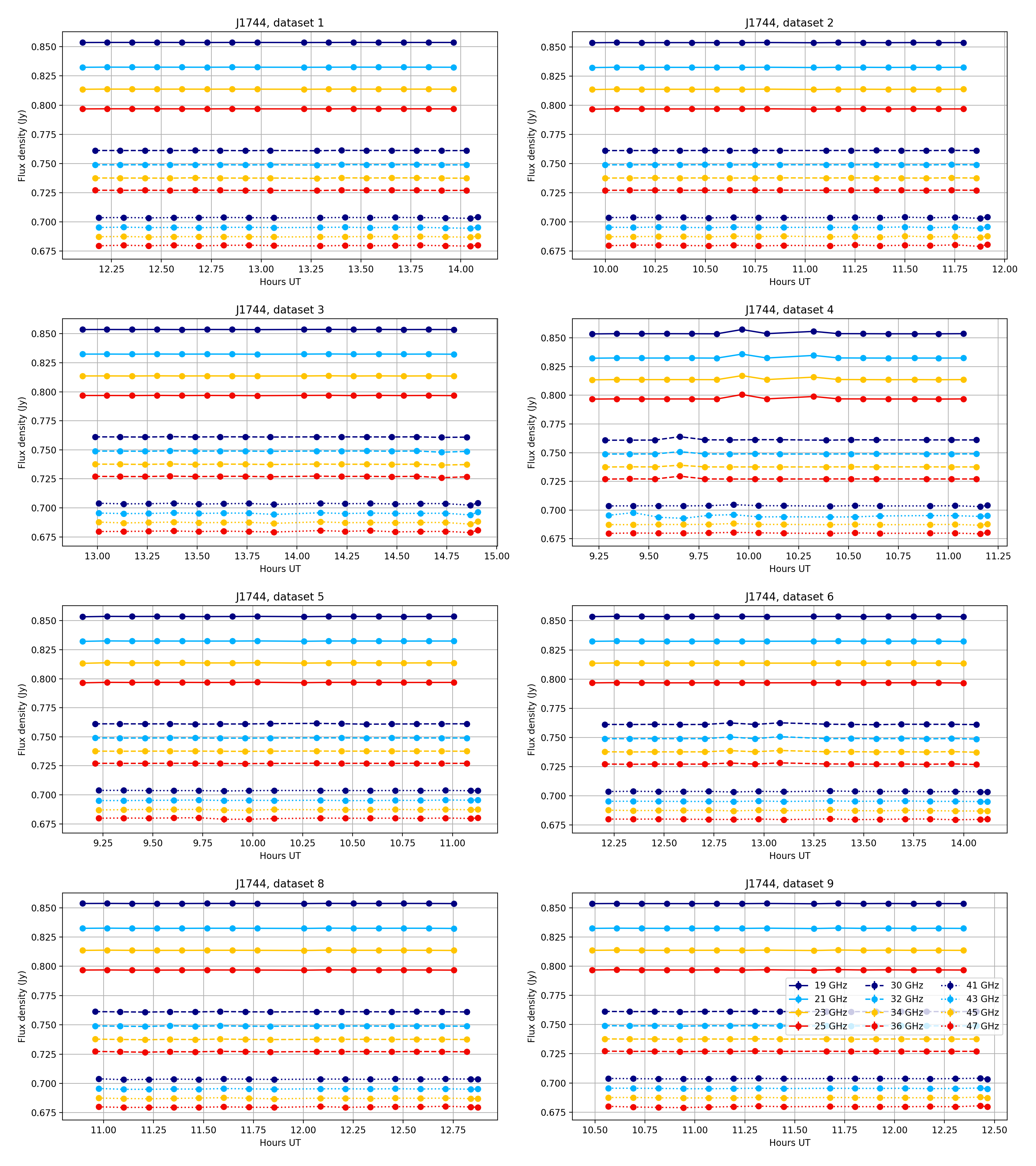}
\caption{Calibrated light curves for J1744-3116 (calibrator), data sets 1-6, 8, and 9. All data is plotted with error bars, which are obscured by the data markers.}
\end{figure*}

\begin{figure*}[h]
\centering
\includegraphics[width=1.0\textwidth]{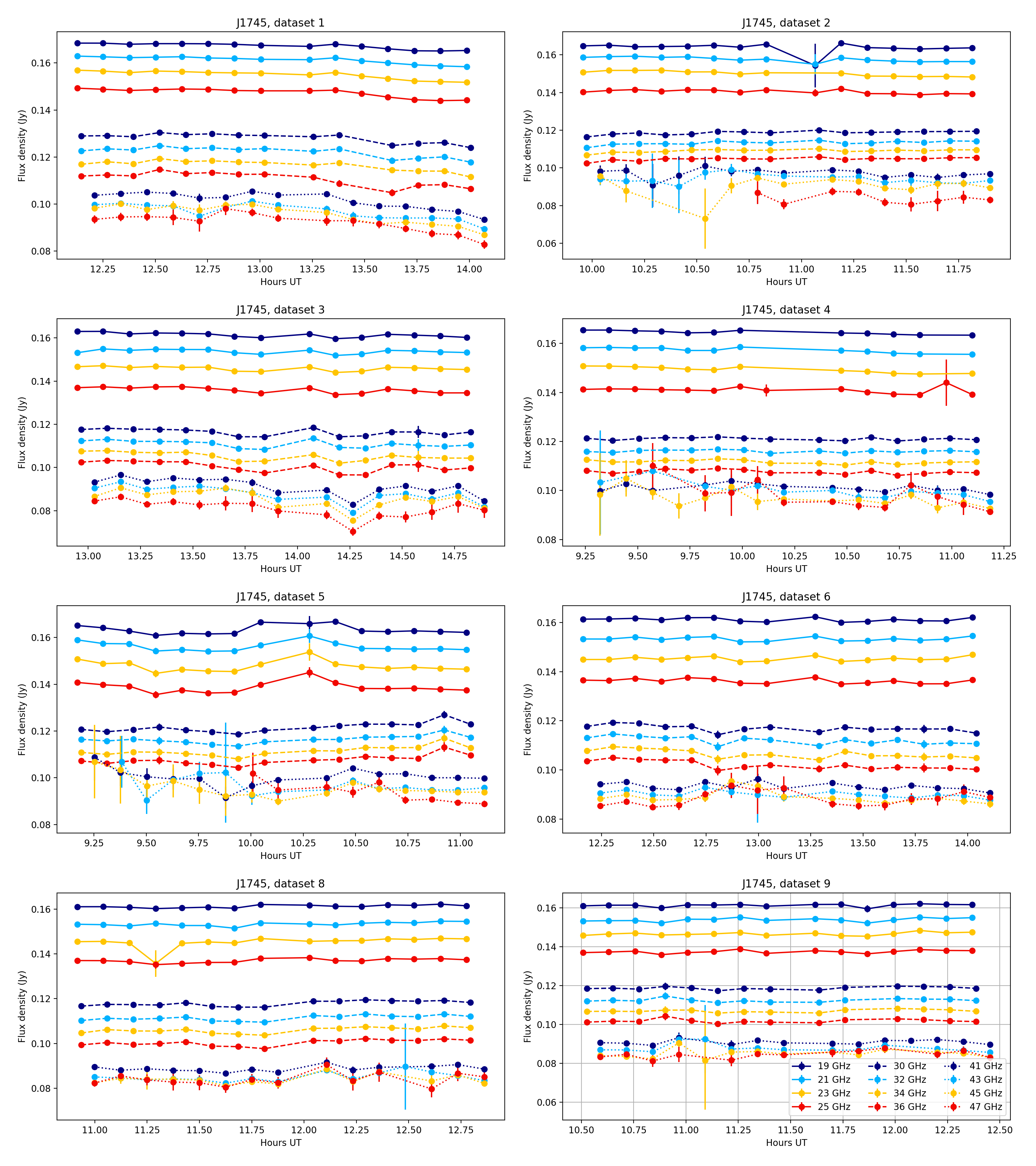}
\caption{Calibrated light curves for J1745-283 (check source), data sets 1-6, 8, and 9. All data is plotted with error bars.}
\end{figure*}

\clearpage
\section*{Appendix B: example correlation results for J1745}

\begin{figure*}[h]
\centering
\includegraphics[width=0.49\textwidth]{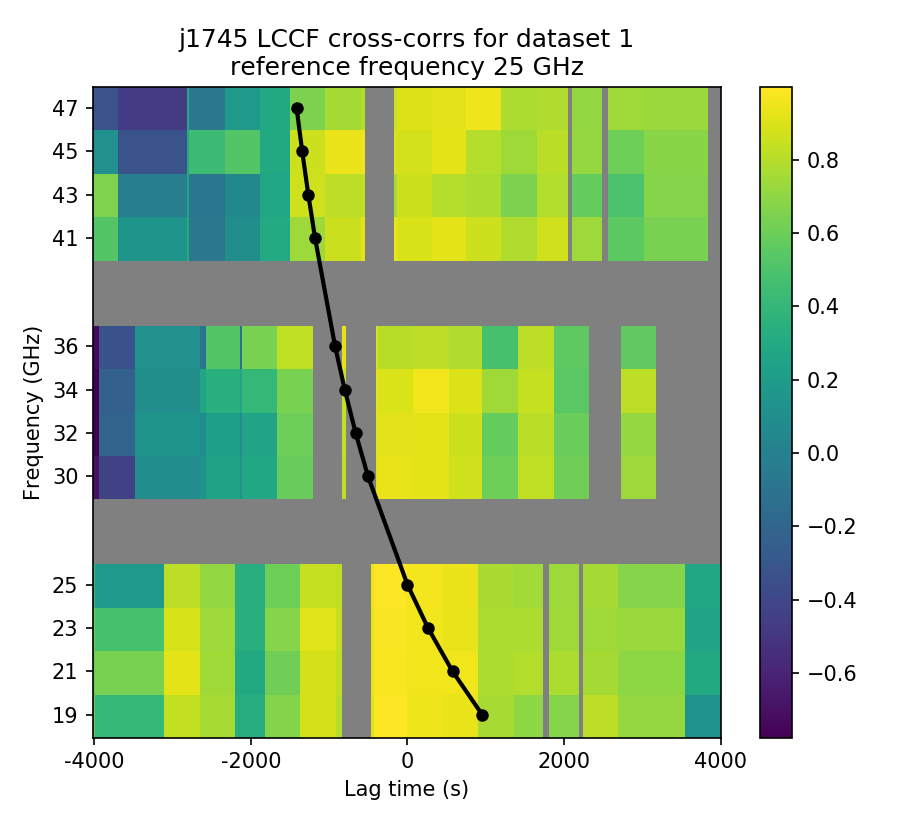}
\includegraphics[width=0.49\textwidth]{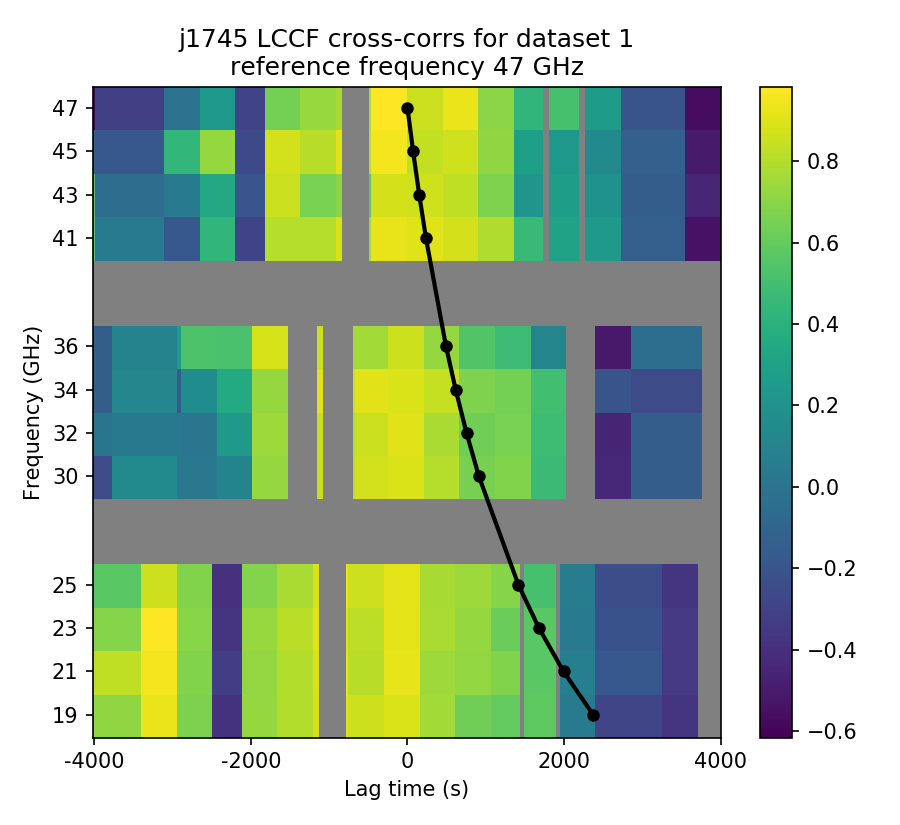}
\includegraphics[width=0.49\textwidth]{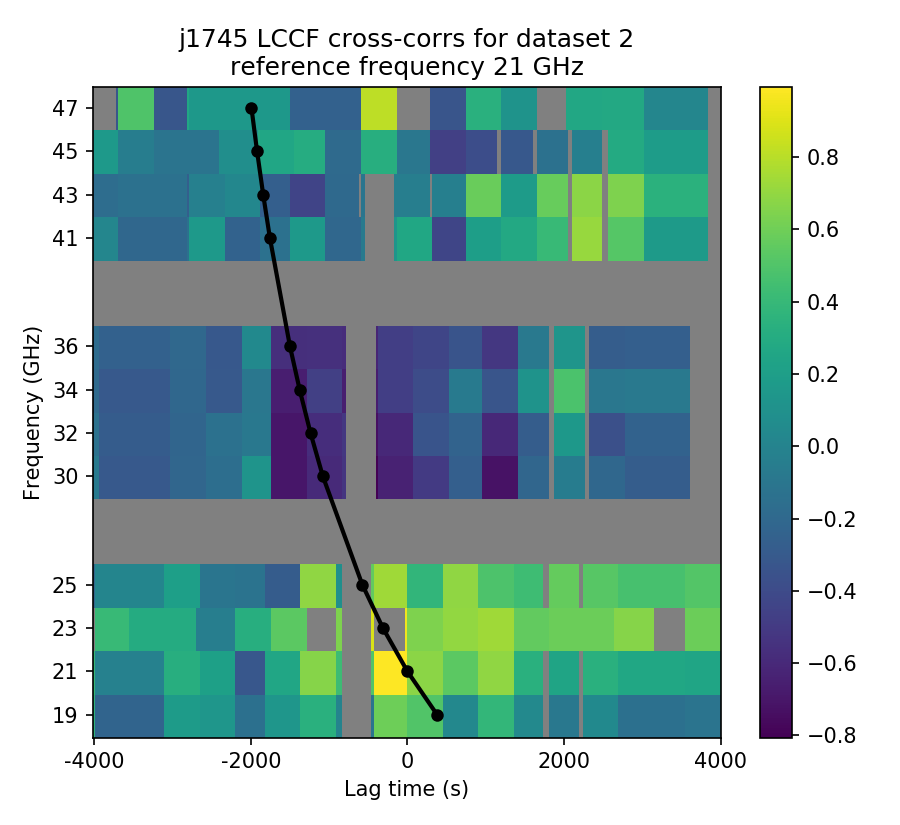}
\includegraphics[width=0.49\textwidth]{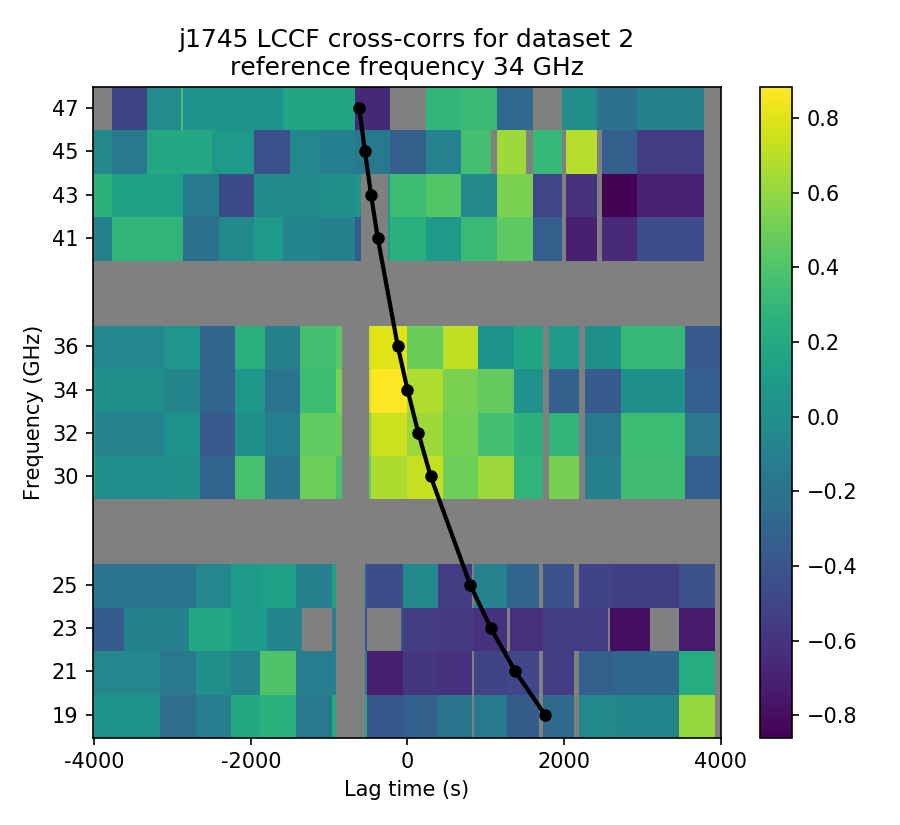}
\caption{Representative cross-correlation plots for check source J1745, two per epoch, for epochs 1 and 2.}
\label{fig:j1745corrs}
\end{figure*}

\begin{figure*}[h]
\centering
\includegraphics[width=0.49\textwidth]{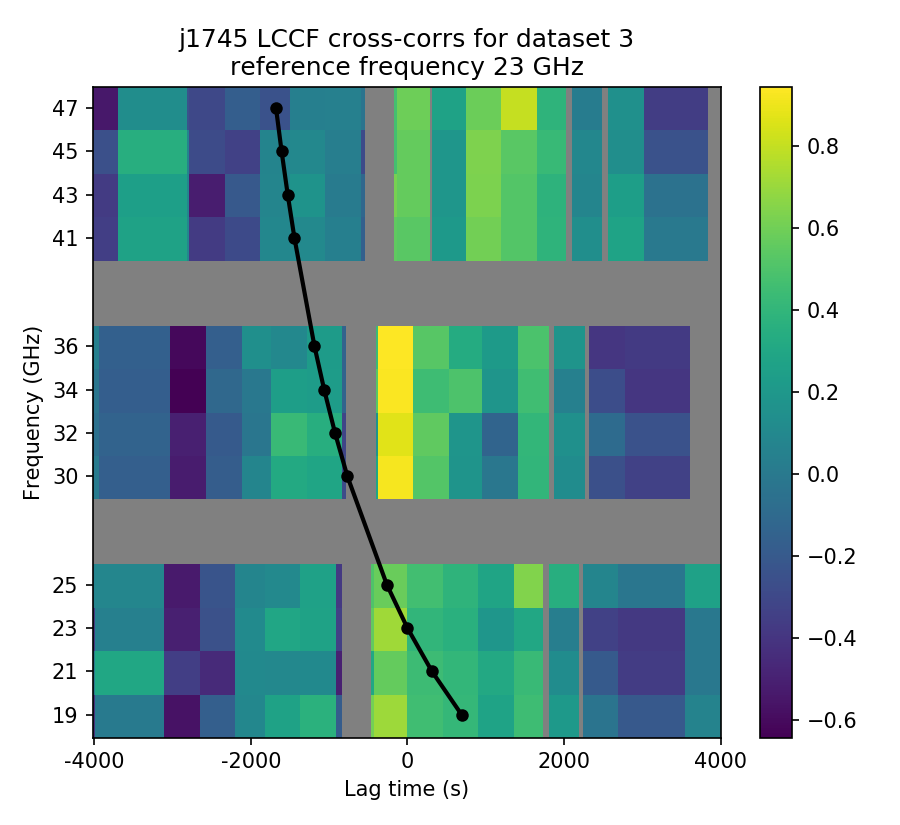}
\includegraphics[width=0.49\textwidth]{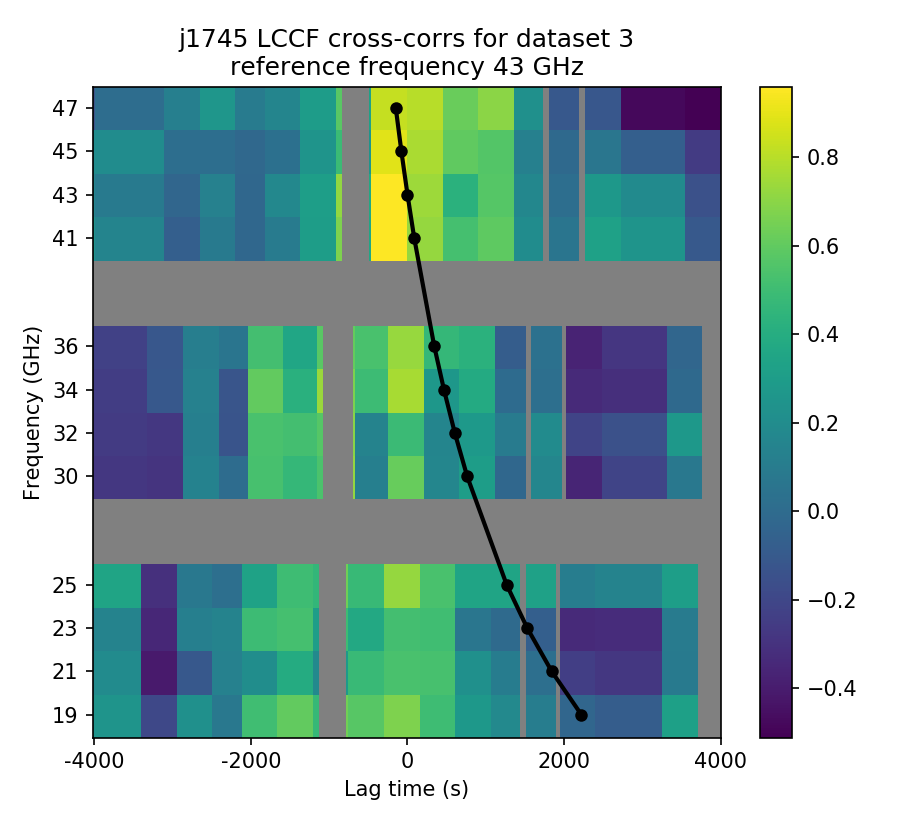}
\includegraphics[width=0.49\textwidth]{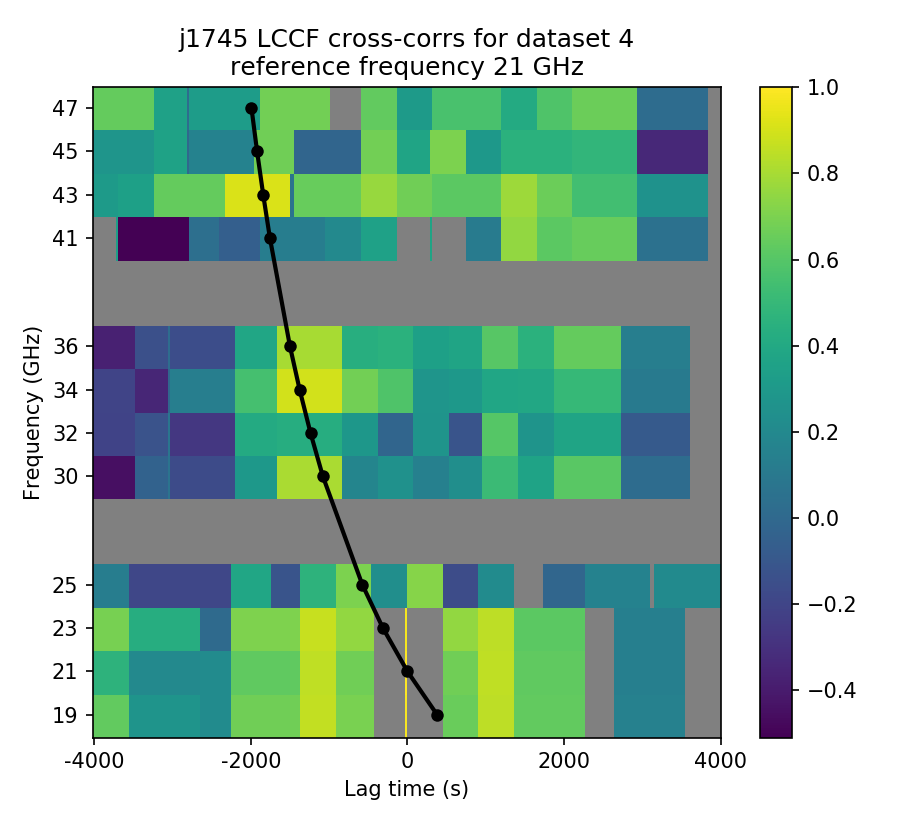}
\includegraphics[width=0.49\textwidth]{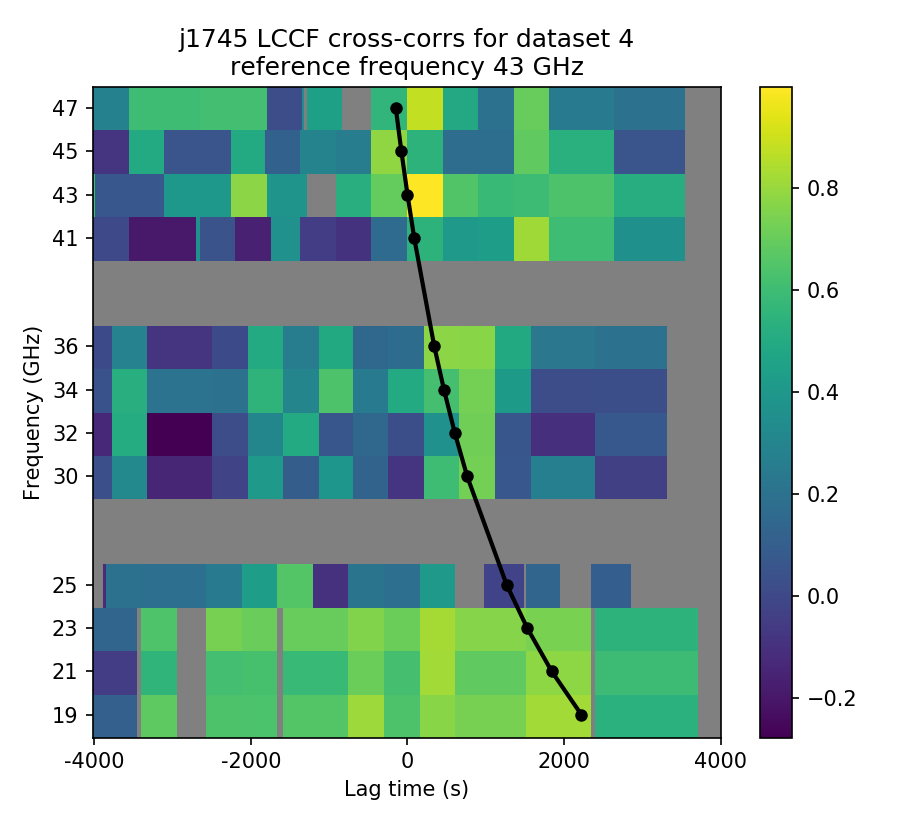}
\caption{Representative cross-correlation plots for check source J1745, two per epoch, for epochs 3 and 4.}
\end{figure*}

\begin{figure*}[h]
\centering
\includegraphics[width=0.49\textwidth]{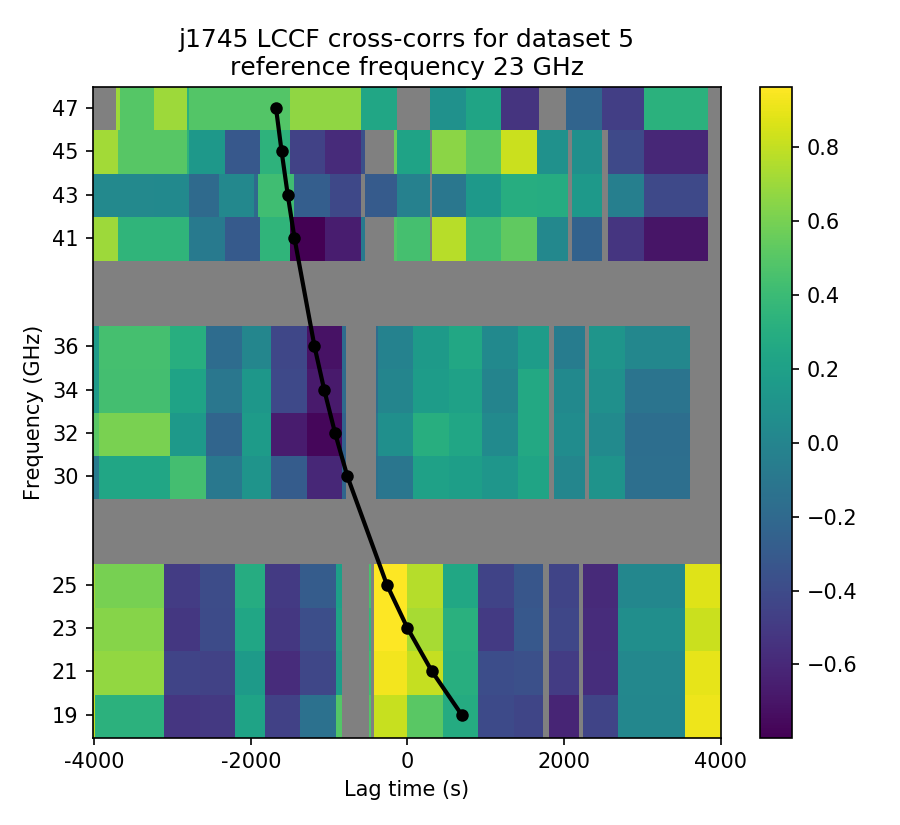}
\includegraphics[width=0.49\textwidth]{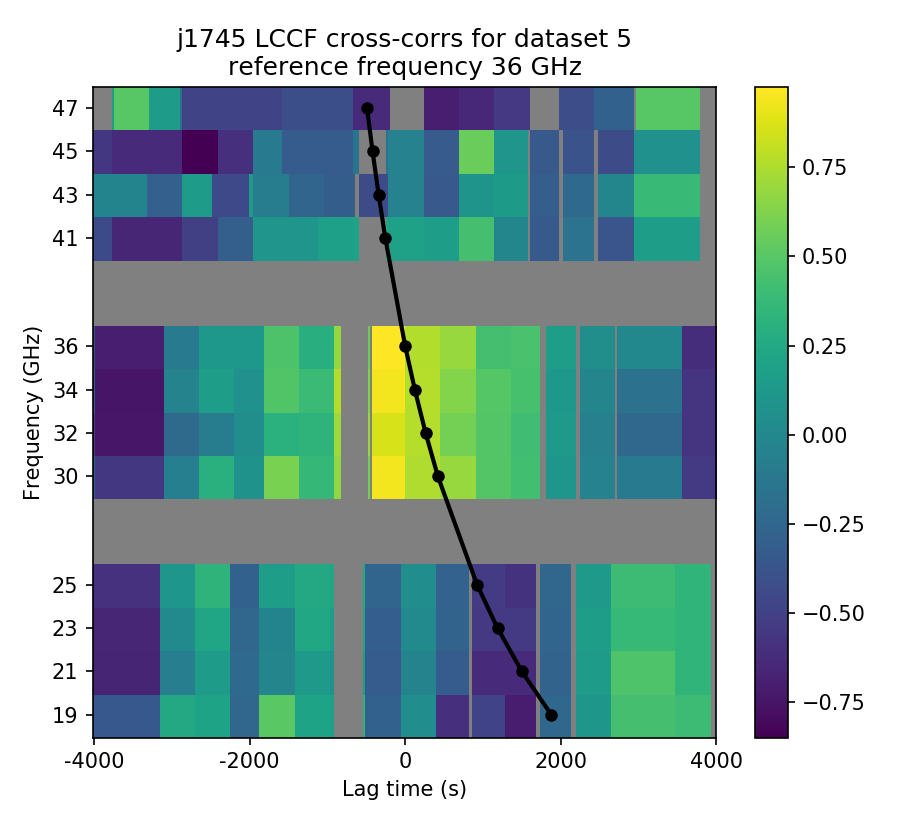}
\includegraphics[width=0.49\textwidth]{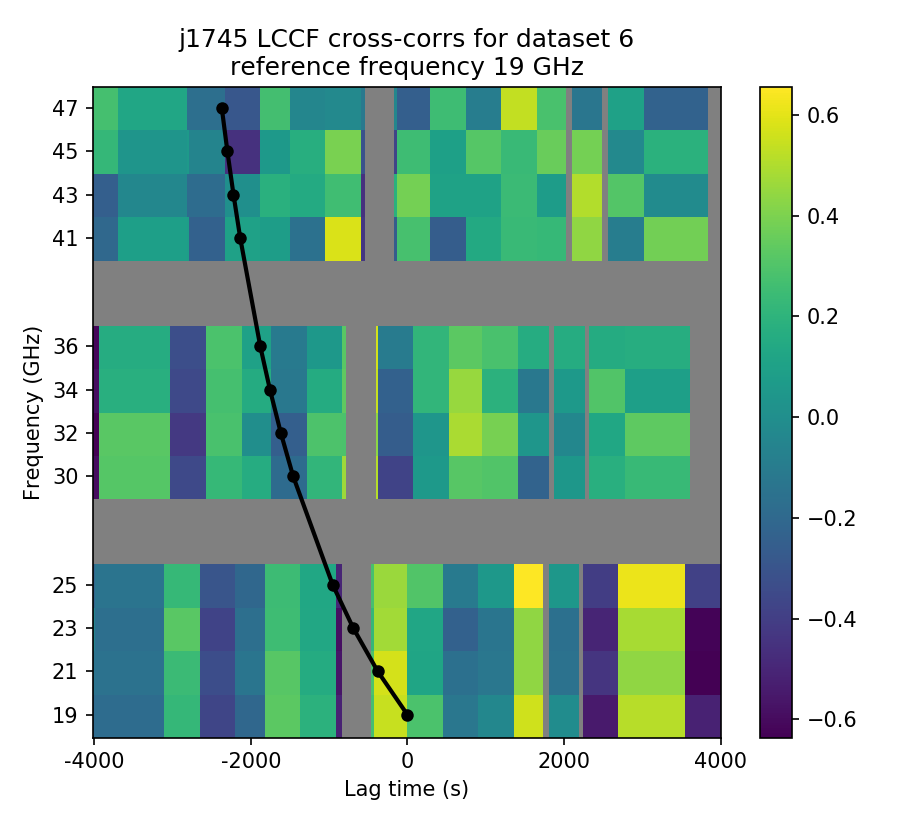}
\includegraphics[width=0.49\textwidth]{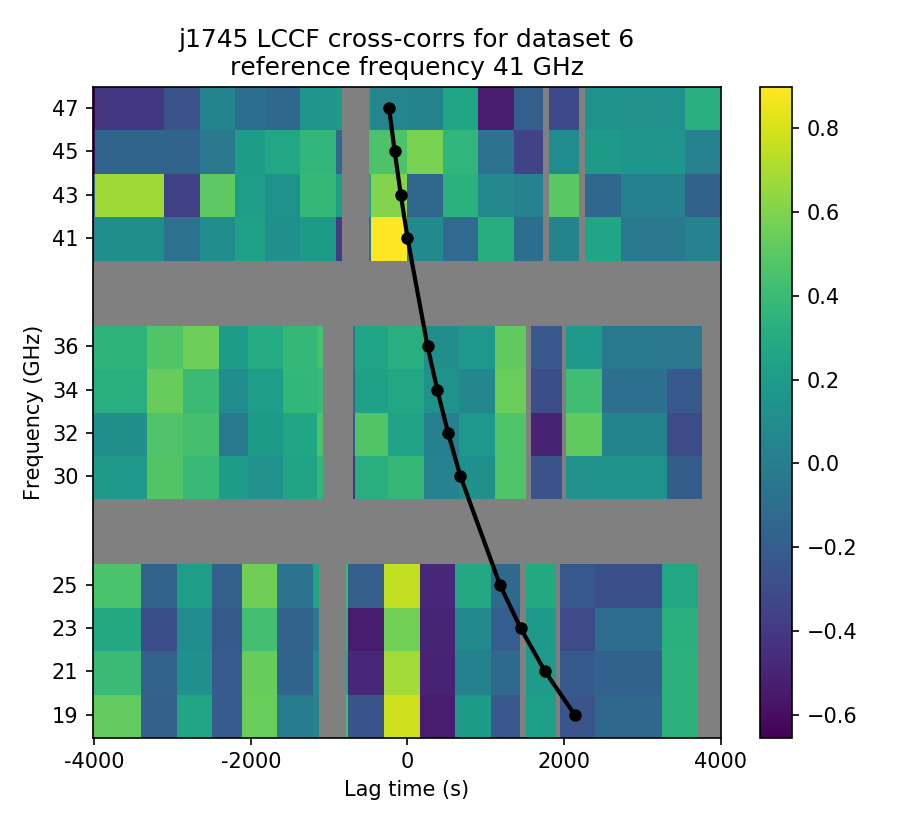}
\caption{Representative cross-correlation plots for check source J1745, two per epoch, for epochs 5 and 6.}
\end{figure*}

\begin{figure*}[h]
\centering
\includegraphics[width=0.49\textwidth]{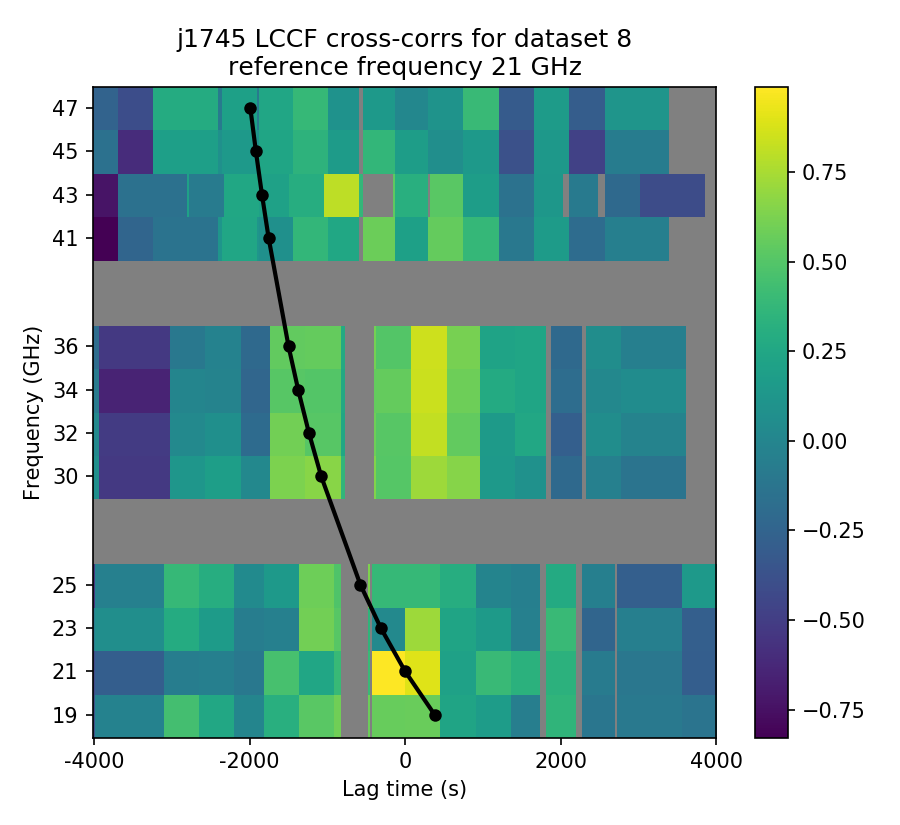}
\includegraphics[width=0.49\textwidth]{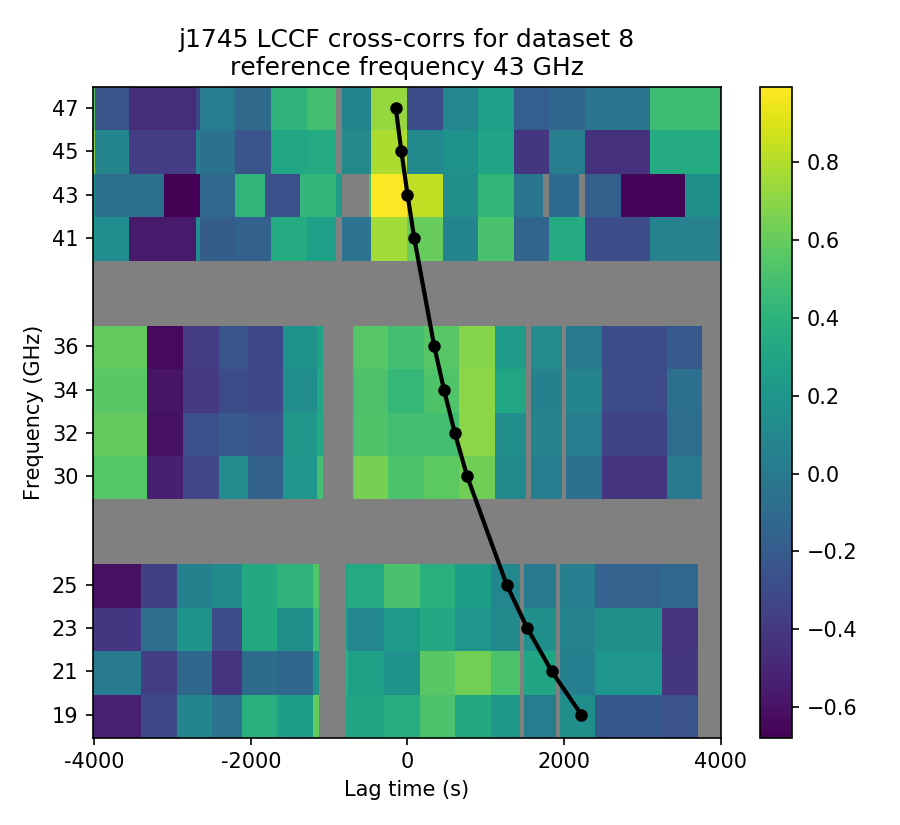}
\includegraphics[width=0.49\textwidth]{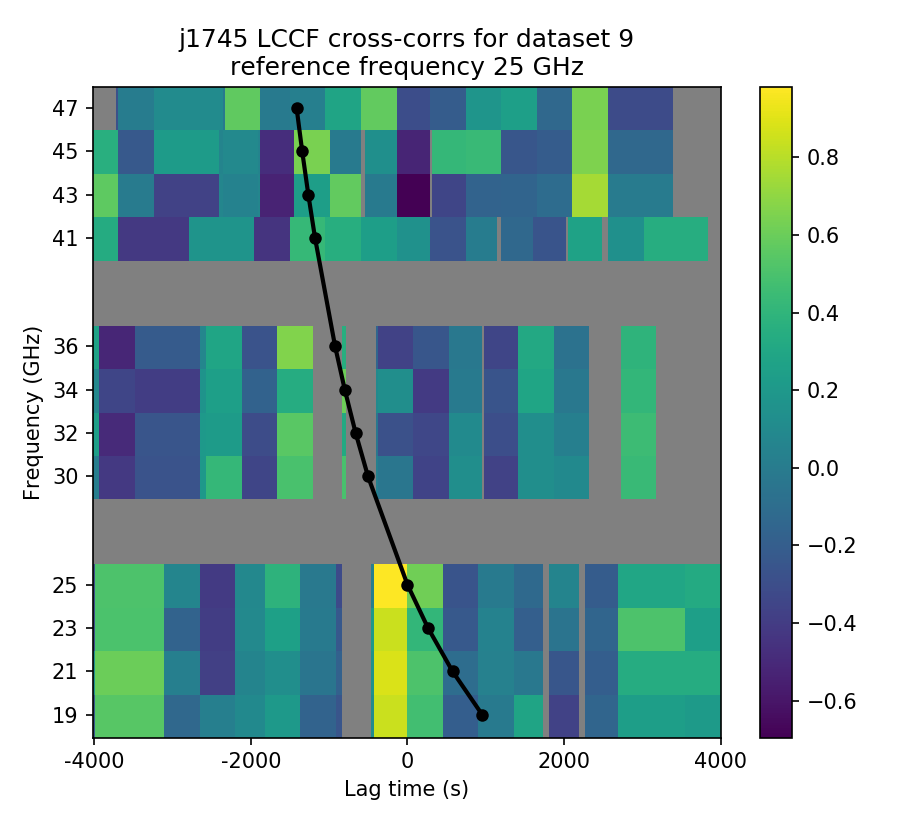}
\includegraphics[width=0.49\textwidth]{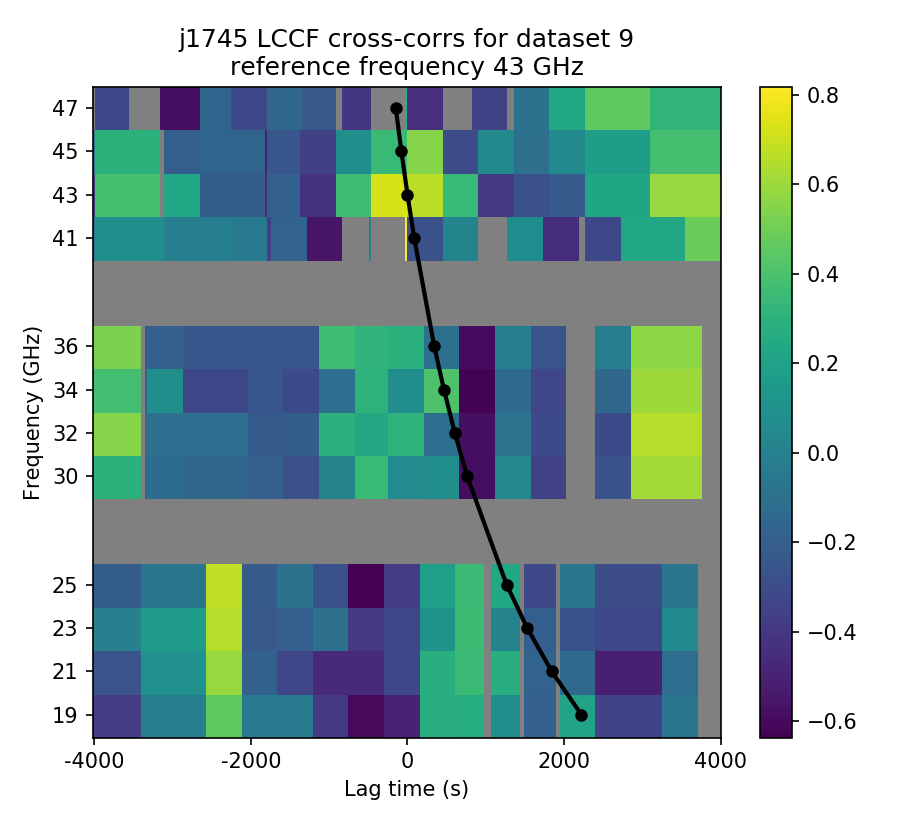}
\caption{Representative cross-correlation plots for check source J1745, two per epoch, for epochs 8 and 9.}
\end{figure*}

\clearpage
\section*{Appendix C: synthetic light curve and correlation results}

\begin{figure*}[h]
\centering
\includegraphics[width=0.43\textwidth]{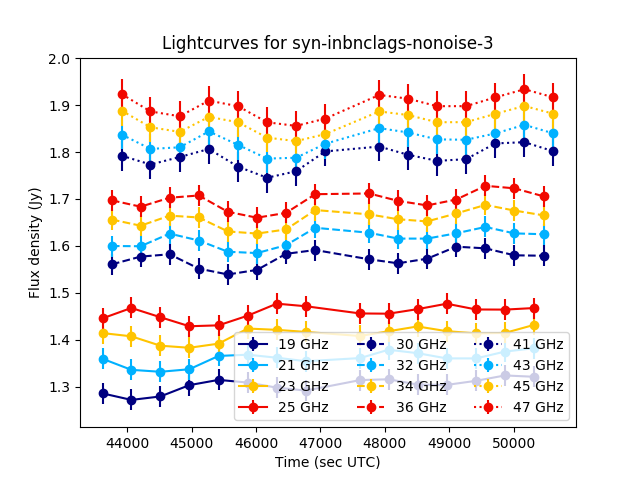}
\hspace{20pt}
\includegraphics[width=0.43\textwidth]{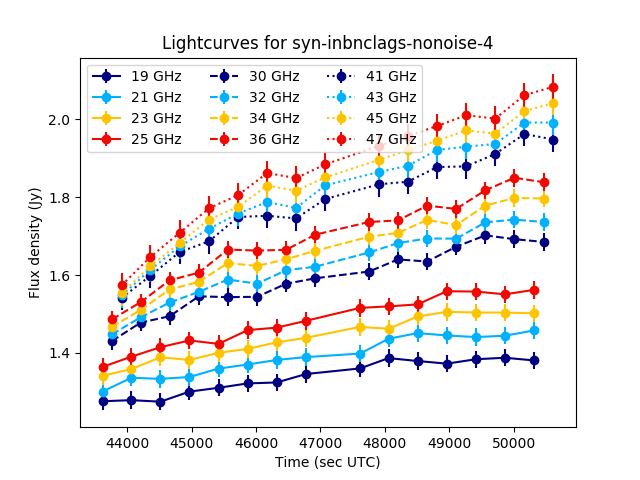}
\includegraphics[width=0.49\textwidth]{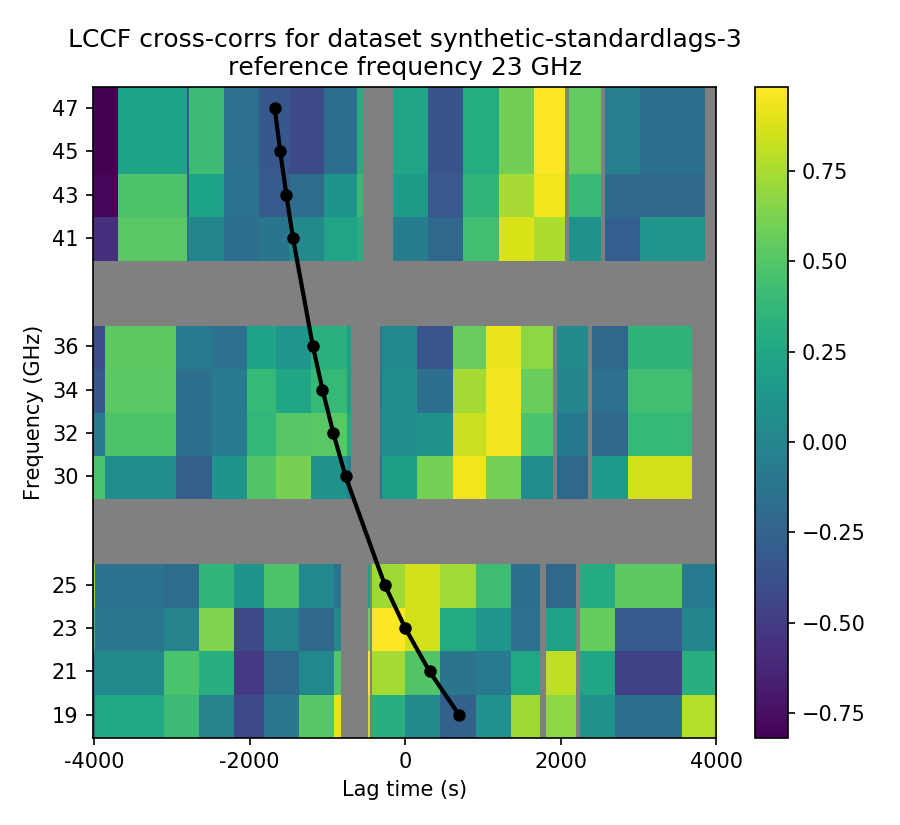}
\includegraphics[width=0.49\textwidth]{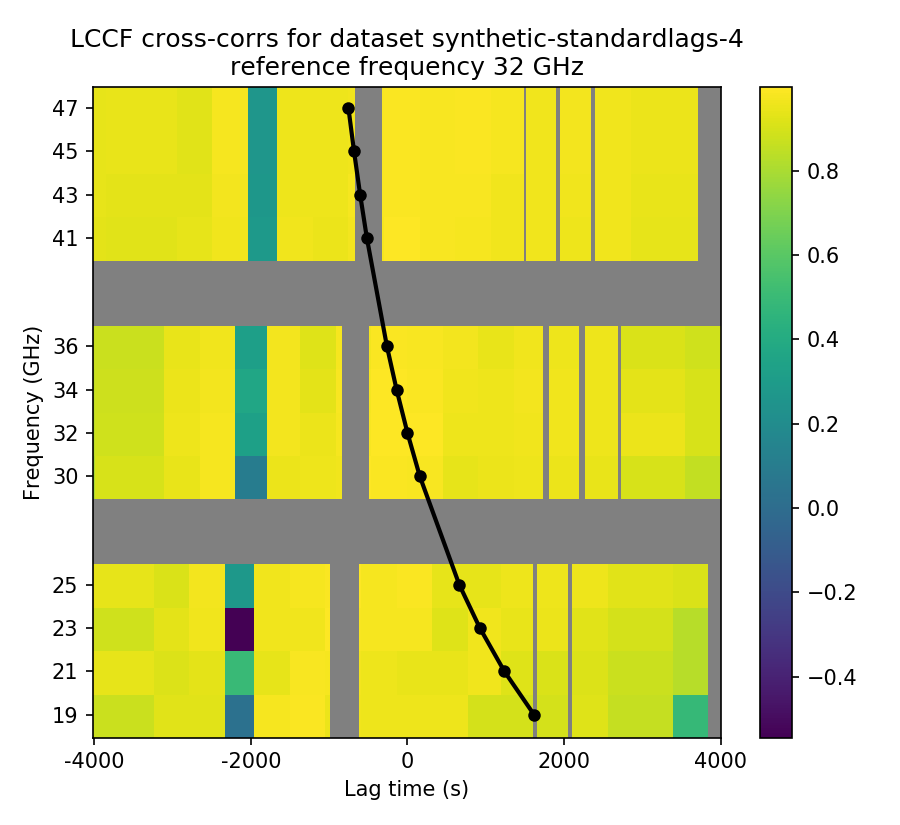}
\includegraphics[width=0.41\textwidth]{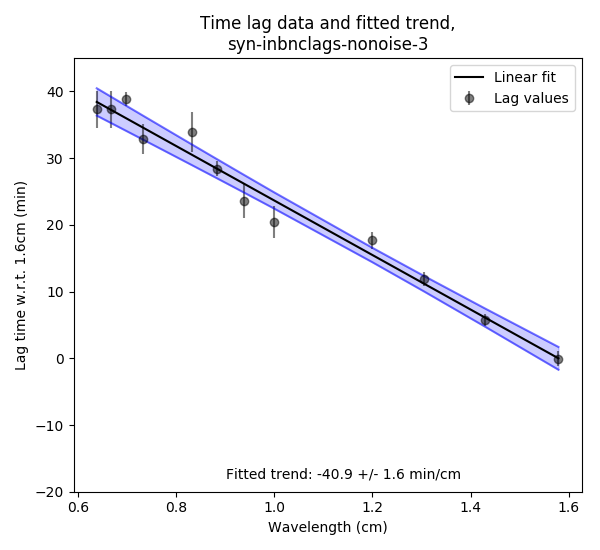}
\hspace{30pt}
\includegraphics[width=0.41\textwidth]{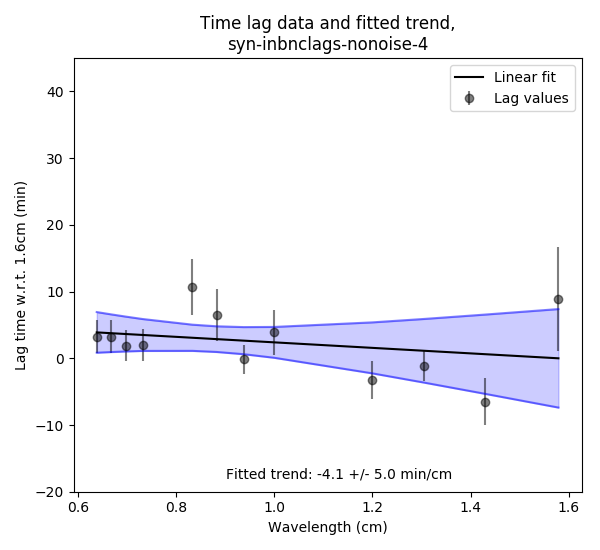}
\caption{Samples of results from synthetic light curves with the inverse time lag relation and no calibration errors. The left column shows data where the inverse time lag trend was successfully recovered from the cross-correlations, the right column shows an example of a case where a time lag relation with a significant error is produced.}
\label{fig:synresults1}
\end{figure*}

\begin{figure*}[h]
\centering
\includegraphics[width=0.43\textwidth]{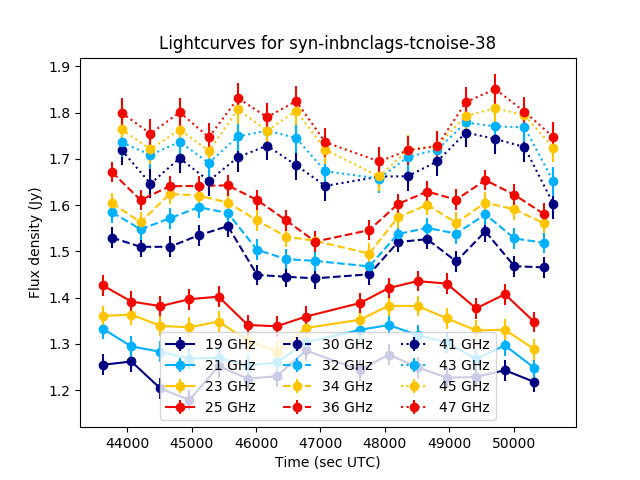}
\hspace{20pt}
\includegraphics[width=0.43\textwidth]{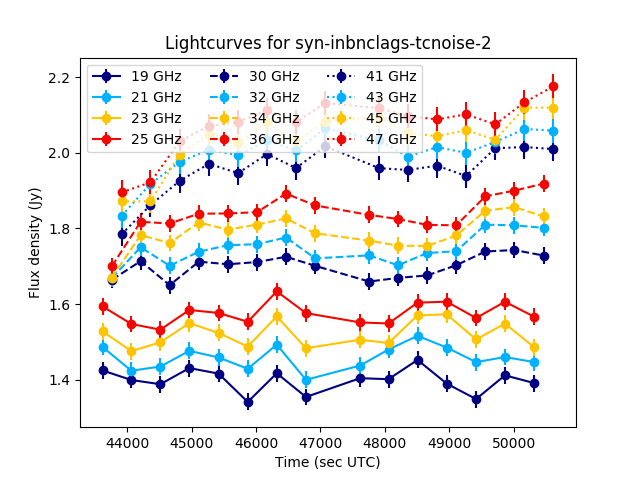}
\includegraphics[width=0.49\textwidth]{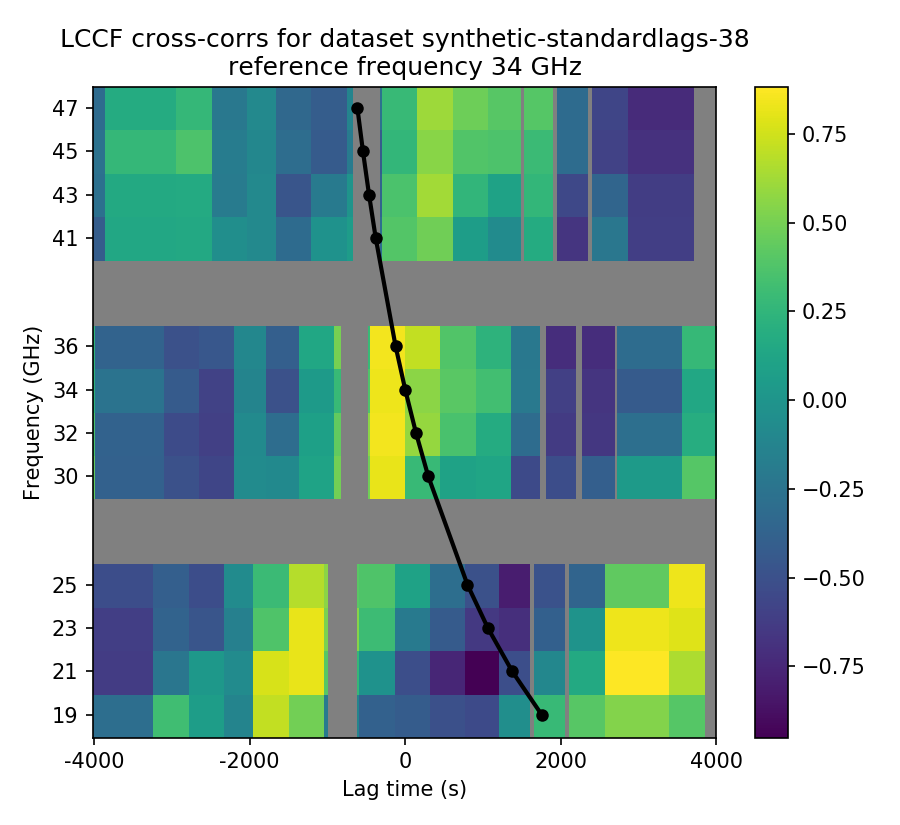}
\includegraphics[width=0.49\textwidth]{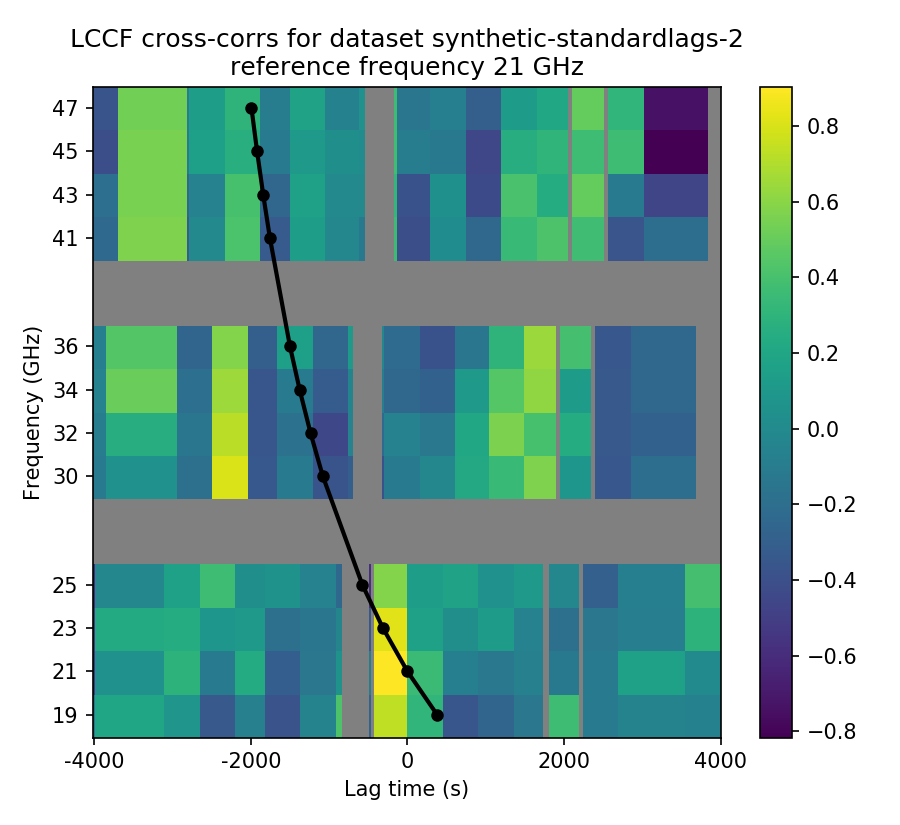}
\includegraphics[width=0.41\textwidth]{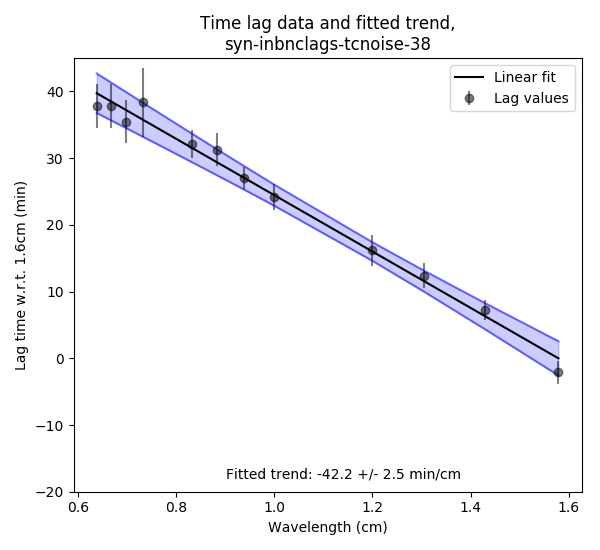}
\hspace{30pt}
\includegraphics[width=0.41\textwidth]{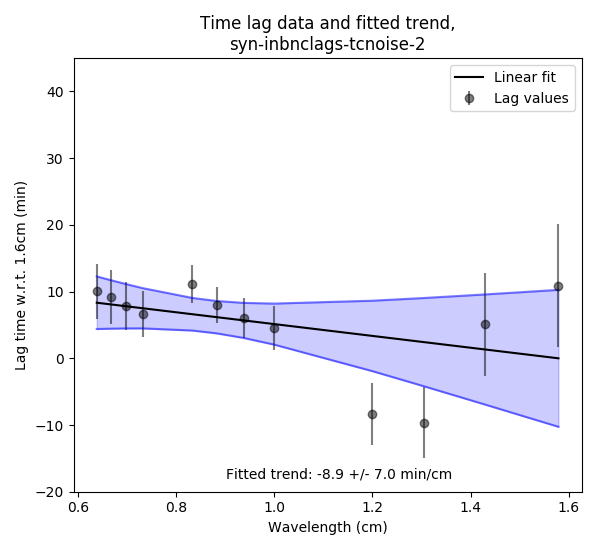}
\caption{Samples of results from synthetic light curves with the inverse time lag relation and with calibration errors included. The left column shows data where the inverse time lag trend was successfully recovered from the cross-correlations, the right column shows an example of a case where a time lag relation with a significant error is produced.}
\label{fig:synresults2}
\end{figure*}

\begin{figure*}[h]
\centering
\includegraphics[width=0.43\textwidth]{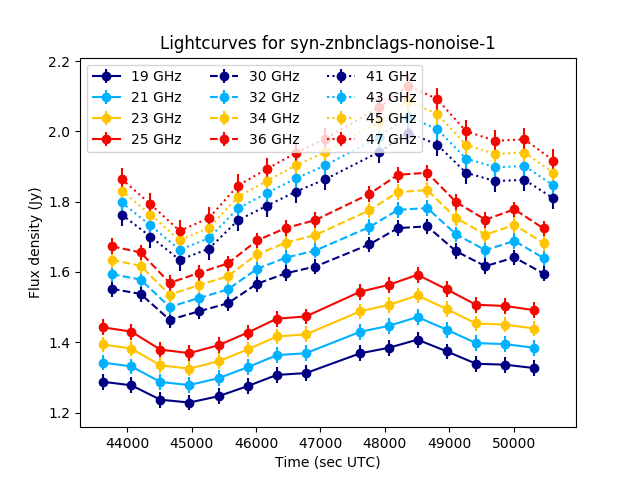}
\hspace{20pt}
\includegraphics[width=0.43\textwidth]{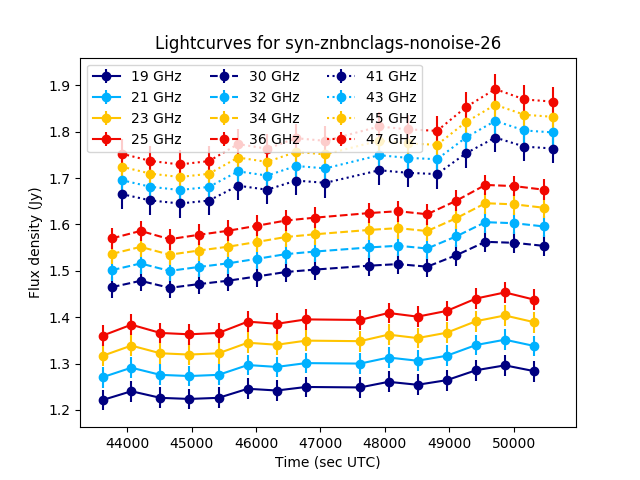}
\includegraphics[width=0.49\textwidth]{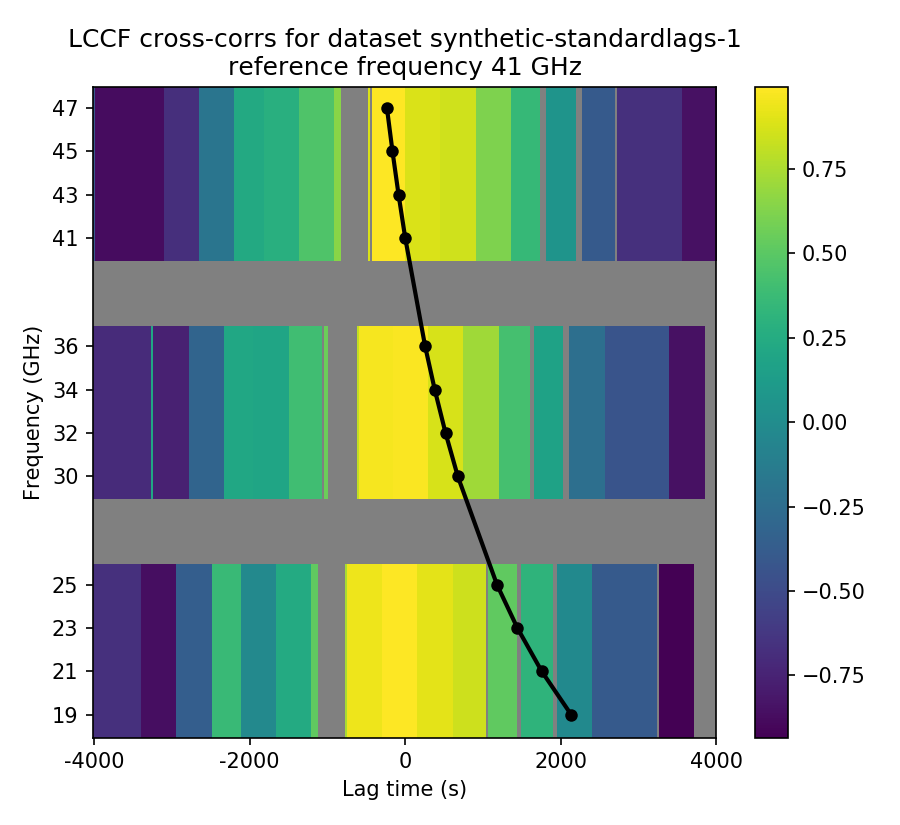}
\includegraphics[width=0.49\textwidth]{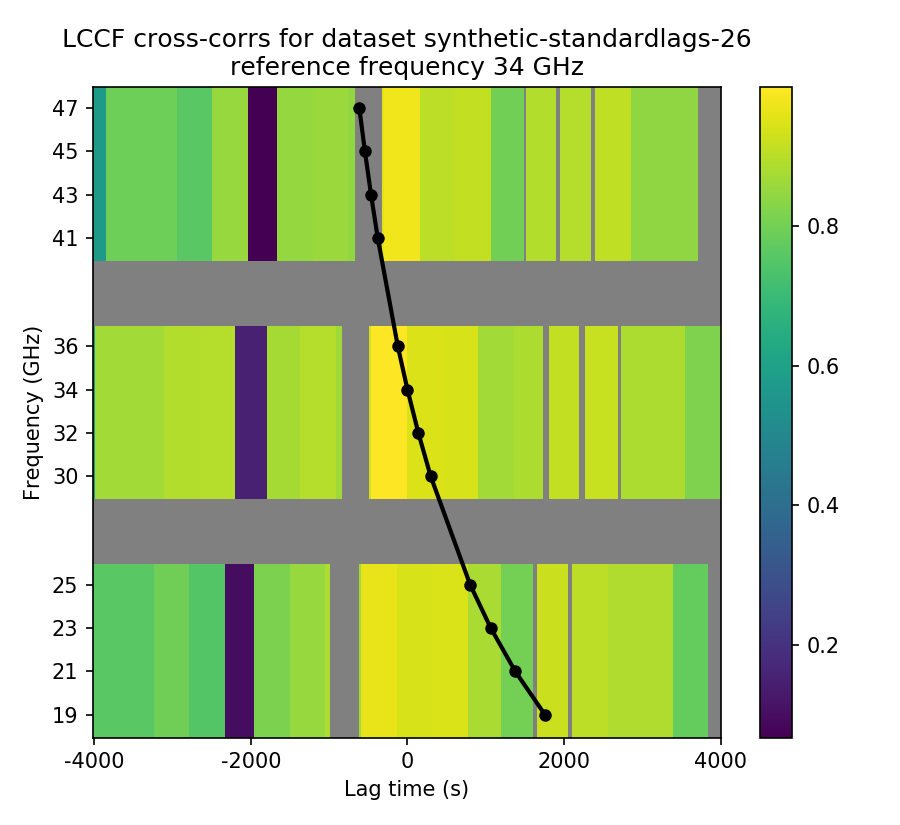}
\includegraphics[width=0.41\textwidth]{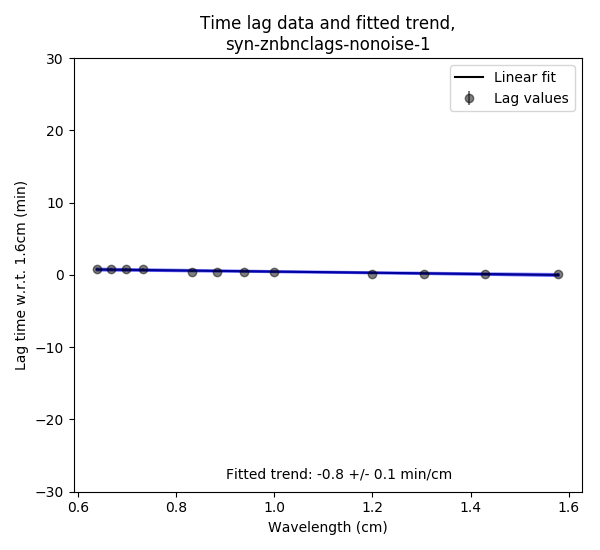}
\hspace{30pt}
\includegraphics[width=0.41\textwidth]{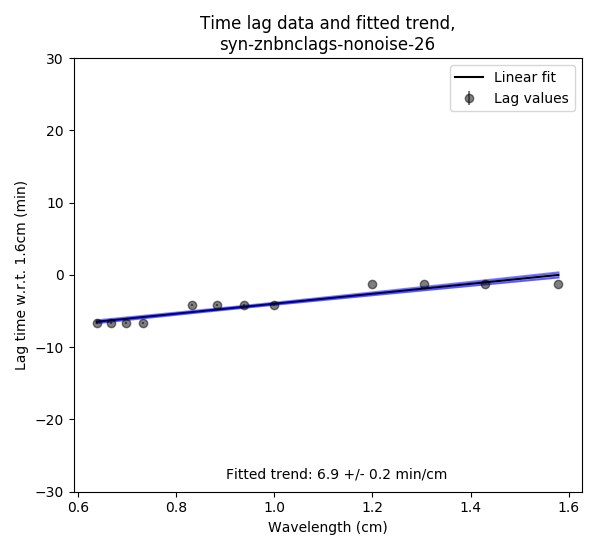}
\caption{Samples of results from synthetic light curves with the zero time lag relation and no calibration errors. The left column shows data where the zero time lag trend was successfully recovered from the cross-correlations, the right column shows an example of a case where a time lag relation with a significant error is produced.}
\label{fig:synresults3}
\end{figure*}

\begin{figure*}[h]
\centering
\includegraphics[width=0.43\textwidth]{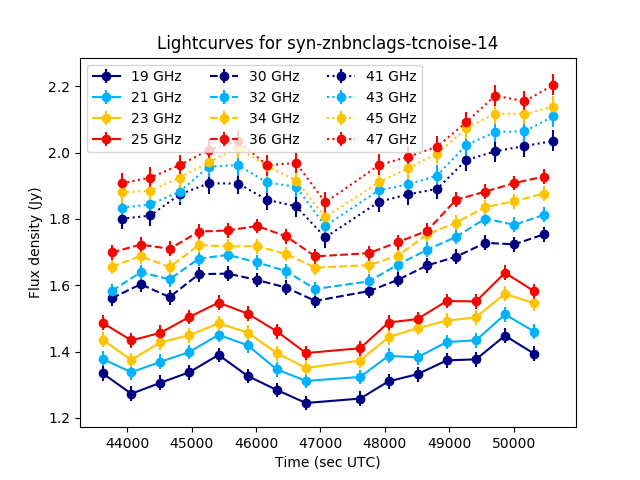}
\hspace{20pt}
\includegraphics[width=0.43\textwidth]{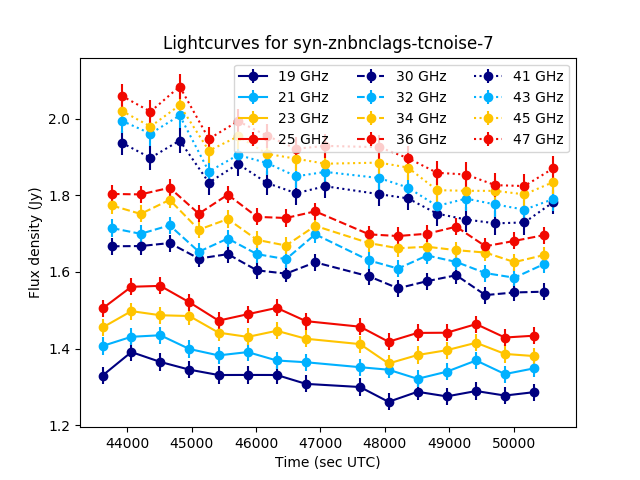}
\includegraphics[width=0.49\textwidth]{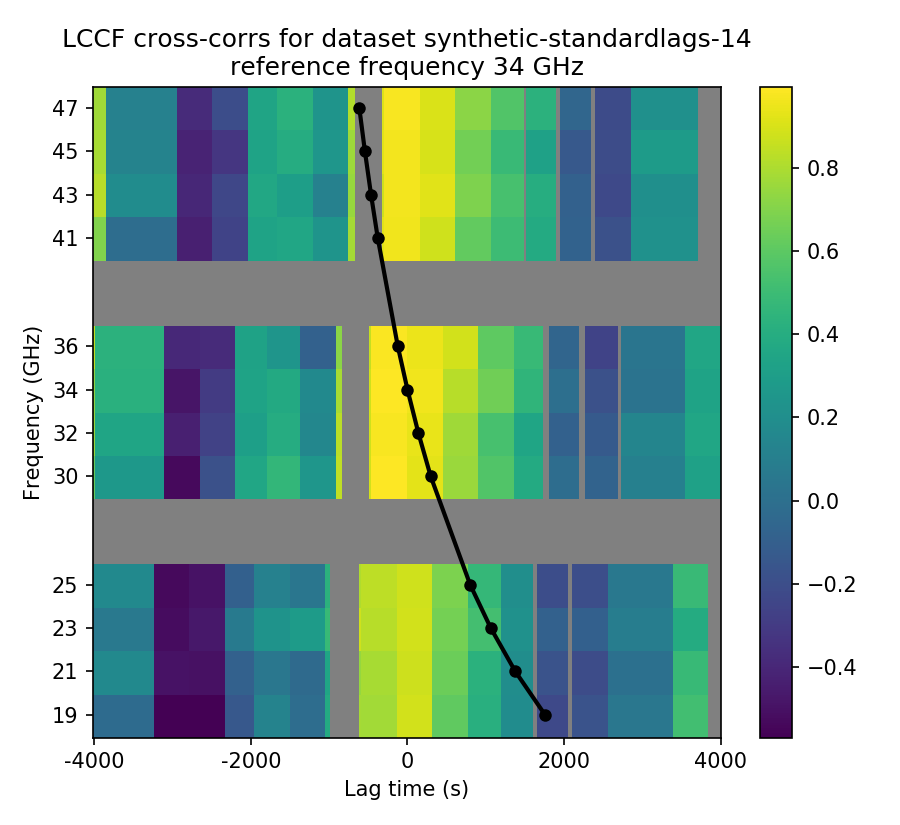}
\includegraphics[width=0.49\textwidth]{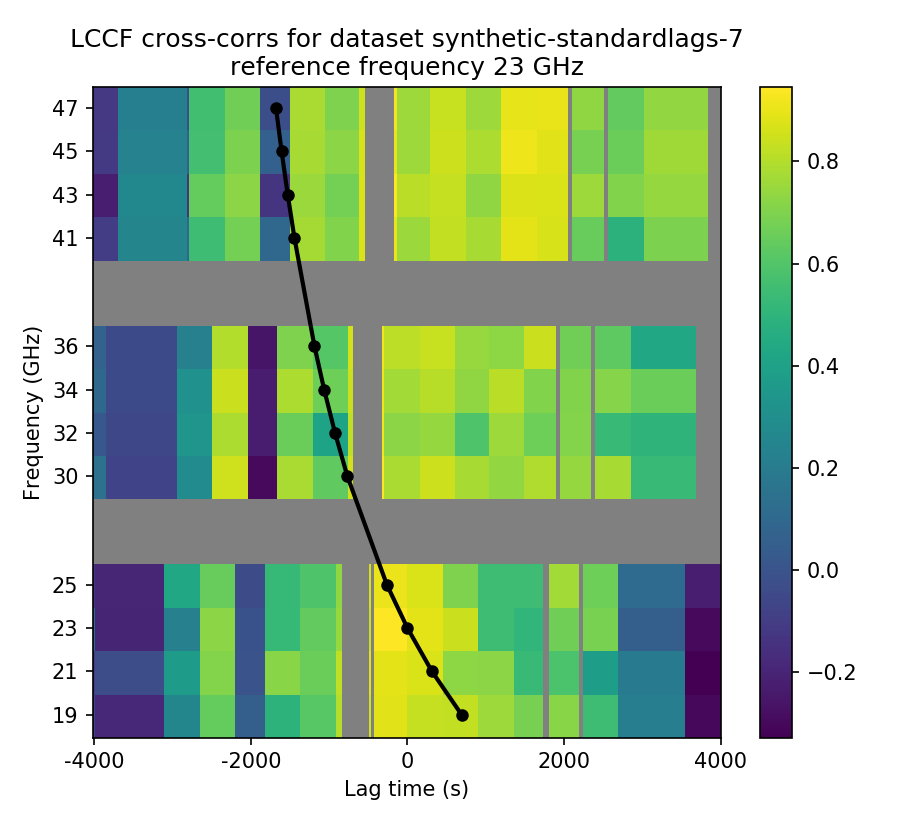}
\includegraphics[width=0.41\textwidth]{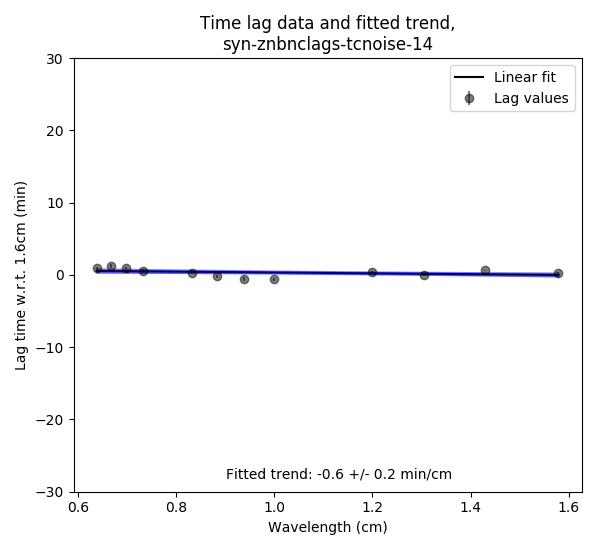}
\hspace{30pt}
\includegraphics[width=0.41\textwidth]{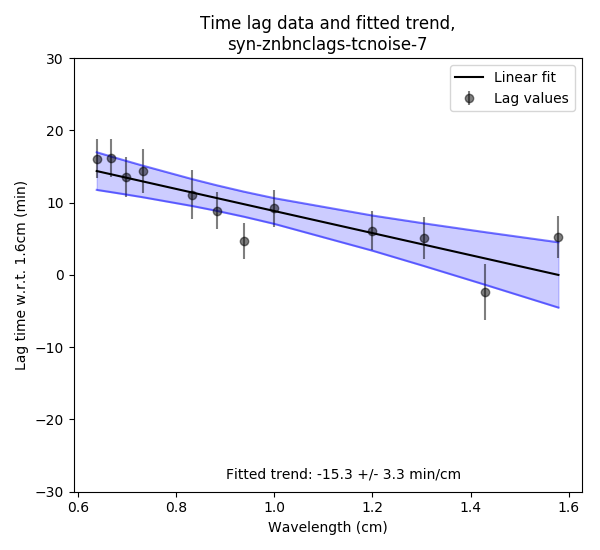}
\caption{Samples of results from synthetic light curves with the zero time lag relation and with calibration errors included. The left column shows data where the zero time lag trend was successfully recovered from the cross-correlations, the right column shows an example of a case where a time lag relation with a significant error is produced.}
\label{fig:synresults4}
\end{figure*}

\begin{figure*}[h]
\centering
\includegraphics[width=0.43\textwidth]{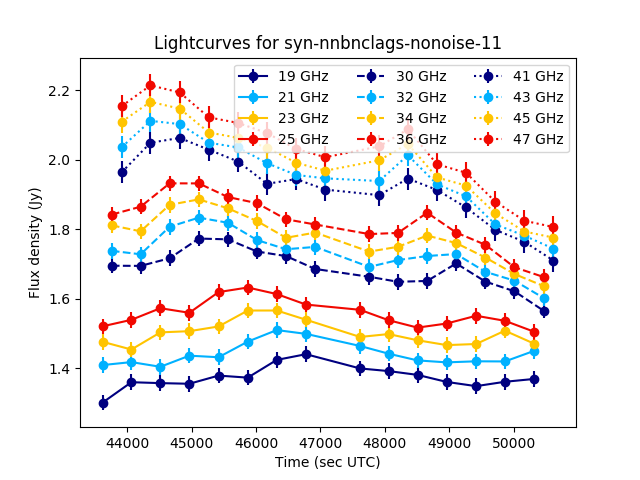}
\hspace{20pt}
\includegraphics[width=0.43\textwidth]{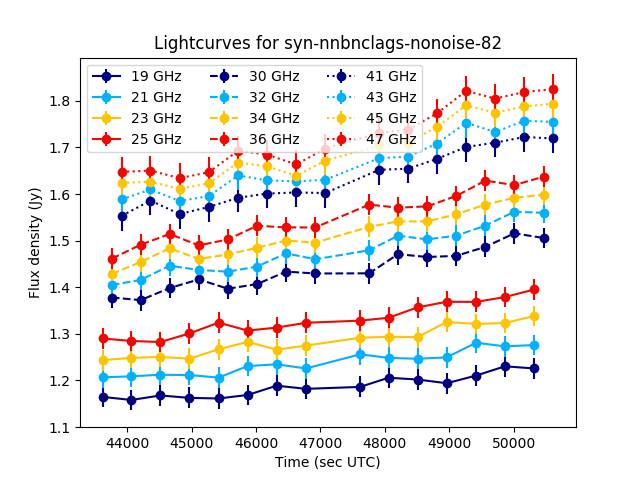}
\includegraphics[width=0.49\textwidth]{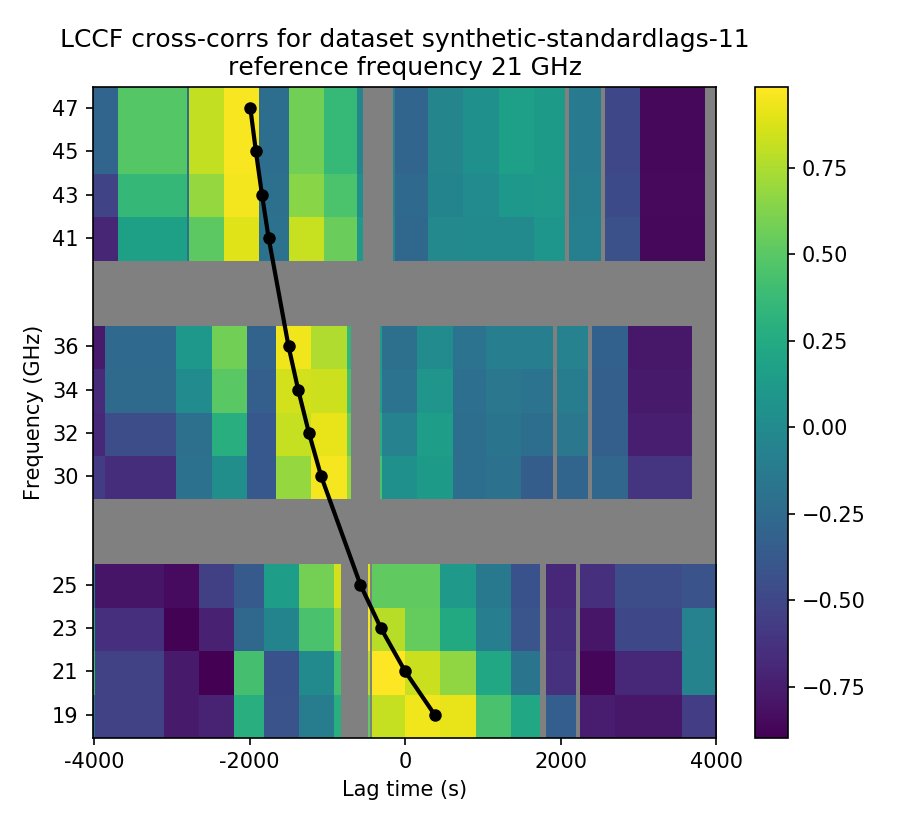}
\includegraphics[width=0.49\textwidth]{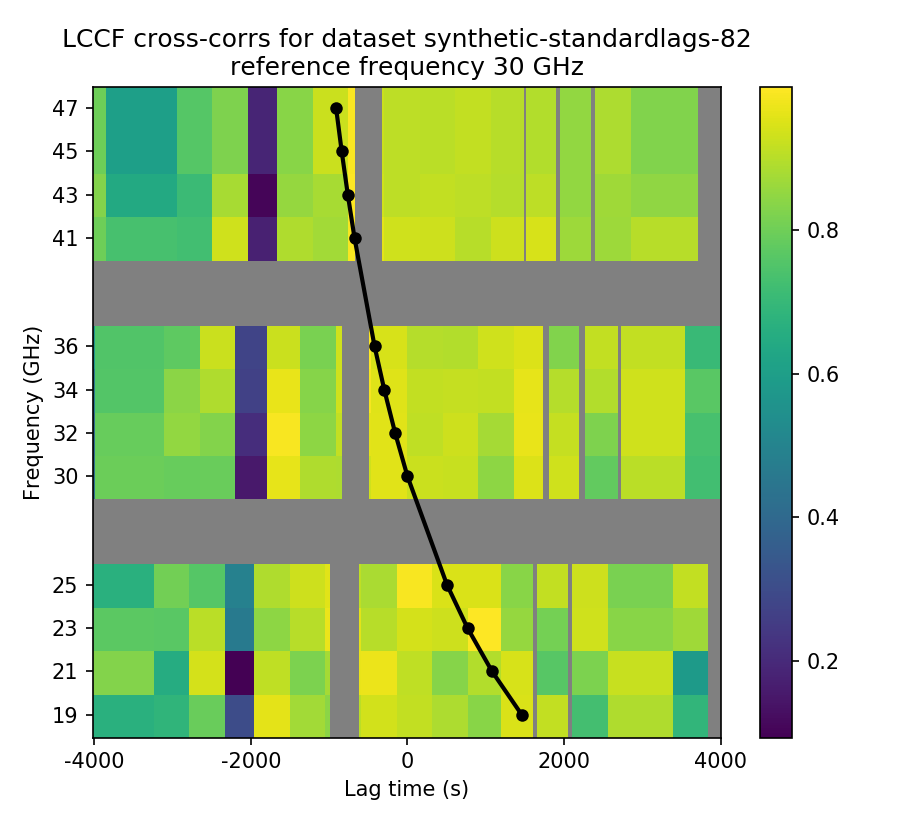}
\includegraphics[width=0.41\textwidth]{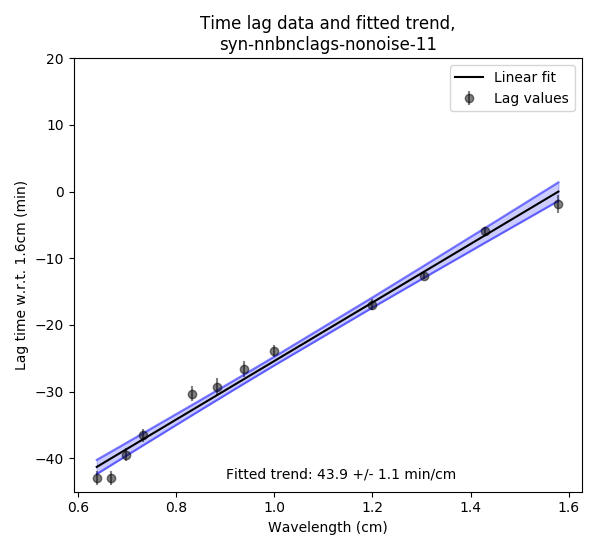}
\hspace{30pt}
\includegraphics[width=0.41\textwidth]{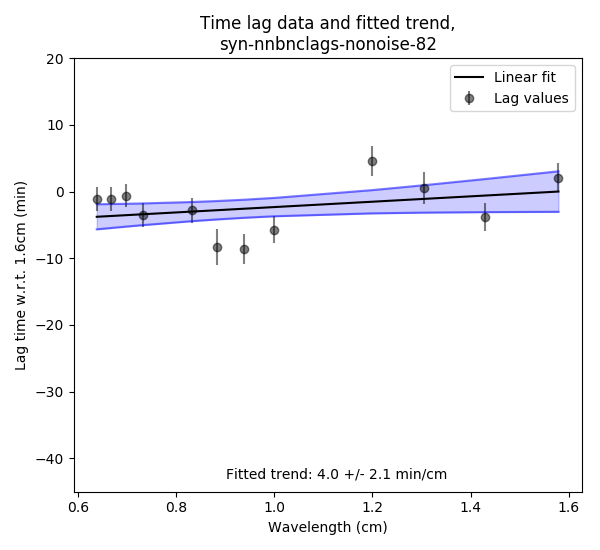}
\caption{Samples of results from synthetic light curves with the nominal time lag relation and no calibration errors. The left column shows data where the nominal time lag trend was successfully recovered from the cross-correlations, the right column shows an example of a case where a time lag relation with a significant error is produced.}
\label{fig:synresults5}
\end{figure*}

\begin{figure*}[h]
\centering
\includegraphics[width=0.43\textwidth]{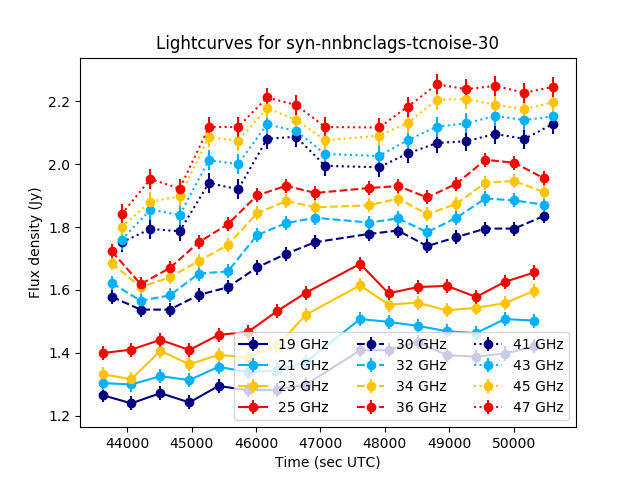}
\hspace{20pt}
\includegraphics[width=0.43\textwidth]{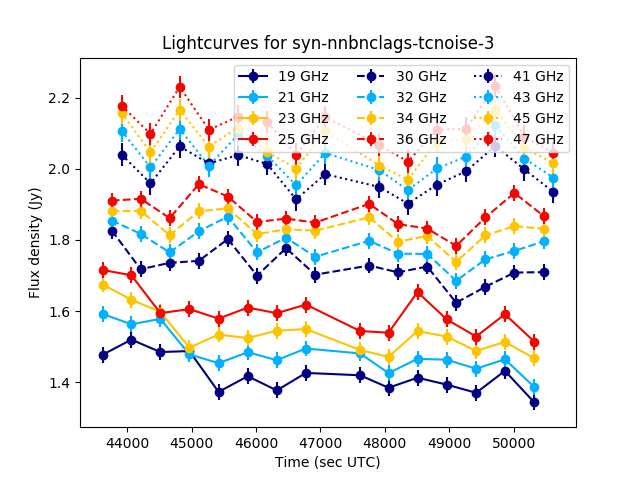}
\includegraphics[width=0.49\textwidth]{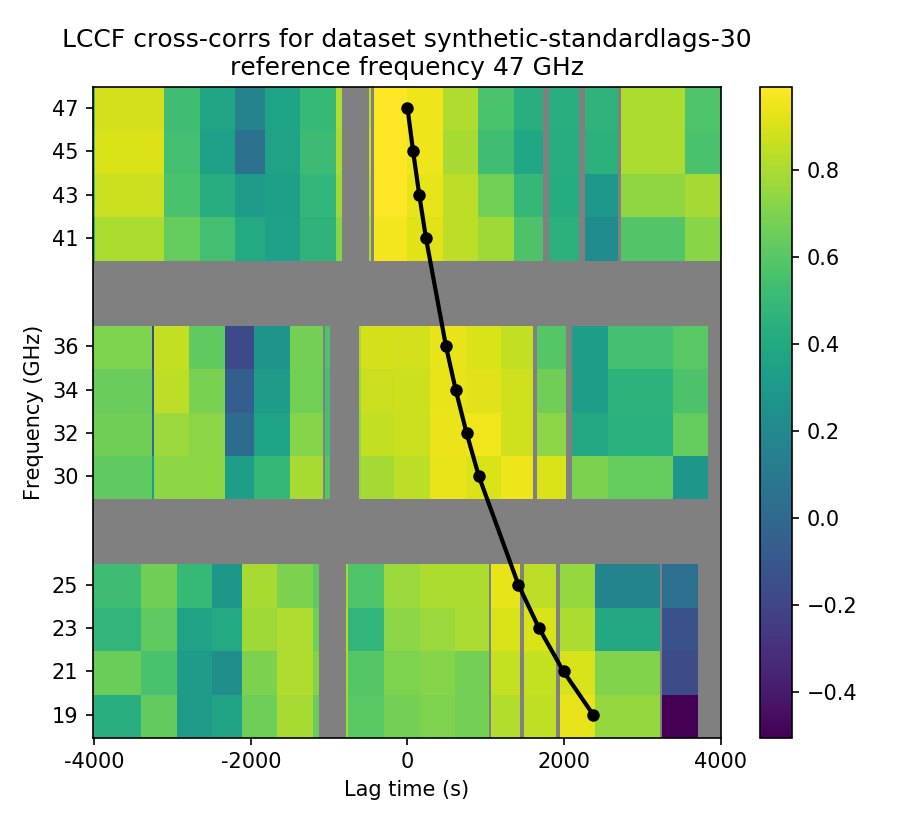}
\includegraphics[width=0.49\textwidth]{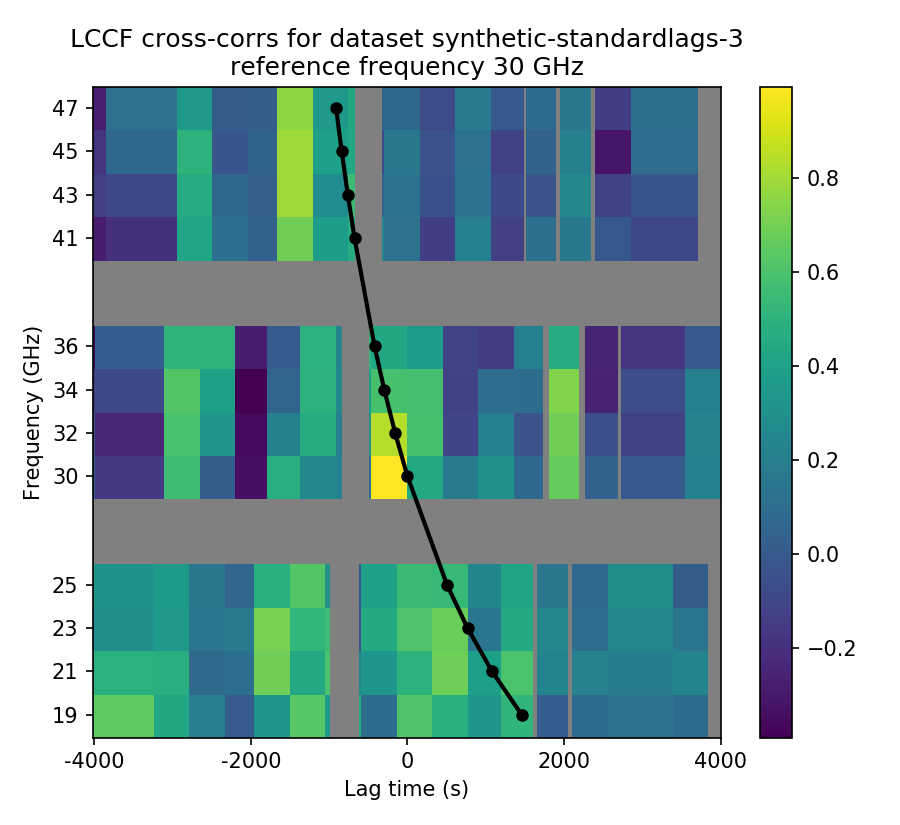}
\includegraphics[width=0.41\textwidth]{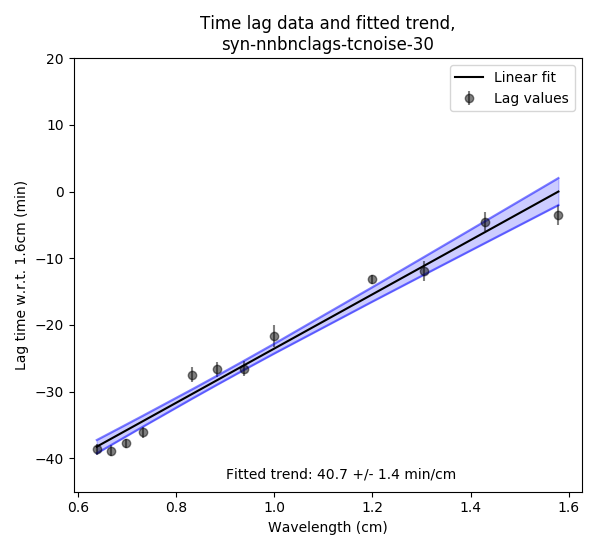}
\hspace{30pt}
\includegraphics[width=0.41\textwidth]{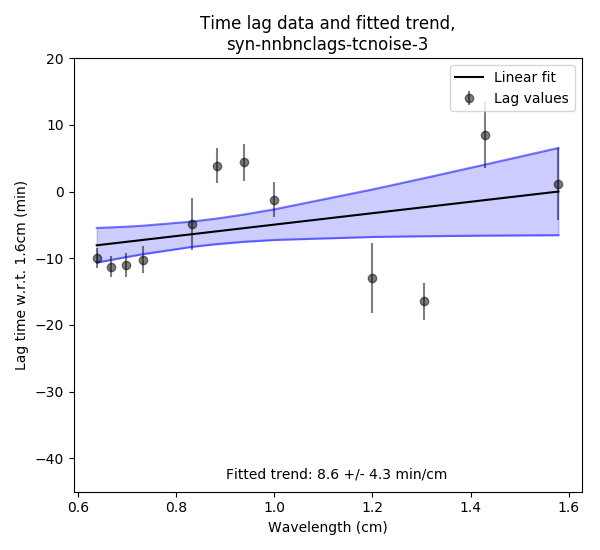}
\caption{Samples of results from synthetic light curves with the nominal time lag relation and with calibration errors included. The left column shows data where the nominal time lag trend was successfully recovered from the cross-correlations, the right column shows an example of a case where a time lag relation with a significant error is produced.}
\label{fig:synresults6}
\end{figure*}

\clearpage
\section*{Appendix D: historical spectral measurements for Sgr\,A*}

\begin{figure*}[h]
\centering
\includegraphics[width=0.9\textwidth]{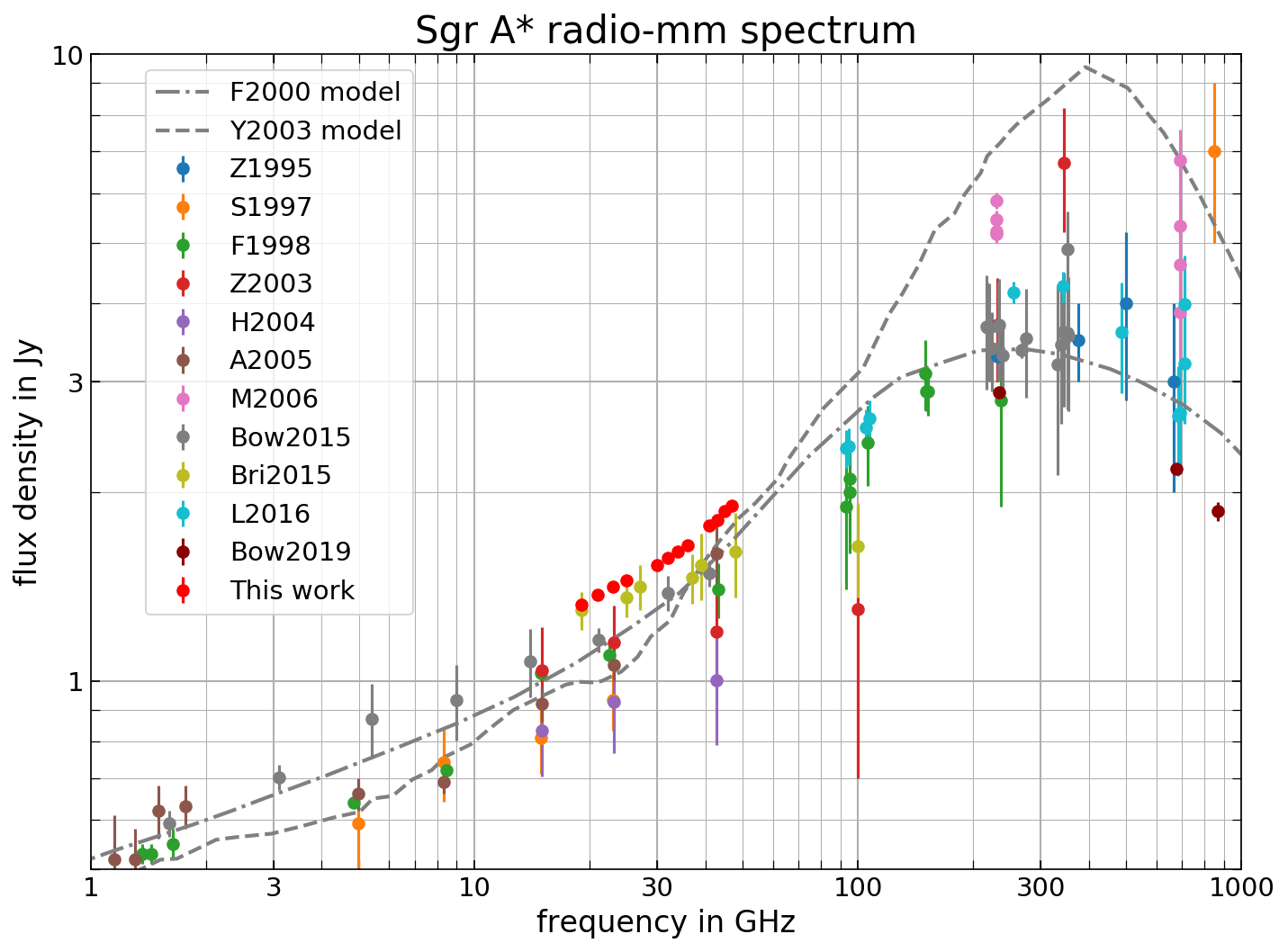}
\caption{The full set of historical data used for the binned plot in Figure \ref{fig:sgra-history}. Measurements are from \citet{Zylka1995, Serabyn1997, Falcke1998, Zhao2003, Herrnstein2004, An2005, Marrone2006, Bower2015, CDB2015, Liu2016, Bower2019}. Theoretical model spectra are from \citet{Falcke2000} and \citet{Yuan2003}.}
\label{fig:sgrafullsed}
\end{figure*}

\clearpage
\section*{Appendix E: Finding the appropriate auto-correlation function for synthetic light curve data}

\begin{figure}[h]
    \centering
    \includegraphics[width=0.49\textwidth]{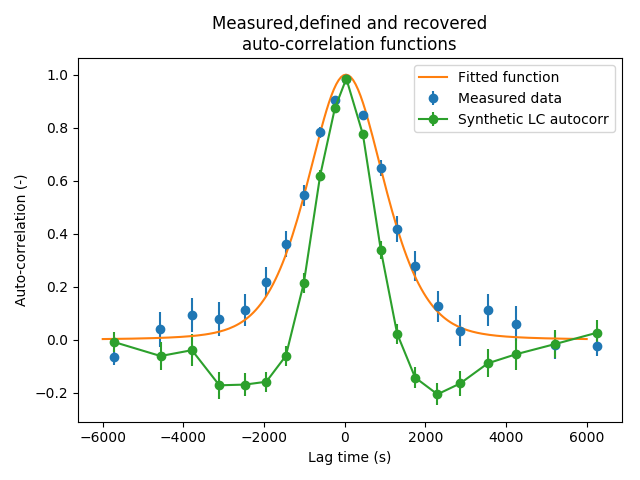}
    \includegraphics[width=0.49\textwidth]{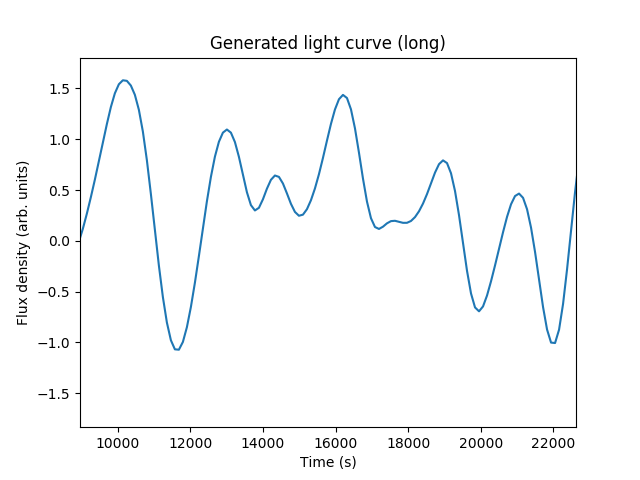}
    \caption{Left: The measured average autocorrelation function from all frequency bands in all data sets combined (separated markers), the functional form fitted to this autocorrelation function (continuous line), and the recovered autocorrelation function from a limited-length synthetic light curve generated using that functional form (connected markers). It is evident that the recovered auto-correlation differs from the prescribed form. Right: An example of a synthetic, longer-term, densely sampled master light curve generated using this initial choice for covariance function. Note that the function values can be negative, as the light curve has not been shifted and scaled to the correct average and variance at this point.}
\label{fig:firstautocorr}
\end{figure}

The synthetic data pipeline starts by defining the form of the auto-correlation function that appears to hold for our measured light curve data. As a starting point, we look at the auto-correlation function we measure for our original light curve data, when we consider the aggregate of light curves from all data sets and all frequency bands. This measured auto-correlation function is depicted in Figure \ref{fig:firstautocorr} using non-connected markers. To check for frequency dependence of the auto-correlation function, we compared the auto-correlation functions for the aggregate light curve data split into separate frequency bands against each other, and their shapes show no detectable frequency dependence outside the expected statistical variance.

We approximate the shape of this auto-correlation function by using a linear combination of a Gaussian (squared exponential) function and a rational quadratic function, both functions that often feature in the description of auto-correlation functions, ensuring that data points separated by progressively larger time lags tend to an uncorrelated state. This function is plotted in Figure \ref{fig:firstautocorr} with a continuous line. We find that its functional form is closely approximated by:

\begin{equation}
\textrm{Cov}(x) = 0.4 \cdot e^{-(x/1500)^2} + 0.6 \cdot \left(1+\left(\frac{x}{1600}\right)^2\right)^{-2},
\end{equation}

where $x$ is the lag time in seconds. The three free parameters that were varied to find the best fit are the strengths of the two components (with the constraint of summing to 1, which is the zero-lag auto-correlation) and the scalings of $x$ in both terms. Using this auto-correlation function as the covariance function for our (normalised) Gaussian process, we generate densely sampled synthetic time series as large as are allowed by our computational resources (see Figure \ref{fig:firstautocorr} for one realisation of such a light curve). These generated light curves are then shifted in time and re-scaled in amplitude to fit with the properties of the flux density for Sgr\,A* at various frequencies. We proceed to sample our synthetic data by applying the sampling cadence from the real-world measurements to a limited segment of this 'master' light curve.

We now check to see if we do indeed get auto-correlation functions from these sampled synthetic light curves that look like the auto-correlation function we expect. We plot the averaged recovered auto-correlation function from many realisations of the synthetic light curve, using the same sampling cadence as our originally measured data (Figure \ref{fig:firstautocorr}, connected markers) together with both the originally measured and the functionally prescribed auto-correlations, and we see that the recovered auto-correlation function is significantly narrower and even dips into negative values (indicating anti-correlation) for a range of time separations. This happens because we measure the auto-correlation of the sampled light curve using the average flux density of the limited time series instead of the 'ideal', long-term flux density average. We therefore tend to see more pairs of points with values of opposite signs (above vs. below the average value) when calculating our auto-correlation function, causing it to have a more rapid falloff with increasing time difference than we would see for longer light curves where the average is closer to the 'true' average. Effectively, we miss a low-frequency component of the auto-correlation function as we only have access to the short timescales when measuring it.

\begin{figure}[h]
    \centering
    \includegraphics[width=0.49\textwidth]{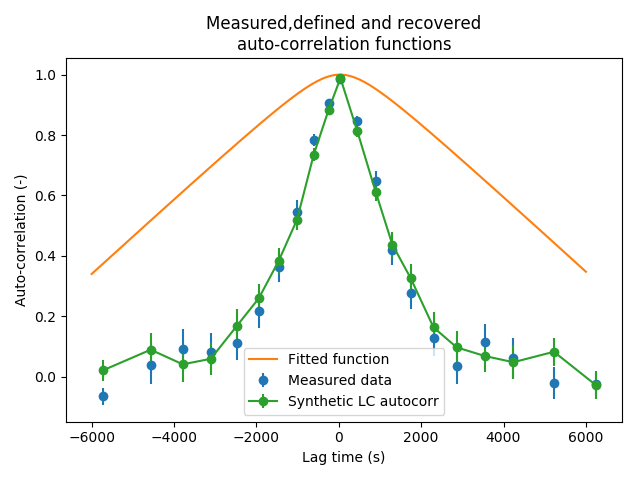}
    \includegraphics[width=0.49\textwidth]{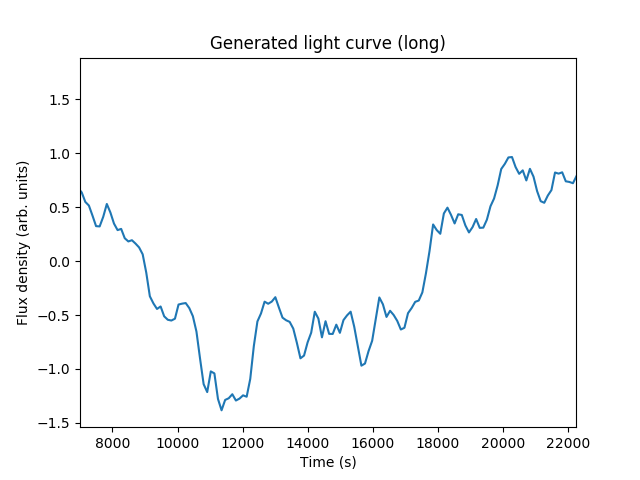}
    \caption{As Figure \ref{fig:firstautocorr}, but with an auto-correlation function chosen to optimise the match between the measured data and the synthetic data. Although the functional form of the auto-correlation function differs significantly from the measured one, the recovered function closely follows the measured function.}
\label{fig:secondautocorr}
\end{figure}

The functional form of the original auto-correlation function to use when generating the synthetic light curve data therefore needs to be chosen in such a way that the recovered shape of this function is the same as for our real measured light curves. To this end, the behaviour of several types of auto-correlation functions was investigated, with the function yielding the best results having the form of a truncated hyperbolic function. Note that this was a heuristic process where we investigated various functional forms in order to closely approach the measured auto-correlation curve with our reconstructed one. The result of this iterative process is shown in Figure \ref{fig:secondautocorr}. The functional form of our final auto-correlation function is given by:

\begin{equation}
\textrm{Cov}(x) = \textrm{max}\left(\frac{11}{10} - \sqrt{\left(\frac{x}{8000}\right)^2 + \frac{1}{100}}\,,0\right),
\end{equation}

where $x$ is again the lag time in seconds, and where the function value has a lower bound of zero imposed on it so that we avoid negative values for the auto-correlation function at large time lags.

With this auto-correlation function, we generate synthetic data that has the appropriate temporal variability characteristics to mimic the behaviour of our measured light curves.

\end{document}